\documentclass[aps,pra,reprint,showpacs]{revtex4-1}
\usepackage{graphics}
\usepackage{epsfig}
\usepackage{amsmath}
\usepackage{amssymb}
\DeclareMathOperator{\tr}{tr}
\DeclareMathOperator{\Det}{Det}
\DeclareMathOperator{\Conv}{Conv}
\DeclareMathOperator{\Extr}{Extr}

\DeclareMathOperator{\Lin}{Lin}
\DeclareMathOperator{\BiLin}{BiLin}

\newcommand{\field}[1]{\mathbb{#1}}
\newcommand{\LieGrp}[1]{\mathrm{#1}}

\newcommand{\ve}[1]{\mathbf{#1}}
\newcommand{\vs}[1]{\boldsymbol{#1}}
\newcommand{\vvs}[1]{\underline{\boldsymbol{#1}}}

\newcommand{\cc}[1]{{#1}^*}

\newcommand{\cmpl}[1]{\overline{#1}}
\newcommand{\pcmpl}[1]{\overline{#1}}

\newcommand{\nsubset}{\not\subset}

\newcommand{\cnvroof}[1]{{#1}^{\cup}}
\newcommand{\cncroof}[1]{{#1}^{\cap}}

\newcommand{\twoprt}[2]{{#1,#2}}

\newcommand{\ee}{\mathrm{e}}

\newcommand{\transp}[1]{#1^\mathrm{t}}
\newcommand{\ptransp}[2]{#1^{\mathrm{t}_{#2}}}

\newcommand{\cket}[1]{\vert #1 \rangle}
\newcommand{\bra}[1]{\langle #1 \vert}
\newcommand{\bracket}[1]{\langle #1 \rangle}
\newcommand{\Id}{\mathrm{I}}
\providecommand{\abs}[1]{\lvert#1\rvert}
\providecommand{\norm}[1]{\lVert#1\rVert}

\newcounter{txtitem}
\renewcommand{\thetxtitem}{(\roman{txtitem})}
\newcommand{\txtitem}{\refstepcounter{txtitem}\thetxtitem}

\begin{document}
%\title{Partial separability revisited}
\title{Partial separability revisited: Necessary and sufficient criteria}
%\title{Partial separability revisited -- extended classification with necessary and sufficient criteria}
\author{Szil\'ard Szalay}
\email{szalay@phy.bme.hu}
\author{Zolt\'an K\"ok\'enyesi}
\email{kokenyesi@phy.bme.hu}
\affiliation{
Department of Theoretical Physics, 
Institute of Physics, 
Budapest University of Technology and Economics, 
H-1111 Budapest, Budafoki \'ut 8, Hungary}
\date{\today}

\begin{abstract}
We extend the classification of mixed states of quantum systems
composed of arbitrary number of subsystems of arbitrary dimensions.
This extended classification is complete in the sense of partial separability
and gives $1+18+1$ partial separability classes in the tripartite case
contrary to a former $1+8+1$.
Then we give necessary and sufficient criteria for these classes,
which make it possible to determine to which class a mixed state belongs. 
These criteria are given by convex roof extensions of functions defined on pure states.
In the special case of three-qubit systems,
we define a different set of such functions
with the help of the Freudenthal triple system approach of three-qubit entanglement.
\end{abstract}
\pacs{
03.65.Ud, % Entanglement and quantum nonlocality (e.g. EPR paradox, Bell's inequalities, GHZ states, etc.)
03.67.Mn  % Entanglement measures, witnesses, and other characterizations
}

\maketitle{}

%*******************************************************************************
%*******************************************************************************
\section{Introduction}
\label{sec:Intro}

The notion of entanglement \cite{BengtssonZyczkowski,Horodecki4} 
was regarded by Schr\"odinger \cite{Schrodinger,Schrodinger2} to be the characteristic trait of quantum mechanics.
It serves as a resource for Quantum Information Theory \cite{NielsenChuang},
a relatively new field of research
dealing with the properties, characterization and applications
(mostly in quantum computation \cite{NielsenChuang})
of the nonlocal behavior of entangled quantum states.

For a multipartite quantum system being in a \emph{pure state},
it is easy to decide, in general, 
which subsystems are entangled with some of the others or, equivalently,
which subsystems can be separated from the others.
For a multipartite quantum system being in a \emph{mixed state},
however, this partial separability problem has not been considered in the full detail yet.
This problem is twofold.
Even if we have the \emph{definitions of the different classes}, 
which is not self-evident at all for more-than-two-partite systems,
\emph{deciding to which class a given state belongs} is also a nontrivial task.
In this paper, we work out solutions for both parts of this problem.
Considering the \emph{first part of this problem}, 
we extend the classification based on $k$-separability and $\alpha_k$-separability 
given by Seevinck and Uffink \cite{SeevinckUffinkMixSep},
which is the extension of the classification dealing only with $\alpha_k$-separability 
given by D\"ur and Cirac \cite{DurCiracTarrach3QBMixSep,DurCiracTarrachBMixSep}.
We discuss in detail the tripartite case,
then give the definitions for systems of arbitrary number of subsystems.
%We consider the \emph{second part of the problem} after a short detour.
Before we outline our solution for the \emph{second part of the problem},
we take a short detour.

If in the \emph{tripartite} case we restrict ourselves to \emph{qubits},
which is a relatively well-understood chapter of the theory of quantum entanglement,
some interesting results are known from the literature.
From the point of view of the present work, the most important ones are the following three.
First, 
\txtitem{} this is the system where the nontrivial structure of entanglement manifested itself for the first time
and it came to light that ``there are different kinds of entanglement'' of \emph{pure states} \cite{DurVidalCiracSLOCC3QB}.
Then, \txtitem{} these different kinds of pure-state entanglement \cite{DurVidalCiracSLOCC3QB}
give rise to classes of mixed state entanglement \cite{Acinetal3QBMixClass}.
On the other hand, 
\txtitem{} \label{text:FTS} recently a beautiful correspondence was found between the three-qubit Hilbert space and a particular FTS (Freudenthal Triple System),
a correspondence which is ``compatible'' with the entanglement of pure three-qubit states \cite{BorstenetalFreudenthal3QBEnt}.
Apart from these three, for the sake of completeness, we have to make mention of
\txtitem{} the famous phenomenon of monogamy of qubit systems, which was revealed first for three-qubit systems \cite{CKWThreetangle}
then shown for multiqubit systems \cite{OsborneVerstraeteMonogamy},
and \txtitem{} the interesting twistor-geometric approach of the entanglement of three-qubit systems \cite{PeterGeom3QBEnt}.

Item \ref{text:FTS} above gives us a hint of an answer to \emph{the second part of the problem} for three-qubit systems.
In the FTS approach of three-qubit entanglement some special quantities have appeared,
from which we gain real valued functions on pure states.
These functions have very useful vanishing properties,
which enable 
their convex roof extensions \cite{BennettetalMixedStates,UhlmannFidelityConcurrence,UhlmannConvRoofs}
to identify all the classes that our extended classification deals with.
On the other hand, it will be possible to define suitable functions 
for the identification of the classes in general,
for subsystems of arbitrary dimensions---moreover, for arbitrary number of subsystems---in another way 
than was done with the FTS approach working only for three qubits.
However, we will keep the considerations coming from the FTS approach,
because these considerations have given us the main ideas,
they have advantages for the case of three-qubits,
and, besides these, they are beautiful and interesting in themselves.

In the bipartite case, a state---either pure or mixed---can be either separable or entangled \cite{WernerSep},
and the vanishing of the convex roof extension of local entropies of pure states
is a \emph{necessary and sufficient criterion} of separability.
For us, this is the archetype of the general method of the detection of convex subsets by convex roof extensions.
However, for more-than-two-partite systems, 
the partial separability properties have a complicated structure,
and, to our knowledge, this method was not used.
Instead of that, 
the usual approach was the use of witness operators, as was done originally for three-qubit systems \cite{Acinetal3QBMixClass},
or other \emph{necessary but not sufficient criteria} for the detection of convex subsets \cite{Horodecki4,BengtssonZyczkowski,SzalaySepCrit,GuhneTothEntDet}.

Before starting, 
we review the classification schemes of states of multipartite quantum systems.
One of the main concepts here was the use of LOCC 
(Local Operations assisted by Classical Communication
\cite{BennettetalEquivalences})
either with certainty or with possibility, %nonzero probability, 
for the purpose of classification.
This concept has turned out to be useful in the restricted case when the input and output states are both pure.
First we recall the classification schemes dealing with LOCC.
For mixed states, only coarse-grained classifications are worked out,
which are recalled as well.

\emph{LOCC classification:}
Two states are equivalent under LOCC---they are in the same LOCC class---by definition
if they can be transformed to each other \emph{with certainty} by the use of LOCC.
For pure states, it turned out that
two states are equivalent under LOCC
if and only if they can be transformed into each other by LU (Local Unitary) transformations \cite{BennettetalEquivalences}.
So, for pure states, this gives the most fine-grained classification scheme imaginable.
Many continuous and discrete parameters are required to label the LOCC classes 
\cite{LindenPopescuOnMultipartEnt,AcinetalGenSchmidt3QB,Acinetal3QBPureCanon,Sudbery3qb,Kempe3qb}.
From the point of view of quantum computational purposes,
two LOCC-equivalent pure states can be used for exactly the same task.
However, to our knowledge, 
there is no such practical criterion of LOCC equivalence for mixed states
as the LU equivalence was for pure states.

\emph{SLOCC classification:}
A coarse-grained classification can be defined if we demand only the possibility of the transformation.
Two states are equivalent under SLOCC (Stochastic LOCC)---they are in the same SLOCC class---by definition
if they can be transformed into each other \emph{with non-zero probability} by the use of LOCC.
For pure states, it turned out that
two states are equivalent under SLOCC
if and only if they can be transformed into each other by LGL (Local General Linear) transformations \cite{DurVidalCiracSLOCC3QB}.
(Sometimes that was called ILO, stands for Invertible Local Operation \cite{DurVidalCiracSLOCC3QB},
but we prefer the uniform naming after the corresponding Lie groups.) 
So this gives a coarse-grained classification scheme for pure states.
In some cases, including the three-qubit case, only countable finite SLOCC classes arise \cite{DurVidalCiracSLOCC3QB}.
From the point of view of quantum computational purposes,
two SLOCC-equivalent pure states can be used for the same task but with a different probability of success.
Again, to our knowledge, 
there is no such practical criterion of SLOCC equivalence for mixed states
as the LGL equivalence was for pure states.

\emph{PS classification (Partial Separability):}
A more coarse-grained classification involves only the partial-separability properties.
This works for both pure and mixed states and gives only countably finite classes in both cases.
We elaborate this classification in detail in this paper for mixed states.
This classification deals with \emph{all the possible kinds of partial separability,}
which are of finite number,
whose special cases are the subsets of $k$-separability and $\alpha_k$-separability \cite{SeevinckUffinkMixSep}.
From the point of view of quantum computational purposes,
however, this classification is a bit too coarse grained,
since it does not make distinction among
pure states contained in
different SLOCC classes but having the same PS properties,
although these states may be suitable for different tasks.

\emph{PSS classification (Partial Separability extended by pure-state SLOCC classes):}
A cure for the problem above is another means of classification,
which was given by Ac\'in \textit{et.~al.}~\cite{Acinetal3QBMixClass}
only for three-qubit states.
Here, the starting point is the \emph{pure}-state SLOCC classes which are of finite number,
and the only difference between the partial separability classes and SLOCC classes
is the split of the three-qubit entangled class into two classes \cite{DurVidalCiracSLOCC3QB}.
The PSS classes arising from these classes for \emph{mixed} states
are the same for biseparability, and only the tripartite entangled set is divided into two classes.
This classification has the advantage of differentiating among different SLOCC classes of pure states,
and also among mixed states depending on which kind of pure entanglement is needed for the preparation of the state.
However, in the majority of the cases there are continuously infinite SLOCC classes of pure states 
labeled by more than one continuous parameter \cite{DurVidalCiracSLOCC3QB,VerstraeteetalSLOCC4QB,ChterentalDjokovicSLOCC4QB},
in which case it is not clear how this classification can be carried out,
if it can be at all.

The organization of this paper is as follows.
In the first half of the paper,
we work out the main concepts on three-qubit states.
%%%%%
In Sec.~\ref{sec:Pure}, we review the SLOCC classification of pure three-qubit states.
We recall the conventional LU invariants (in Sec.~\ref{subsec:Pure:ClassLU})
and the LSL tensors (Local Special Linear) of the FTS approach (in Sec.~\ref{subsec:Pure:ClassLSL})
by which the SLOCC classes can be identified.
Then we obtain a new set of LU invariants (in Sec.~\ref{subsec:Pure:NewInvs}) 
being necessary later for mixed states.
%%%%%
In Sec.~\ref{sec:Mixed}, we elaborate the 
PSS classification (which contains also the PS classification)
for mixed three-qubit states.
We define the PS(S) subsets (in Sec.~\ref{subsec:Mixed:Subsets})
and PS(S) classes (in Sec.~\ref{subsec:Mixed:Classes}).
Then we give the functions for the identification of the PS(S) classes
(in Sec.~\ref{subsec:Mixed:CRoof}).
%%%%%
In Sec.~\ref{sec:Xmpl}, we demonstrate the nonemptiness of some of the new classes
for the three-qubit case
%by explicit (in subsection \ref{subsec:Xmpl:Explicit})
%and numerical examples (in subsection \ref{subsec:Xmpl:Numerical}).
by explicit examples.
%%%%%
In Sec.~\ref{sec:GenThreePart}, we generalize the functions for the case of three subsystems of arbitrary dimensions.
First we see how far the method coming from the FTS approach can go (in Sec.~\ref{subsec:GenThreePart:FTS});
then we formulate a more general set of functions working without limitations (in Sec.~\ref{subsec:GenThreePart:nFTS}).
%%%%%
In Sec.~\ref{sec:Gen}, we generalize the construction for the case of arbitrary number of subsystems of arbitrary dimensions.
We work out the labeling of the PS subsets (in Sec.~\ref{subsec:Gen:PSsubsets})
along with the PS classes 
and give a general conjecture about their nonemptiness (in Sec.~\ref{subsec:Gen:PSclasses}).
Then we construct 
the functions identifying the PS subsets and classes 
with the minimal requirements (in Sec.~\ref{subsec:Gen:Indicators}),
as well as with stronger requirements leading to entanglement-monotone functions 
(in Sec.~\ref{subsec:Gen:monIndicators}).
%%%%%
In Sec.~\ref{sec:Sum}, we give a summary, some remarks, and open questions.
%%%%%
Some technicalities about the new set of three-qubit LU invariants
and proofs of some statements about the general construction
are left to Appendixes \ref{appsec:explicit}
and \ref{appsec:Gen}.

%*******************************************************************************
%*******************************************************************************
\section{Pure three-qubit states}
\label{sec:Pure}

Before starting,
we set some conventions that are very convenient for the tripartite case.
The labels of the subsystems are the numbers $1$, $2$, and $3$,
while the letters $a$, $b$, and $c$ are variables 
taking their values in the set of labels $\{1,2,3\}$.
When $a$, $b$, and $c$ appear together in a formula,
they form a partition of $\{1,2,3\}$,
so they take always different values,
and \emph{the formula is understood for all the different values of these variables automatically.}
(However, sometimes a formula is symmetric under the interchange of two such variables
in which case we keep only one of the identical formulas.)

The Hilbert space of a three-qubit system is
$\mathcal{H}=\mathcal{H}^1\otimes\mathcal{H}^2\otimes\mathcal{H}^3$,
where, after the choice of an orthonormal basis $\{\cket{0},\cket{1}\}\subset\mathcal{H}^a$, $\mathcal{H}^a\cong\field{C}^2$.
The $\cket{\psi}\in\mathcal{H}$ state vectors
are not required to be normalized in this section,
so the $0\in\mathcal{H}$ zero vector is also allowed.
(The physical states arise, however, from normalized vectors.)

%*******************************************************************************
\subsection{SLOCC classification by LU-invariants}
\label{subsec:Pure:ClassLU}

It is a well-known and celebrated result of D\"ur, Vidal, and Cirac \cite{DurVidalCiracSLOCC3QB} that
``three qubits can be entangled in two inequivalent ways.'' 
More fully, there are $1+1+3+1+1$ three-qubit SLOCC classes,
that is, subsets invariant under LGL transformations.
\begin{itemize}
\item $\mathcal{V}^\text{Null}$ (class Null): The zero-vector of $\mathcal{H}$.
\item $\mathcal{V}^{1|2|3}$ (class $1|2|3$): These vectors are fully separable, which are of the form
$\cket{\psi_1}\otimes\cket{\psi_2}\otimes\cket{\psi_3}$.
\item $\mathcal{V}^{a|bc}$ (three biseparable classes $a|bc$),
for example,
$\cket{\psi_1}\otimes\cket{\psi_{23}}\in\mathcal{V}^{1|23}$, where 
$\cket{\psi_{23}}\neq \cket{\psi_2}\otimes\cket{\psi_3}$.
For such $\cket{\psi_{23}}$, a representative element is the standard B (Bell) state,
\begin{subequations}
\begin{equation}
\label{eq:B}
\cket{\text{B}}=\frac{1}{\sqrt{2}}\bigl(\cket{00}+\cket{11}\bigr).
\end{equation}
\item $\mathcal{V}^\text{W}$ (Class W): 
This is the first class of genuine tripartite entanglement,
when no subsystem can be separated from the others.
A representative element is the standard W state,
\begin{equation}
\label{eq:W}
\cket{\text{W}}=\frac{1}{\sqrt{3}}\bigl(\cket{100}+\cket{010}+\cket{001}\bigr).
\end{equation}
\item $\mathcal{V}^\text{GHZ}$ (Class GHZ): 
This is the second class of genuine tripartite entanglement,
the class of Greenberger-Horne-Zeilinger-type entanglement.
A representative element is the standard GHZ state,
\begin{equation}
\label{eq:GHZ}
\cket{\text{GHZ}}=\frac{1}{\sqrt{2}}\bigl(\cket{000}+\cket{111}\bigr).
\end{equation}
\end{subequations}
\end{itemize}
Formally speaking,
these classes define disjoint, LGL-invariant subsets of $\mathcal{H}$,
and cover $\mathcal{H}$ entirely:
$\mathcal{H}=
\mathcal{V}^\text{Null}
\cup\mathcal{V}^{1|2|3}
\cup\mathcal{V}^{1|23}
\cup\mathcal{V}^{2|13}
\cup\mathcal{V}^{3|12}
\cup\mathcal{V}^\text{W}
\cup\mathcal{V}^\text{GHZ}$.
Except $\mathcal{V}^\text{Null}$, these classes are not closed. 
For the partial separability issues, we define
$\mathcal{V}^{123}=\mathcal{V}^\text{W}\cup\mathcal{V}^\text{GHZ}$.

For any $\cket{\psi}\in\mathcal{H}$,
it can be determined to which class $\cket{\psi}$ belongs
by the vanishing of the following quantities:
the norm,
\begin{subequations}
%\label{eq:LUdets}
\label{eq:pureLUinvs}
\begin{equation}
\label{eq:pureLUinvs:n}
n(\psi)= \norm{\psi}^2,
\end{equation}
the local entropies,
\begin{equation}
\label{eq:pureLUinvs:sa}
s_a(\psi)= 4 \det \bigl[\tr_{bc} \bigl(\cket{\psi}\bra{\psi}\bigr)\bigr],
\end{equation}
[here we use a normalized quantum-Tsallis entropy of parameter $2$
(see in Sec.~\ref{subsec:GenThreePart:FTS}),
although every entropy does the job,
since they vanish only for pure density matrices]
and the three-tangle,
\begin{equation}
\label{eq:pureLUinvs:tau}
\tau(\psi)=4\abs{\Det(\psi)},
\end{equation}
\end{subequations}
which is given by Cayley's hyperdeterminant $\Det(\psi)$ 
\cite{CayleyHDet,GelfandetalDiscriminants,CKWThreetangle}.
All of these quantities are LU invariants,
[which is $\LieGrp{U}(2)^{\times3}$ in this case,]
moreover, $n$ is invariant under the larger group $\LieGrp{U}(8)$
and $\tau$ under $\bigl[\LieGrp{U}(1)\times\LieGrp{SL}(2,\field{C})\bigr]^{\times3}
\cong\LieGrp{U}(1)\times\LieGrp{SL}(2,\field{C})^{\times3}$.
It follows from the invariance properties and other observations \cite{DurVidalCiracSLOCC3QB}
that the SLOCC classes of pure three-qubit states
can be determined by the vanishing of these quantities
in the way which can be seen in Table \ref{tab:pureSLOCC}.
\begin{table}
\begin{tabular}{c||c|ccc|c}
Class                     & $n(\psi)$  & $s_1(\psi)$ & $s_2(\psi)$ & $s_3(\psi)$ & $\tau(\psi)$ \\
\hline
\hline
$\mathcal{V}^\text{Null}$ & $=0$       & $=0$        & $=0$        & $=0$        & $=0$         \\
\hline
$\mathcal{V}^{1|2|3}$     & $>0$       & $=0$        & $=0$        & $=0$        & $=0$         \\
\hline
$\mathcal{V}^{1|23}$      & $>0$       & $=0$        & $>0$        & $>0$        & $=0$         \\
$\mathcal{V}^{2|13}$      & $>0$       & $>0$        & $=0$        & $>0$        & $=0$         \\
$\mathcal{V}^{3|12}$      & $>0$       & $>0$        & $>0$        & $=0$        & $=0$         \\
\hline
$\mathcal{V}^\text{W}$    & $>0$       & $>0$        & $>0$        & $>0$        & $=0$         \\
$\mathcal{V}^\text{GHZ}$  & $>0$       & $>0$        & $>0$        & $>0$        & $>0$   
\end{tabular}
\caption{SLOCC classes of three-qubit state vectors
identified by the vanishing of LU-invariants (\ref{eq:pureLUinvs}).}
\label{tab:pureSLOCC}
\end{table}

Our aim is the characterization of the \emph{mixed states}
by the vanishing of some quantities,
in a similar way
that the conditions in Table \ref{tab:pureSLOCC} for the quantities in (\ref{eq:pureLUinvs}) characterize the \emph{pure states.}
To obtain such a characterization scheme,
we need, 
on the one hand, the generalization of classes defined somehow,
and, on the other hand, a suitable set of quantities which
are vanishing for some classes determined somehow and nonvanishing for the others.
These two issues are strongly related,
and it will turn out that 
we can define a set of quantities which suits well the classification given by Seevinck and Uffink \cite{SeevinckUffinkMixSep},
but a ``more complete'' set of quantities suits well an extended but still relevant classification,
which are elaborated in Sec.~\ref{sec:Mixed}.

%*******************************************************************************
\subsection{SLOCC classification by LSL-covariants}
\label{subsec:Pure:ClassLSL}

In \cite{BorstenetalFreudenthal3QBEnt},
Borsten \textit{et.~al.}~revealed a very elegant correspondence
between 
the three-qubit Hilbert space $\mathcal{H}\cong\field{C}^2\otimes\field{C}^2\otimes\field{C}^2$
and 
the FTS (Freudenthal Triple System) $\mathfrak{M}(\mathcal{J})\cong\field{C}\oplus\field{C}\oplus\mathcal{J}\oplus\mathcal{J}$
over the cubic Jordan algebra $\mathcal{J}\cong\field{C}\oplus\field{C}\oplus\field{C}$.
The fundamental point of this correspondence is
that the automorphism group of this FTS is
$\LieGrp{Aut}[\mathfrak{M}(\field{C}\oplus\field{C}\oplus\field{C})]=\LieGrp{SL}(2,\field{C})^{\times3}$,
which is just the relevant LSL subgroup 
of $\LieGrp{GL}(2,\field{C})^{\times3}$, the LGL-group of SLOCC equivalence for three-qubit pure states.
(This group-theoretical coincidence arises only in the three-qubit case.)
It has been shown \cite{BorstenetalFreudenthal3QBEnt} that
the vectors of \emph{different SLOCC classes of entanglement} in the three-qubit Hilbert space
are in one-to-one correspondence with
the elements of \emph{different rank} in the FTS.
The rank of an element of an FTS
is characterized by the vanishing of some associated elements, 
which are covariant---maybe invariant---under the action of the automorphism group,
resulting in conditions for the SLOCC classes in the Hilbert-space by the vanishing or non vanishing of
$\LieGrp{SL}(2,\field{C})^{\times3}$ tensors.
Hence, this classification is manifestly invariant under SLOCC equivalence \cite{BorstenetalFreudenthal3QBEnt},
which cannot be seen directly in the conventional classification,
since the $s_a$ local entropies are scalars only under $\LieGrp{U}(2)^{\times3}$.
(However, the invariance of the vanishing of the functions $s_a$ follows easily from the fact 
that the local rank is invariant under invertible transformations \cite{DurVidalCiracSLOCC3QB}.)

Let the three-qubit state $\cket{\psi}\in\mathcal{H}$ be expressed 
in the computational basis $\{\cket{ijk}=\cket{i}\otimes\cket{j}\otimes\cket{k}\}$
as
\begin{equation*}
\cket{\psi}=\sum_{i,j,k=0}^1\psi^{ijk}\cket{ijk}.
\end{equation*}
We can assign an element $\psi\in\mathfrak{M}(\mathcal{\field{C}\oplus\field{C}\oplus\field{C}})$ to this
and calculate some associated quantities
needed for the identification of its rank.
Here we list these quantities 
in the form in which we use them:
\begin{subequations}
\label{eq:tensors}
\begin{align}
%\Upsilon_\phi(\psi) &= 3T(\psi,\psi,\phi) + \{\psi,\phi\}\psi\\
\label{eq:tensors:Upsilon}
\begin{split}
[\Upsilon_\phi(\psi)]^{ijk} = 
 &-\varepsilon_{ll'}\varepsilon_{mm'}\varepsilon_{nn'}\psi^{imn}\psi^{lm'n'}\phi^{l'jk}\\
 &-\varepsilon_{mm'}\varepsilon_{nn'}\varepsilon_{ll'}\psi^{ljn}\psi^{l'mn'}\phi^{im'k}\\
 &-\varepsilon_{nn'}\varepsilon_{ll'}\varepsilon_{mm'}\psi^{lmk}\psi^{l'm'n}\phi^{ijn'},
\end{split}\\
\label{eq:tensors:gamma1}
[\gamma_1(\psi)]^{ii'}=&\; \varepsilon_{jj'}\varepsilon_{kk'}\psi^{ijk}\psi^{i'j'k'},\\
\label{eq:tensors:gamma2}
[\gamma_2(\psi)]^{jj'}=&\; \varepsilon_{kk'}\varepsilon_{ii'}\psi^{ijk}\psi^{i'j'k'},\\
\label{eq:tensors:gamma3}
[\gamma_3(\psi)]^{kk'}=&\; \varepsilon_{ii'}\varepsilon_{jj'}\psi^{ijk}\psi^{i'j'k'},\\
\label{eq:tensors:T}
\begin{split}
[T(\psi,\psi,\psi)]^{ijk}
=&-\varepsilon_{ll'}\varepsilon_{mm'}\varepsilon_{nn'}\psi^{imn}\psi^{lm'n'}\psi^{l'jk}\\
=&-\varepsilon_{mm'}\varepsilon_{nn'}\varepsilon_{ll'}\psi^{ljn}\psi^{l'mn'}\psi^{im'k}\\
=&-\varepsilon_{nn'}\varepsilon_{ll'}\varepsilon_{mm'}\psi^{lmk}\psi^{l'm'n}\psi^{ijn'},
\end{split}\\
\begin{split}
\label{eq:tensors:q}
q(\psi) =&\;
\varepsilon_{ii'}\varepsilon_{jj'}
\varepsilon_{kk'}\varepsilon_{ll'}
\varepsilon_{mm'}\varepsilon_{nn'}\\
&\qquad\times\psi^{ikl}\psi^{jk'l'}\psi^{i'mn}\psi^{j'm'n'}.
\end{split}
\end{align}
\end{subequations}
(For the basic definitions 
of Jordan algebras, Freudenthal triple systems
and the operations and maps defined on them,
see in \cite{BorstenetalFreudenthal3QBEnt} and in the references therein.)
Here the summation for the pairs of indices occurring upstairs and downstairs 
are understood, and
\begin{equation*}
\varepsilon_{ii'}=\begin{bmatrix}        
0&1\\  
-1&0
\end{bmatrix}
\end{equation*}
is the matrix of the 
$\LieGrp{Sp}(1)\cong\LieGrp{SL}(2)$-invariant non-degenerate antisymmetric bilinear form:
Since $\transp{M}\varepsilon M= \varepsilon \det(M)$,
index contraction by $\varepsilon$ is invariant under $\LieGrp{SL}(2,\field{C})$ transformations.
This shows that
if we regard $\psi$ and $\phi$ as tensors that
transform as a $(\mathbf{2},\mathbf{2},\mathbf{2})$
under $\LieGrp{SL}(2,\field{C})^{\times3}$,
then so do $\Upsilon_\phi(\psi)$ and $T(\psi,\psi,\psi)$,
while $\gamma_1(\psi)$, $\gamma_2(\psi)$, and $\gamma_3(\psi)$, being symmetric, transform as 
$(\mathbf{3},\mathbf{1},\mathbf{1})$,
$(\mathbf{1},\mathbf{3},\mathbf{1})$, and
$(\mathbf{1},\mathbf{1},\mathbf{3})$, respectively,
and $q(\psi)$ transforms as $(\mathbf{1},\mathbf{1},\mathbf{1})$;
that is, it is scalar.
[Note that for any $2\times2$ matrix $M$, the determinant
$2\det(M)=\varepsilon_{ii'}\varepsilon_{jj'}M^{ij}M^{i'j'}$,
so $2\det[\gamma_a(\psi)]=q(\psi)$.]

The main result of \cite{BorstenetalFreudenthal3QBEnt} is 
that the conditions for the SLOCC classes can be formulated by the vanishing of these tensors
in the way which can be seen in Table \ref{tab:pureSLOCC2}.
%%%%%%%%%%%%%%%%%%%%%%%%%%%%%%%%%%%%%%%%
\begin{table}
\begin{tabular}{c||c|c|ccc|c|c}
Class  & $\psi$  & $\Upsilon_\phi(\psi)$    & $\gamma_1(\psi)$ & $\gamma_2(\psi)$ & $\gamma_3(\psi)$ &  $T(\psi,\psi,\psi)$  & $q(\psi)$ \\
\hline
\hline
$\mathcal{V}^\text{Null}$ & $=0$    & $=0,\forall\phi$         & $=0$     & $=0$     & $=0$     & $=0$     & $=0$ \\
\hline
$\mathcal{V}^{1|2|3}$     & $\neq0$ & $=0,\forall\phi$         & $=0$     & $=0$     & $=0$     & $=0$     & $=0$ \\
\hline
$\mathcal{V}^{1|23}$      & $\neq0$ & $\neq0,\exists\phi$      & $\neq0$  & $=0$     & $=0$     & $=0$     & $=0$ \\
$\mathcal{V}^{2|13}$      & $\neq0$ & $\neq0,\exists\phi$      & $=0$     & $\neq0$  & $=0$     & $=0$     & $=0$ \\
$\mathcal{V}^{3|12}$      & $\neq0$ & $\neq0,\exists\phi$      & $=0$     & $=0$     & $\neq0$  & $=0$     & $=0$ \\
\hline
$\mathcal{V}^\text{W}$    & $\neq0$ & $\neq0,\exists\phi$      & $\neq0$  & $\neq0$  & $\neq0$  & $\neq0$  & $=0$ \\
$\mathcal{V}^\text{GHZ}$  & $\neq0$ & $\neq0,\exists\phi$      & $\neq0$  & $\neq0$  & $\neq0$  & $\neq0$  & $\neq0$
\end{tabular}
\caption{SLOCC classes of three-qubit state vectors
 identified by the vanishing of LSL-covariants (\ref{eq:tensors}).}
\label{tab:pureSLOCC2}
\end{table}
%%%%%%%%%%%%%%%%%%%%%%%%%%%%%%%%%%%%%%%%
In the light of the conditions by the norm and the four determinants, (see in Table \ref{tab:pureSLOCC}),
this scheme constructed by seven quantities seems to be redundant.
Indeed, it is redundant for pure states,
%but it will turn out that this shows us the way towards the generalization for mixed states.
but it will turn out that this way leads to the generalization for mixed states.

%*******************************************************************************
\subsection{SLOCC classification by a new set of LU-invariants}
\label{subsec:Pure:NewInvs}

To follow this way,
we need quantities which can be extended from pure states to mixed states
by the convex roof construction \cite{BennettetalMixedStates,UhlmannFidelityConcurrence,UhlmannConvRoofs}.
There is no natural ordering on the tensors of (\ref{eq:tensors}) %$\mathfrak{M}(\mathcal{J})$,
so convex roof construction does not work directly for them,
but we can form quantities from them taking values in the field of real numbers.
During this, 
we  lose the covariance under $\LieGrp{SL}(2,\field{C})^{\times3}$,
but gain the invariance under the group $\LieGrp{U}(2)^{\times3}$.

Returning from the FTS language to the Hilbert space language,
we have ``state vectors''
\begin{align*}
\cket{\Upsilon_\phi(\psi)}&=\sum_{i,j,k=0}^1[\Upsilon_\phi(\psi)]^{ijk}\cket{ijk}\in\mathcal{H},\\
\cket{T(\psi,\psi,\psi)}&=\sum_{i,j,k=0}^1[T(\psi,\psi,\psi)]^{ijk}\cket{ijk}\in\mathcal{H},
\end{align*}
and ``local operators''
\begin{align*}
\gamma_1(\psi)\varepsilon&=\sum_{i,i'=0}^1[\gamma_1(\psi)\varepsilon]^i_{\;i'}\cket{i}\bra{i'}\in\Lin(\mathcal{H}^1),\\
\gamma_2(\psi)\varepsilon&=\sum_{j,j'=0}^1[\gamma_2(\psi)\varepsilon]^j_{\;j'}\cket{j}\bra{j'}\in\Lin(\mathcal{H}^2),\\
\gamma_3(\psi)\varepsilon&=\sum_{k,k'=0}^1[\gamma_3(\psi)\varepsilon]^k_{\;k'}\cket{k}\bra{k'}\in\Lin(\mathcal{H}^3),
\end{align*}
associated with $\cket{\psi}\in\mathcal{H}$ through (\ref{eq:tensors}).
These are just computational auxiliaries,
not state vectors and local operators in the ordinary sense,
because they depend nonlinearly on the state vector $\cket{\psi}$.
[Note that 
$\varepsilon   \in \mathcal{H}^{a*}\otimes\mathcal{H}^{a*}
              \cong\Lin(\mathcal{H}^a\to\mathcal{H}^{a*})
              \cong\BiLin(\mathcal{H}^a\times\mathcal{H}^a\to\field{C})$,
while
$\gamma_a(\psi)\in  \mathcal{H}^a\otimes\mathcal{H}^a
              \cong\Lin(\mathcal{H}^{a*}\to\mathcal{H}^a)$,
so $\gamma_a(\psi)\varepsilon\in\Lin(\mathcal{H}^a\to\mathcal{H}^a)$.]

Now, the vanishing conditions of the tensors (\ref{eq:tensors}) in Table \ref{tab:pureSLOCC2} 
can be reformulated.
Clearly, $\psi=0$ if and only if $\norm{\psi}^2=0$.
Taking a look at $\Upsilon_\phi(\psi)$ in (\ref{eq:tensors:Upsilon}) it turns out that
$\Upsilon_\phi(\psi)$ can be written in the Hilbert space language as
\begin{equation*}
\cket{\Upsilon_\phi(\psi)} = Y(\psi) \cket{\phi}
\end{equation*}
with the ``operator''
\begin{equation*}
Y(\psi) = 
-\gamma_1(\psi)\varepsilon\otimes\Id\otimes\Id
-\Id\otimes\gamma_2(\psi)\varepsilon\otimes\Id
-\Id\otimes\Id\otimes\gamma_3(\psi)\varepsilon.
\end{equation*}
%(In the Hilbert-space language, $\varepsilon=\varepsilon_{ii'}\cket{i}\bra{i'}$.)
Using this,
the vanishing condition of $\Upsilon_\phi(\psi)$ for all $\phi$,
\begin{equation*}
\begin{split}
\cket{\Upsilon_\phi(\psi)}=0\;\;\forall\cket{\phi}\quad
&\Longleftrightarrow\quad Y(\psi) \cket{\phi}=0\;\;\forall\cket{\phi}\\
&\Longleftrightarrow\quad Y(\psi)=0\\
&\Longleftrightarrow\quad \norm{Y(\psi)}^2=0 \;\; \text{for any norm},
\end{split}
\end{equation*}
so we can eliminate the quantors and $\phi$ from the condition.
Using the usual complex matrix $2$-norm $\norm{M}^2=\tr(M^\dagger M)$,
we have
\begin{equation*}
\norm{Y(\psi)}^2=4\bigl(\norm{\gamma_1(\psi)}^2 + \norm{\gamma_2(\psi)}^2 + \norm{\gamma_3(\psi)}^2\bigr).
\end{equation*}
This formula has a remarkable structure, namely
if we note that $s_a(\psi)=\norm{\gamma_b(\psi)}^2 + \norm{\gamma_c(\psi)}^2$
%(see in \cite{BorstenetalFreudenthal3QBEnt},
%%%%%%%%%%%%%%%%
%\footnote{Unfortunately,
%there are some misprints in the constant factors
%given in \cite{BorstenetalFreudenthal3QBEnt}.
%(6):  $K=3\tr(\rho_A\otimes\rho_B\rho_{AB})-\tr(\rho_A^3)-\tr(\rho_B^3)=\dots$
%(29): $S_A=\tr\gamma^{B\dagger}\gamma^B+\tr\gamma^{C\dagger}\gamma^C$,
%(30): $\tr\gamma^{A\dagger}\gamma^A=\frac12[S_B+S_C-S_A]$,
%(33): $||T||^2 =\frac{2}{3}(K-\abs{\psi}^6)+\frac{1}{4}\abs{\psi}^2(S_A+S_B+S_C)$.
%%%(33): $\rangle T\vert T\langle =\frac{2}{3}(K-\abs{\psi}^6)+\frac{1}{4}\abs{\psi}^2(S_A+S_B+S_C)$.
%% bug: labjegyzetbe nem engedi ezeket
%These have no influence on the results of the paper,
%we list them only to prevent misunderstandings.}%
%%%%%%%%%%%%%%%%
%)
and $\gamma_a(\psi)=0$ if and only if $\norm{\gamma_a(\psi)}^2=0$.
Now turn to the vanishing of $T(\psi,\psi,\psi)$, given in (\ref{eq:tensors:T}).
Again, this vanishes if and only if its norm $\norm{T(\psi,\psi,\psi)}^2$ does.
This can be calculated by the use of the form
\begin{equation*}
\begin{split}
\cket{T(\psi,\psi,\psi)} 
&= -\gamma_1(\psi)\varepsilon\otimes\Id\otimes\Id \cket{\psi}\\
&= -\Id\otimes\gamma_2(\psi)\varepsilon\otimes\Id \cket{\psi}\\
&= -\Id\otimes\Id\otimes\gamma_3(\psi)\varepsilon \cket{\psi}\\
&= \frac13 Y(\psi) \cket{\psi}.
\end{split}
\end{equation*}
(The quantity $\norm{T(\psi,\psi,\psi)}^2$ also appears in the
twistor-geometric approach of three-qubit entanglement, 
it is proportional to $\omega_{\text{ABC}}$ in \cite{PeterGeom3QBEnt}.)
About the scalar $q$, note that $q(\psi)=-2\Det(\psi)$ \cite{CKWThreetangle},
and it vanishes if and only if the three-tangle (\ref{eq:pureLUinvs:tau}) does.

Summarizing the observations above,
it is useful to define the following set of real-valued functions on $\mathcal{H}$:
\begin{subequations}
\label{eq:newPureLUinvs}
\begin{align}
n(\psi)     &= \norm{\psi}^2,\\
\label{eq:newPureLUinvs:y}
y(\psi)     &= \frac23\bigl(g_1(\psi) + g_2(\psi) + g_3(\psi)\bigr),\\
\label{eq:newPureLUinvs:sa}
s_a(\psi)   &= g_b(\psi) + g_c(\psi),\\
\label{eq:newPureLUinvs:ga}
g_a(\psi)   &= \norm{\gamma_a(\psi)}^2,\\
\label{eq:newPureLUinvs:t}
t(\psi)     &= 4\norm{T(\psi,\psi,\psi)}^2,\\
\label{eq:newPureLUinvs:tau2}
\tau^2(\psi)&= 4\abs{q(\psi)}^2.
\end{align}
\end{subequations}
[The explicit forms of these functions
and their relations to other important quantities
can be found in the Appendixes 
\ref{appsubsec:explicit:stdLU}, \ref{appsubsec:explicit:LUcanon}, and \ref{appsubsec:explicit:WoottersConc}.
The constant factors have been chosen so that 
$0\leq y(\psi),s_a(\psi),g_a(\psi),t(\psi),\tau^2(\psi) \leq1$ for normalized states,
which is shown in Appendix \ref{appsubsec:explicit:ranges}.]
These quantities are obtained by index contraction of $\psi^{ijk}$s and complex conjugated $\cc{(\psi^{i'j'k'})}$s by $\delta_{ii'}$s
from the tensors in (\ref{eq:tensors}), which were obtained by index contraction of $\psi^{ijk}$s and $\psi^{i'j'k'}$s by $\varepsilon_{ii'}$s.
From the contractions of free indices of the tensors in (\ref{eq:tensors}), 
we have $U^\dagger\delta U= \delta$ for $U\in\LieGrp{U}(2)$.
From the contractions inside the tensors of (\ref{eq:tensors}),
we have $\transp{U}\varepsilon U= \varepsilon \det(U)$
but for every factor $\det(U)$ there is a conjugated $\cc{\det(U)}=1/\det(U)$
from $U^\dagger\varepsilon \cc{U}= \varepsilon \cc{\det(U)}$.
Consequently, all the functions in (\ref{eq:newPureLUinvs}) are LU invariant,
while their vanishings are still LSL invariant.
(Again, $n$ is invariant under the larger group $\LieGrp{U}(8)$,
and $\tau^2$ under $\left[\LieGrp{U}(1)\times\LieGrp{SL}(2,\field{C})\right]^{\times3}$.)

Now, the conditions for the SLOCC classes 
by the vanishing of the tensors in (\ref{eq:tensors})
(see in Table \ref{tab:pureSLOCC2})
can be reformulated 
by the vanishing of the functions in (\ref{eq:newPureLUinvs})
in the way which can be seen in Table \ref{tab:pureSLOCC3}.
%%%%%%%%%%%%%%%%%%%%%%%%%%%%%%%%%%%%%%%%
\begin{table*}
\begin{tabular}{c||c|c|ccc|ccc|c|c}
Class  & $n(\psi)$ & $y(\psi)$ & $s_1(\psi)$ & $s_2(\psi)$ & $s_3(\psi)$ & $g_1(\psi)$ & $g_2(\psi)$ & $g_3(\psi)$ & $t(\psi)$ & $\tau^2(\psi)$  \\
\hline
\hline
$\mathcal{V}^\text{Null}$ & $=0$     & $=0$     & $=0$     & $=0$     & $=0$     & $=0$     & $=0$     & $=0$     & $=0$     & $=0$ \\
\hline
$\mathcal{V}^{1|2|3}$     & $>0$     & $=0$     & $=0$     & $=0$     & $=0$     & $=0$     & $=0$     & $=0$     & $=0$     & $=0$ \\
\hline
$\mathcal{V}^{1|23}$      & $>0$     & $>0$     & $=0$     & $>0$     & $>0$     & $>0$     & $=0$     & $=0$     & $=0$     & $=0$ \\
$\mathcal{V}^{2|13}$      & $>0$     & $>0$     & $>0$     & $=0$     & $>0$     & $=0$     & $>0$     & $=0$     & $=0$     & $=0$ \\
$\mathcal{V}^{3|12}$      & $>0$     & $>0$     & $>0$     & $>0$     & $=0$     & $=0$     & $=0$     & $>0$     & $=0$     & $=0$ \\
\hline
$\mathcal{V}^\text{W}$    & $>0$     & $>0$     & $>0$     & $>0$     & $>0$     & $>0$     & $>0$     & $>0$     & $>0$     & $=0$ \\
$\mathcal{V}^\text{GHZ}$  & $>0$     & $>0$     & $>0$     & $>0$     & $>0$     & $>0$     & $>0$     & $>0$     & $>0$     & $>0$    
\end{tabular}
\caption{SLOCC classes of three-qubit state vectors
identified by the vanishing of the pure-state indicator functions 
given in (\ref{eq:newPureLUinvs}).}
\label{tab:pureSLOCC3}
\end{table*}
%%%%%%%%%%%%%%%%%%%%%%%%%%%%%%%%%%%%%%%%
We call the functions in (\ref{eq:newPureLUinvs}) \emph{pure state indicator functions} for the three-qubit case.
We will give the exact definition of indicator functions for the general case later (in Sec.~\ref{subsec:Gen:Indicators}),
until that point we just use this name 
for non-negative functions having the vanishing properties given in Table~\ref{tab:pureSLOCC3}. 
Although this scheme constructed by ten quantities is even more redundant than the previous two,
but it will turn out that these ten indicator functions (\ref{eq:newPureLUinvs})
will be necessary in the case of mixed states.
Moreover, investigating Table~\ref{tab:pureSLOCC3}, we can catch all the ideas leading to the general construction.

%*******************************************************************************
%*******************************************************************************
\section{Mixed three-qubit states}
\label{sec:Mixed}

Here we recall and extend the 
PSS classification for three qubits.
%in parallel with the
%PS classification for three subsystems of arbitrary dimensions.
The main concept here, first given in \cite{DurCiracTarrach3QBMixSep,DurCiracTarrachBMixSep},
then used and extended in \cite{SeevinckUffinkMixSep,Acinetal3QBMixClass}, 
is that we define a density matrix to be the element of a class
according to whether it can or cannot be mixed by the use of
pure states of some given kinds.

%*******************************************************************************
\subsection{Convex subsets}
\label{subsec:Mixed:Subsets}

Let us introduce some convenient notations.
The set of states $\mathcal{D}\equiv\mathcal{D}(\mathcal{H})\subset\Lin(\mathcal{H})$
is the convex body of positive semidefinite operators of unit trace acting on $\mathcal{H}$, 
while the set of pure states $\mathcal{P}\subset\mathcal{D}$
is the set of extremal points of $\mathcal{D}$,
which are the projection operators of rank $1$.
(For the sake of simplicity, we have restricted ourselves to the operators of unit trace, that is, density matrices,
in spite of the fact that the construction could be extended for the whole cone of positive semidefinite operators.)
Disjoint subsets in $\mathcal{P}$
given by unit vectors of different SLOCC classes are
\begin{subequations}
\label{eq:Psets}
\begin{align}
\mathcal{P}^{1|2|3}       &=\bigl\{\cket{\psi}\bra{\psi}\;\big\vert\; \cket{\psi}\in \mathcal{V}^{1|2|3},\; \norm{\psi}^2=1 \bigr\},\\
\mathcal{P}^{a|bc}        &=\bigl\{\cket{\psi}\bra{\psi}\;\big\vert\; \cket{\psi}\in \mathcal{V}^{a|bc},\; \norm{\psi}^2=1 \bigr\},\\
\mathcal{P}^\text{W}    &=\bigl\{\cket{\psi}\bra{\psi}\;\big\vert\; \cket{\psi}\in \mathcal{V}^\text{W},\; \norm{\psi}^2=1 \bigr\},\\
\mathcal{P}^\text{GHZ}  &=\bigl\{\cket{\psi}\bra{\psi}\;\big\vert\; \cket{\psi}\in \mathcal{V}^\text{GHZ},\; \norm{\psi}^2=1 \bigr\},
\end{align}
which cover $\mathcal{P}$ entirely:
$\mathcal{P}=\mathcal{P}^{1|2|3}\cup\mathcal{P}^{1|23}\cup\mathcal{P}^{2|13}\cup\mathcal{P}^{3|12}
\cup\mathcal{P}^\text{W}\cup\mathcal{P}^\text{GHZ}$.
Besides these, if only partial separability properties are considered,
define
\begin{equation}
\mathcal{P}^{123}          =\bigl\{\cket{\psi}\bra{\psi}\;\big\vert\; \cket{\psi}\in \mathcal{V}^{123},\; \norm{\psi}^2=1 \bigr\},
\end{equation}
\end{subequations}
so $\mathcal{P}^{123}=\mathcal{P}^\text{W}\cup\mathcal{P}^\text{GHZ}$.
Except $\mathcal{P}^{1|2|3}$, none of the above sets are closed.

\begin{figure}
 \includegraphics{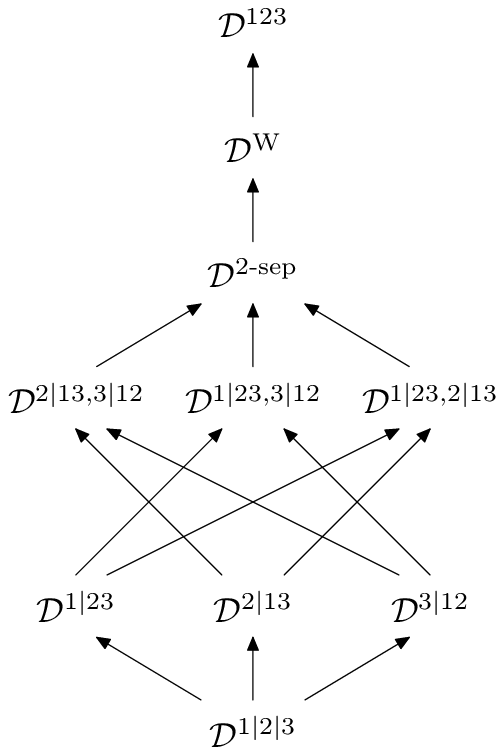}
 \caption{Inclusion hierarchy of the PSS sets $\mathcal{D}^{\dots}$ given in (\ref{eq:Dsets}).}
\label{fig:incl}
\end{figure}

The notion of $k$-separability and $\alpha_k$-separability \cite{SeevinckUffinkMixSep},
and the relevant classes of \cite{Acinetal3QBMixClass} for three-qubit systems
can be formulated as the convex hulls of some of the sets (\ref{eq:Psets}).
%%%%%%
The \emph{$3$-separable states} ($\mathcal{D}^\text{$3$-sep}$), 
or, equivalently $1|2|3$-separable states ($\mathcal{D}^{1|2|3}$) 
can be mixed from the pure states of $\mathcal{P}^{1|2|3}$, i.e., they are fully separable.
%%%%%%
The \emph{$a|bc$-separable states} ($\mathcal{D}^{a|bc}$)
can be written in the form $\sum_ip_i\varrho_{a,i}\otimes\varrho_{bc,i}$,
[$\varrho_{a,i}\in\mathcal{D}(\mathcal{H}^a)$,
$\varrho_{bc,i}\in\mathcal{D}(\mathcal{H}^b\otimes\mathcal{H}^c)$],
where we demand only the split between $a$ and $bc$,
but s split between $b$ and $c$ can also occur in the pure-state decompositions,
so they can be mixed from the pure states of $\mathcal{P}^{1|2|3}$ and $\mathcal{P}^{a|bc}$.
%%%%%%
The \emph{$2$-separable states}, also called \emph{biseparable states} ($\mathcal{D}^\text{$2$-sep}$) 
are of the form $\sum_ip_i\varrho_{a_i,i}\otimes\varrho_{b_ic_i,i}$,
so they can be mixed from the pure states of $\mathcal{P}^{1|2|3}$,
$\mathcal{P}^{1|23}$, $\mathcal{P}^{2|13}$, and $\mathcal{P}^{3|12}$.
These states are also of relevance because, although they are not separable under any $a|bc$ split,
there is no need of genuine three-qubit entangled pure state to mix them \cite{SeevinckUffinkMixSep}.
%%%%%%
From the point of view of convex hulls of extremal points,
it can be seen better than originally in \cite{SeevinckUffinkMixSep} that we can define
three new partial separability sets ``between'' the $a|bc$-separable
and $2$-separable ones.
For example, the \emph{$2|13$-$3|12$-separable states} ($\mathcal{D}^{\twoprt{2|13}{3|12}}$)
are the states which
can be mixed from the pure states of $\mathcal{P}^{1|2|3}$, $\mathcal{P}^{2|13}$, and $\mathcal{P}^{3|12}$.
States of this kind are also of relevance,
since there is no need of $1|23$-separable pure states to mix them,
that is, entanglement between the $2$ and the $3$ subsystems.
%%%%%%
Beyond these, 
we use the set of \emph{W-states} \cite{Acinetal3QBMixClass} ($\mathcal{D}^\text{W}$)
which can be expressed as the mixture of the pure states $\mathcal{P}^{1|2|3}$,
$\mathcal{P}^{1|23}$, $\mathcal{P}^{2|13}$, $\mathcal{P}^{3|12}$,
and $\mathcal{P}^\text{W}$,
so there is no need for pure states of GHZ type to mix them,
and the set of \emph{GHZ states} ($\mathcal{D}^\text{GHZ}$)
or, equivalently, $1$-separable ($\mathcal{D}^\text{$1$-sep}$),
or $123$-separable states ($\mathcal{D}^{123}$),
which is equal to the full set of states ($\mathcal{D}$).
Summarizing, we have the following 
%\emph{subsets of partial separability}, 
%(or PS subsets, for short),
\emph{PSS subsets} in $\mathcal{D}$
arising as convex hulls of pure states of given kinds:
\begin{subequations}
\label{eq:Dsets}
\begin{align}
%\label{eq:Dsets:3-sep}
%\mathcal{D}^\text{$3$-sep} &= \Conv\bigl( 
\label{eq:Dsets:1|2|3}
\mathcal{D}^{1|2|3} &= \Conv\bigl( 
\mathcal{P}^{1|2|3}\bigr)\equiv\mathcal{D}^\text{$3$-sep},\\
\mathcal{D}^{a|bc}  &= \Conv\bigl(
\mathcal{P}^{1|2|3}\cup\mathcal{P}^{a|bc}\bigr),\\
\label{eq:Dsets:bcacab}
%\mathcal{D}^{\substack{b|ac\\c|ab}}  &= \Conv \bigl(
\mathcal{D}^\twoprt{b|ac}{c|ab}  &= \Conv \bigl(
\mathcal{P}^{1|2|3}\cup\mathcal{P}^{b|ac}\cup\mathcal{P}^{c|ab}\bigr),\\
\label{eq:Dsets:2-sep}
\mathcal{D}^\text{$2$-sep}  &= \Conv \bigl(
\mathcal{P}^{1|2|3}\cup\mathcal{P}^{1|23}\cup\mathcal{P}^{2|13}\cup\mathcal{P}^{3|12}\bigr),\\
\label{eq:Dsets:W}
\begin{split}
\mathcal{D}^\text{W}  &= \Conv \bigl(
\mathcal{P}^{1|2|3}\cup\mathcal{P}^{1|23}\cup\mathcal{P}^{2|13}\cup\mathcal{P}^{3|12}\\
&\qquad\qquad\cup\mathcal{P}^\text{W}\bigr),
\end{split}\\
%\label{eq:Dsets:1-sep}
\label{eq:Dsets:123}
\begin{split}
\mathcal{D}^{123}  &= \Conv \bigl(
\mathcal{P}^{1|2|3}\cup\mathcal{P}^{1|23}\cup\mathcal{P}^{2|13}\cup\mathcal{P}^{3|12}\\
&\qquad\qquad\cup\underbrace{\mathcal{P}^\text{W}\cup\mathcal{P}^\text{GHZ}}_{\mathcal{P}^{123}}\bigr)
\equiv\mathcal{D}^\text{$1$-sep}\equiv\mathcal{D}.
\end{split}
\end{align}
\end{subequations}
These sets are convex and they
contain each other in a hierarchic way,
which is illustrated in Fig.~\ref{fig:incl}.
%%%%%%%%%%%%%%%%%%%%%%%%%%%%%%%%%%%%%%%%
%%%%%%%%%%%%%%%%%%%%%%%%%%%%%%%%%%%%%%%%

From an abstract point of view,
we form the convex hulls of \emph{closed} sets \cite{Acinetal3QBMixClass},
and the convex hulls of \emph{all the possible closed sets}
arising from the unions of the $\mathcal{P}^{\dots}$ sets (\ref{eq:Psets}) of extremal points
are listed in (\ref{eq:Dsets}) above.
We mean the PSS classification
involving the PSS subsets (\ref{eq:Dsets:1|2|3})--(\ref{eq:Dsets:123})
[and the PS classification 
involving the PS subsets (\ref{eq:Dsets:1|2|3})--(\ref{eq:Dsets:2-sep}) and (\ref{eq:Dsets:123})]
to be \emph{complete} in \emph{this} sense.
As special, noncomplete cases, 
we get back the classification involving only the sets 
$\mathcal{D}^\text{$k$-sep}$ and $\mathcal{D}^{\alpha_k}$ 
(for any $k$-partite split $\alpha_k$)
obtained by Seevinck and Uffink \cite{SeevinckUffinkMixSep},
the classification involving only the sets $\mathcal{D}^{\alpha_k}$
obtained by D\"ur and Cirac \cite{DurCiracTarrach3QBMixSep,DurCiracTarrachBMixSep}
and also the classification involving only the sets $\mathcal{D}^\text{$k$-sep}$ and $\mathcal{D}^\text{W}$,
obtained by Ac\'in, Bru\ss{}, Lewenstein and Sanpera \cite{Acinetal3QBMixClass}.

%*******************************************************************************
\subsection{Classes}
\label{subsec:Mixed:Classes}

Now, we determine the \emph{PSS classes} of three-qubit mixed states.
The abstract definition of these classes \cite{SeevinckUffinkMixSep}
is that they are the possible nontrivial intersections of the $\mathcal{D}^{\dots}$ convex subsets listed in (\ref{eq:Dsets}).
Since we want to deal also with the sets $\mathcal{D}^{\twoprt{b|ac}{c|ab}}$,
we cannot draw an expressive ``onionlike'' figure as was done in \cite{SeevinckUffinkMixSep}
for the sets $\mathcal{D}^{1|2|3}$, $\mathcal{D}^{a|bc}$, and $\mathcal{D}^\text{$2$-sep}$.
We have to proceed in a formal manner.

If we have the sets $A_1,A_2,\dots,A_n$,
all of their possible intersections can be constructed
as the intersections for all $i$ the set $A_i$ or its complement $\cmpl{A}_i$.
We have $10$ PSS subsets $\mathcal{D}^{\dots}$,
so we can formally write $2^{10}=1024$ possible intersections in this way.
If $B\subseteq A$, then $B\cap \cmpl{A}=\emptyset$,
so some intersections will be automatically empty
(\emph{``empty by construction''})
and, 
using the inclusion hierarchy of PSS subsets in Fig.~\ref{fig:incl},
we write only the intersections which are ``not empty by construction.''
The number of these will turn out to be only $21$.
(Again, if $B\subseteq A$, then $B\cap A=B$ and $\cmpl{B}\cap \cmpl{A}=\cmpl{A}$,
so we can write these $21$ classes as intersection sequences
much shorter than $10$ terms.)
Since the appearance of the $\mathcal{D}^{\twoprt{b|ac}{c|ab}}$-type sets in the intersections
makes the meaning of the classes a little bit involved,
we write out the list of the PSS classes with detailed explanations.

First, the class
\begin{subequations}
\label{eq:Classes}
\begin{equation}
%\mathcal{C}^3=\mathcal{D}^\text{$3$-sep}
\mathcal{C}^3=\mathcal{D}^{1|2|3}
\end{equation}
is the set of fully separable states.

Then come the $18$ classes of $2$-separable entangled states,
that is, the subsets in $\mathcal{D}^\text{$2$-sep}\setminus\mathcal{D}^{1|2|3}$.
%%%%%%
The first one of them is
\begin{equation}
\begin{split}
\mathcal{C}^{2.8}&=\cmpl{\mathcal{D}^{1|2|3}}\cap\mathcal{D}^{1|23}\cap\mathcal{D}^{2|13}\cap\mathcal{D}^{3|12}\\
&=\bigl(\mathcal{D}^{1|23}\cap\mathcal{D}^{2|13}\cap\mathcal{D}^{3|12}\bigr)\setminus\mathcal{D}^{1|2|3},
\end{split}
\end{equation}
which is the set of states 
which can    be written as $1|23$-separable states
(i.e.,convex combinations of $\mathcal{P}^{1|2|3}$ and $\mathcal{P}^{1|23}$ pure states;
the formation is not unique)
and can also be written as $2|13$-separable states
and can also be written as $3|12$-separable states
but cannot   be written as $1|2|3$-separable states.
%States of this class is sometimes called semiseparable. ..... cite ???
The existence of such states was counterintuitive
because, for pure states, if a tripartite pure state is separable under any $a|bc$ bipartition, then it is fully separable.
For mixed states, however, explicit examples can be constructed \cite{BennettetalUPB,Acinetal3QBMixClass},
which can be written in the form $\sum_ip_i\varrho_{a,i}\otimes\varrho_{bc,i}$ for any $a|bc$ bipartition,
but cannot be written in the form $\sum_ip_i\varrho_{1,i}\otimes\varrho_{2,i}\otimes\varrho_{3,i}$.
Alternatively, we can say that
states of this class can not be mixed without the use of bipartite entanglement,
but they can be mixed by the use of bipartite entanglement
inside only one bipartite subsystem; it does not matter which one.
(This is class $2.8$ in \cite{SeevinckUffinkMixSep}.)
%%%%%%
The next three classes are
\begin{equation}
\begin{split}
\mathcal{C}^{2.7.a}&=\cmpl{\mathcal{D}^{a|bc}}\cap\mathcal{D}^{b|ac}\cap\mathcal{D}^{c|ab}\\
&=\bigl(\mathcal{D}^{b|ac}\cap\mathcal{D}^{c|ab}\bigr)\setminus\mathcal{D}^{a|bc}.
\end{split}
\end{equation}
For example, $\mathcal{C}^{2.7.1}$
is the set of states which can be written as $2|13$-separable states
and can also be written as $3|12$-separable states
but cannot   be written as $1|23$-separable states.
Alternatively, we can say that
states of this class cannot be mixed by the use of bipartite entanglement only inside the $23$ subsystem,
but they can be mixed by the use of bipartite entanglement
inside either the $12$ or the $13$ subsystems;
both of them are equally suitable.
(These three classes are classes $2.7$, $2.6$, and $2.5$ in \cite{SeevinckUffinkMixSep}.)
%%%%%%
The next three classes are
\begin{equation}
\begin{split}
\mathcal{C}^{2.6.a}&=\mathcal{D}^{a|bc}\cap\cmpl{\mathcal{D}^{b|ac}}\cap\cmpl{\mathcal{D}^{c|ab}}\cap\mathcal{D}^{\twoprt{b|ac}{c|ab}}\\
&= \mathcal{D}^{a|bc}\cap\bigl[\mathcal{D}^{\twoprt{b|ac}{c|ab}}\setminus\bigl(\mathcal{D}^{b|ac}\cup\mathcal{D}^{c|ab} \bigr)  \bigr].
\end{split}
\end{equation}
For example, $\mathcal{C}^{2.6.1}$
is the set of states which can be written as $1|23$-separable states
and can also be written as states of a new kind:
where the state can be written as
$2|13$-$3|12$-separable states which are neither $2|13$-separable nor $3|12$-separable.
And this is the novelty here.
Alternatively, we can say that
to mix a state of this class
we need bipartite entanglement 
either inside the $23$ subsystem,
or inside both of the $12$ and the $13$ subsystems.
(The latter seems like a roundabout connecting the $2$ and $3$ subsystems through the $1$ subsystem.)
%%%%%%
The next three classes are
\begin{equation}
\begin{split}
\mathcal{C}^{2.5.a}&=\mathcal{D}^{a|bc}\cap\cmpl{\mathcal{D}^{b|ac}}\cap\cmpl{\mathcal{D}^{c|ab}}\cap\cmpl{\mathcal{D}^{\twoprt{b|ac}{c|ab}}}\\
&\equiv\mathcal{D}^{a|bc}\cap\cmpl{\mathcal{D}^{\twoprt{b|ac}{c|ab}}}
=\mathcal{D}^{a|bc}\setminus\mathcal{D}^{\twoprt{b|ac}{c|ab}}.
\end{split}
\end{equation}
For example, $\mathcal{C}^{2.5.1}$
is the set of states which can be written as $1|23$-separable states
but cannot  be written as $2|13$-$3|12$-separable states.
Alternatively, we can say that
states of this class 
%an not be mixed without the use of bipartite entanglement inside $23$ subsystem,
%ut in this case no bipartite entanglement inside the other two bipartite subsystems are needed, contrary to $\mathcal{C}^{2.7.1}$.
%n the other hand, it 
cannot be mixed by the use of bipartite entanglement only inside both of the $12$ and $13$ subsystems, contrary to $\mathcal{C}^{2.6.1}$.
(The roundabout does not exist here.)
(The unions $\mathcal{C}^{2.6.a}\cup\mathcal{C}^{2.5.a}=\mathcal{D}^{a|bc}\cap\cmpl{\mathcal{D}^{b|ac}}\cap\cmpl{\mathcal{D}^{c|ab}}$
are classes $2.4$, $2.3$, and $2.2$ in \cite{SeevinckUffinkMixSep}.)
%%%%%%
The next class is
\begin{equation}
\begin{split}
\mathcal{C}^{2.4}=&\cmpl{\mathcal{D}^{1|23}}\cap\cmpl{\mathcal{D}^{2|13}}\cap\cmpl{\mathcal{D}^{3|12}}\\
&\cap\mathcal{D}^{\twoprt{2|13}{3|12}}\cap\mathcal{D}^{\twoprt{1|23}{3|12}}\cap\mathcal{D}^{\twoprt{1|23}{2|13}}\\
=&\bigl(\mathcal{D}^{\twoprt{2|13}{3|12}}\cap\mathcal{D}^{\twoprt{1|23}{3|12}}\cap\mathcal{D}^{\twoprt{1|23}{2|13}}\bigr)\\
&\setminus\bigl(\mathcal{D}^{1|23}\cup\mathcal{D}^{2|13}\cup\mathcal{D}^{3|12}\bigr)\\
=&\bigl[\mathcal{D}^{\twoprt{2|13}{3|12}}\setminus\bigl(\mathcal{D}^{2|13}\cup\mathcal{D}^{3|12} \bigr)  \bigr]\\
 &\cap\bigl[\mathcal{D}^{\twoprt{1|23}{3|12}}\setminus\bigl(\mathcal{D}^{1|23}\cup\mathcal{D}^{3|12} \bigr)  \bigr]\\
 &\cap\bigl[\mathcal{D}^{\twoprt{1|23}{2|13}}\setminus\bigl(\mathcal{D}^{1|23}\cup\mathcal{D}^{2|13} \bigr)  \bigr],
\end{split}
\end{equation}
which is the set of states 
which can be mixed by the use of bipartite entanglement inside any two bipartite subsystems,
but cannot be mixed by the use of bipartite entanglement inside only one bipartite subsystem.
%%%%%%
The next three classes are
\begin{equation}
\begin{split}
\mathcal{C}^{2.3.a}=&\cmpl{\mathcal{D}^{a|bc}}\cap\cmpl{\mathcal{D}^{\twoprt{b|ac}{c|ab}}}\cap
\mathcal{D}^{\twoprt{a|bc}{c|ab}}\cap\mathcal{D}^{\twoprt{a|bc}{b|ac}}\\
=&\Bigl[\bigl[\mathcal{D}^{\twoprt{a|bc}{c|ab}}\setminus\bigl(\mathcal{D}^{c|ab}\cup\mathcal{D}^{a|bc} \bigr)  \bigr]\\
&\cap   \bigl[\mathcal{D}^{\twoprt{a|bc}{b|ac}}\setminus\bigl(\mathcal{D}^{a|bc}\cup\mathcal{D}^{b|ac} \bigr)  \bigr]\Bigr]
\setminus\mathcal{D}^{\twoprt{b|ac}{c|ab}}.
\end{split}
\end{equation}
For example, $\mathcal{C}^{2.3.1}$
is the set of states
which can be mixed by the use of bipartite entanglement inside the $23$ subsystem
together with bipartite entanglement inside either the $12$ or the $13$ subsystems,
but cannot be mixed by the use of bipartite entanglement inside the $12$ \emph{and} the $13$ subsystems only.
(Note that mixing by the use of only one kind of bipartite entanglement has already been excluded.)
%%%%%%
The next three classes are
\begin{equation}
\begin{split}
\mathcal{C}^{2.2.a}&=\mathcal{D}^{\twoprt{b|ac}{c|ab}}\cap\cmpl{\mathcal{D}^{\twoprt{a|bc}{c|ab}}}\cap\cmpl{\mathcal{D}^{\twoprt{a|bc}{b|ac}}}\\
&=\mathcal{D}^{\twoprt{b|ac}{c|ab}}\setminus\bigl(\mathcal{D}^{\twoprt{a|bc}{c|ab}}\cup\mathcal{D}^{\twoprt{a|bc}{b|ac}}\bigr).
\end{split}
\end{equation}
For example, $\mathcal{C}^{2.3.1}$
is the set of states
which can be mixed by the use of bipartite entanglement inside both the $12$ and the $13$ subsystems together,
but cannot be mixed by the use of bipartite entanglement inside the $23$ subsystem
together with bipartite entanglement inside only one of the $12$ or the $13$ subsystems.
%%%%%%
The next class is
\begin{equation}
\begin{split}
\mathcal{C}^{2.1}&=\cmpl{\mathcal{D}^{\twoprt{2|13}{3|12}}}\cap\cmpl{\mathcal{D}^{\twoprt{1|23}{3|12}}}\cap\cmpl{\mathcal{D}^{\twoprt{1|23}{2|13}}}\cap\mathcal{D}^\text{$2$-sep}\\
&=\mathcal{D}^\text{$2$-sep}\setminus\bigl(\mathcal{D}^{\twoprt{2|13}{3|12}}\cup\mathcal{D}^{\twoprt{1|23}{3|12}}\cup \mathcal{D}^{\twoprt{1|23}{2|13}}\bigr),
\end{split}
\end{equation}
which is the set of states
which can be mixed by the use of bipartite entanglement inside all the three bipartite subsystems,
but cannot be mixed by the use of bipartite entanglement inside only two (or one) bipartite subsystems.
(The union
$\mathcal{C}^{2.4}
\cup\mathcal{C}^{2.3.1}\cup\mathcal{C}^{2.3.2}\cup\mathcal{C}^{2.3.3}
\cup\mathcal{C}^{2.2.1}\cup\mathcal{C}^{2.2.2}\cup\mathcal{C}^{2.2.3}
\cup\mathcal{C}^{2.1}= \mathcal{D}^\text{$2$-sep}\setminus\bigl(\mathcal{D}^{1|23}\cup\mathcal{D}^{2|13}\cup\mathcal{D}^{3|12}\bigr)$
is class $2.1$ in \cite{SeevinckUffinkMixSep}.)

Then come the $2$ classes of states containing genuine tripartite entanglement \cite{Acinetal3QBMixClass},
that is, the subsets in $\mathcal{D}\setminus\mathcal{D}^\text{$2$-sep}$.
%%%%%%
The class
\begin{equation}
\mathcal{C}^\text{W}=\cmpl{\mathcal{D}^\text{$2$-sep}}\cap\mathcal{D}^\text{W}
=\mathcal{D}^\text{W}\setminus\mathcal{D}^\text{$2$-sep}
\end{equation}
is the set of states which cannot be mixed
without the use of some tripartite entangled pure states,
but there is no need for GHZ type entanglement \cite{Acinetal3QBMixClass}.
%%%%%%
The class
\begin{equation}
%\mathcal{C}^\text{GHZ}=\cmpl{\mathcal{D}^\text{W}}\cap\mathcal{D}^\text{$1$-sep}
%=\mathcal{D}^\text{$1$-sep}\setminus\mathcal{D}^\text{W}
\mathcal{C}^\text{GHZ}=\cmpl{\mathcal{D}^\text{W}}\cap\mathcal{D}^{123}
=\mathcal{D}^{123}\setminus\mathcal{D}^\text{W}
\end{equation}
is the set of states which cannot be mixed 
without the use of GHZ type entanglement.
All the above classes are PSS classes.
For the PS classification define the class of states containing genuine tripartite entanglement
instead of $\mathcal{C}^\text{W}$ and $\mathcal{C}^\text{GHZ}$:
\begin{equation}
%\mathcal{C}^1=\mathcal{C}^\text{W}\cup\mathcal{C}^\text{GHZ}=\mathcal{D}^\text{$1$-sep}\setminus\mathcal{D}^\text{$2$-sep}.
\mathcal{C}^1=\mathcal{C}^\text{W}\cup\mathcal{C}^\text{GHZ}=\mathcal{D}^{123}\setminus\mathcal{D}^\text{$2$-sep}.
\end{equation}
\end{subequations}

Except $\mathcal{C}^3$,
the $\mathcal{C}^\text{\dots}$ PS(S) classes above are neither convex nor closed,
but, by construction, they cover $\mathcal{D}$ entirely.
Unfortunately, we cannot draw an onionlike figure 
illustrating these classes, like the one in \cite{SeevinckUffinkMixSep}
(maybe it could be drawn in three dimensions);
we only summarize these $1+18+1+1$ classes in Table~\ref{tab:mixClasses}.
%%%%%%%%%%%%%%%%%%%%%%%%%%%%%%%%%%%%%%%%
\begin{table*}
\begin{tabular}{cc||c|ccc|ccc|c|cc||ccc}
PSS Class  & 
PS Class  & 
$\mathcal{D}^{1|2|3}$ & 
$\mathcal{D}^{a|bc}$ & 
$\mathcal{D}^{b|ac}$ & 
$\mathcal{D}^{c|ab}$ & 
$\mathcal{D}^{\twoprt{b|ac}{c|ab}}$ & 
$\mathcal{D}^{\twoprt{a|bc}{c|ab}}$ &
$\mathcal{D}^{\twoprt{a|bc}{b|ac}}$ &
$\mathcal{D}^\text{$2$-sep}$ &
$\mathcal{D}^\text{W}$ &
$\mathcal{D}^{123}$ &
%$\mathcal{D}^\text{GHZ}$ &
in \cite{SeevinckUffinkMixSep} &
in \cite{DurCiracTarrachBMixSep} &
in \cite{Acinetal3QBMixClass} \\
%$\mathcal{D}^\text{$1$-sep}$  \\
\hline
\hline
%                                                     || 1|2|3     |  a|bc         b|ac         c|ab      | b|ac/c|ab    c|ab/a|bc    a|bc/b|ac  |  2-sep     |  W         |  GHZ
$\mathcal{C}^3$           & $\mathcal{C}^3$           & $\subset$  & $\subset$  & $\subset$  & $\subset$  & $\subset$  & $\subset$  & $\subset$  & $\subset$  & $\subset$  & $\subset$  & 3       & 5       & S \\
\hline
$\mathcal{C}^{2.8}$       & $\mathcal{C}^{2.8}$       & $\nsubset$ & $\subset$  & $\subset$  & $\subset$  & $\subset$  & $\subset$  & $\subset$  & $\subset$  & $\subset$  & $\subset$  & 2.8     & 4       & B \\
$\mathcal{C}^{2.7.a}$     & $\mathcal{C}^{2.7.a}$     & $\nsubset$ & $\nsubset$ & $\subset$  & $\subset$  & $\subset$  & $\subset$  & $\subset$  & $\subset$  & $\subset$  & $\subset$  & 2.7,6,5 & 3.3,2,1 & B \\
$\mathcal{C}^{2.6.a}$     & $\mathcal{C}^{2.6.a}$     & $\nsubset$ & $\subset$  & $\nsubset$ & $\nsubset$ & $\subset$  & $\subset$  & $\subset$  & $\subset$  & $\subset$  & $\subset$  & 2.4,3,2 & 2.3,2,1 & B \\
$\mathcal{C}^{2.5.a}$     & $\mathcal{C}^{2.5.a}$     & $\nsubset$ & $\subset$  & $\nsubset$ & $\nsubset$ & $\nsubset$ & $\subset$  & $\subset$  & $\subset$  & $\subset$  & $\subset$  & 2.4,3,2 & 2.3,2,1 & B \\
$\mathcal{C}^{2.4}$       & $\mathcal{C}^{2.4}$       & $\nsubset$ & $\nsubset$ & $\nsubset$ & $\nsubset$ & $\subset$  & $\subset$  & $\subset$  & $\subset$  & $\subset$  & $\subset$  & 2.1     & 1       & B \\
$\mathcal{C}^{2.3.a}$     & $\mathcal{C}^{2.3.a}$     & $\nsubset$ & $\nsubset$ & $\nsubset$ & $\nsubset$ & $\nsubset$ & $\subset$  & $\subset$  & $\subset$  & $\subset$  & $\subset$  & 2.1     & 1       & B \\
$\mathcal{C}^{2.2.a}$     & $\mathcal{C}^{2.2.a}$     & $\nsubset$ & $\nsubset$ & $\nsubset$ & $\nsubset$ & $\subset$  & $\nsubset$ & $\nsubset$ & $\subset$  & $\subset$  & $\subset$  & 2.1     & 1       & B \\
$\mathcal{C}^{2.1}$       & $\mathcal{C}^{2.1}$       & $\nsubset$ & $\nsubset$ & $\nsubset$ & $\nsubset$ & $\nsubset$ & $\nsubset$ & $\nsubset$ & $\subset$  & $\subset$  & $\subset$  & 2.1     & 1       & B \\
\hline
$\mathcal{C}^\text{W}$    & $\mathcal{C}^1$           & $\nsubset$ & $\nsubset$ & $\nsubset$ & $\nsubset$ & $\nsubset$ & $\nsubset$ & $\nsubset$ & $\nsubset$ & $\subset$  & $\subset$  & 1       & 1       & W \\
$\mathcal{C}^\text{GHZ}$  & $\mathcal{C}^1$           & $\nsubset$ & $\nsubset$ & $\nsubset$ & $\nsubset$ & $\nsubset$ & $\nsubset$ & $\nsubset$ & $\nsubset$ & $\nsubset$ & $\subset$  & 1       & 1       & GHZ
\end{tabular}
\caption{PSS classes of mixed three-qubit states
and PS classes of mixed tripartite states.
Additionally, we show the 
classifications obtained by
Seevinck and Uffink \cite{SeevinckUffinkMixSep},
D\"ur, Cirac and Tarrach \cite{DurCiracTarrachBMixSep},
and Ac\'in, Bru\ss{}, Lewenstein and Sanpera \cite{Acinetal3QBMixClass}.} 
\label{tab:mixClasses}
\end{table*}
%%%%%%%%%%%%%%%%%%%%%%%%%%%%%%%%%%%%%%%%
The non-emptiness of the PS(S) classes above 
is not obvious,
since it depends on the arrangement of different kinds of extremal points.
(We know only that they are not empty \emph{by construction}.)
This issue has not been handled yet,
but experiences in the geometry of mixed states \cite{BengtssonZyczkowski} 
suggest that 
the arrangement of different kinds of extremal points 
leading to some empty classes
would be very implausible.

%*******************************************************************************
\subsection{Convex roof quantities}
\label{subsec:Mixed:CRoof}

As a next step, we obtain indicator functions on mixed states
from the pure-state indicator functions (\ref{eq:newPureLUinvs})
by convex roof construction \cite{BennettetalMixedStates,UhlmannFidelityConcurrence,UhlmannConvRoofs}.
In general, let
\begin{equation*}
f:\mathcal{P}\longrightarrow \field{R}
\end{equation*}
be a continuous function.
Then its convex roof extension is defined as
\begin{equation}
\label{eq:cnvroofext}
\begin{split}
\cnvroof{f}&:\mathcal{D}\longrightarrow \field{R},\\
%\cnvroof{f}(\varrho)=\min\{\sum_i p_i f(\psi_i)\mid 0\leq p_i,\;\sum_i p_i=1,\;\sum_i p_i \cket{\psi_i}\bra{\psi_i}=\varrho,\}
\cnvroof{f}&(\varrho)=\min  \sum_i p_i f(\psi_i),
\end{split}
\end{equation}
where the minimization
takes place on all pure-state decompositions of $\varrho$:
$0\leq p_i$, $\sum_i p_i=1$, $\sum_i p_i \cket{\psi_i}\bra{\psi_i}=\varrho$.
The existence of the minimum is crucial for our construction.
It follows from the Schr\"odinger mixture theorem \cite{SchrodingerMixtureThm},
also known as the Gisin-Hughston-Jozsa-Wootters lemma \cite{GisinMixtureThm,HughstonJozsaWoottersMixtureThm}, that
the decompositions for $m$ pure states 
are labeled by the elements of the \emph{compact} complex manifold,
called Stiefel manifold, $\mathrm{St}_d(\field{C}^m)= \LieGrp{U}(m)/\LieGrp{U}(m-d)$,
where $d=\dim\mathcal{H}$ \cite{BengtssonZyczkowski}.
The Carath\'eodory theorem ensures that we need only \emph{finite} $m$, 
or to be more precise $m\leq d^2$, shown by Uhlmann \cite{UhlmannOptimalDecomp}.
These observations guarantee the existence of the minimum in (\ref{eq:cnvroofext}).

Now, it is easy to prove the following 
necessary and sufficient conditions for the PSS subsets (\ref{eq:Dsets})
given by the convex roof extension of the indicator functions (\ref{eq:newPureLUinvs}):
\begin{subequations}
\label{eq:vanishing}
\begin{align}
\label{eq:vanishing:y}
\varrho&\in\mathcal{D}^{1|2|3}&
\quad&\Longleftrightarrow&\quad \cnvroof{y}(\varrho)&=0,\\
\label{eq:vanishing:sa}
\varrho&\in\mathcal{D}^{a|bc}&
\quad&\Longleftrightarrow&\quad \cnvroof{s}_a(\varrho)&=0,\\
\label{eq:vanishing:ga}
\varrho&\in\mathcal{D}^{\twoprt{b|ac}{c|ab}}&
\quad&\Longleftrightarrow&\quad \cnvroof{g}_a(\varrho)&=0,\\
\label{eq:vanishing:t}
\varrho&\in\mathcal{D}^\text{$2$-sep}&
\quad&\Longleftrightarrow&\quad \cnvroof{t}(\varrho)&=0,\\
\varrho&\in\mathcal{D}^\text{W}&
\quad&\Longleftrightarrow&\quad \cnvroof{{\tau^2}}(\varrho)&=0.
\end{align}
\end{subequations}
To see the \textit{$\Rightarrow$ implications,}
observe that all the $\mathcal{D}^\text{\dots}$ PSS subsets are the convex hulls of
such pure states [see in (\ref{eq:Dsets})] for which the given functions
vanish [see in Table \ref{tab:pureSLOCC3}].
Since these functions can take only non-negative values,
the minimum in the convex roof extension is zero.
To see the \textit{$\Leftarrow$ implications,}
note that
if the convex roof extension of a non-negative function vanishes
then there exists a decomposition for pure states for which the function vanishes.
Again, the vanishing of a given function
singles out the pure states [see in Table \ref{tab:pureSLOCC3}]
from which the states of the given $\mathcal{D}^\text{\dots}$ PSS subset can be mixed [see in (\ref{eq:Dsets})].

The necessary and sufficient conditions for the PSS subsets (\ref{eq:vanishing})
yields necessary and sufficient conditions for the PSS classes,
and we can fill out Table \ref{tab:mixIdent} 
for the identification of the PSS classes of Table \ref{tab:mixClasses}, given for mixed states,
similar to Table \ref{tab:pureSLOCC3}, given for pure states.
%%%%%%%%%%%%%%%%%%%%%%%%%%%%%%%%%%%%%%%%
\begin{table*}
\begin{tabular}{c||c|ccc|ccc|c|c}
Class % & $\cnvroof{n}(\varrho)$ 
& $\cnvroof{y}(\varrho)$ & 
$\cnvroof{s}_a(\varrho)$ & $\cnvroof{s}_b(\varrho)$ & $\cnvroof{s}_c(\varrho)$ & 
$\cnvroof{g}_a(\varrho)$ & $\cnvroof{g}_b(\varrho)$ & $\cnvroof{g}_c(\varrho)$ & 
$\cnvroof{t}(\varrho)$ & $\cnvroof{{\tau^2}}(\varrho)$  \\
\hline
\hline
%                         | y        | s_a        s_b        s_c      | g_a        g_b        g_c      | t        | tau
$\mathcal{C}^3$           & $=0$     & $=0$     & $=0$     & $=0$     & $=0$     & $=0$     & $=0$     & $=0$     & $=0$ \\
\hline
$\mathcal{C}^{2.8}$       & $>0$     & $=0$     & $=0$     & $=0$     & $=0$     & $=0$     & $=0$     & $=0$     & $=0$ \\
$\mathcal{C}^{2.7.a}$     & $>0$     & $>0$     & $=0$     & $=0$     & $=0$     & $=0$     & $=0$     & $=0$     & $=0$ \\
$\mathcal{C}^{2.6.a}$     & $>0$     & $=0$     & $>0$     & $>0$     & $=0$     & $=0$     & $=0$     & $=0$     & $=0$ \\
$\mathcal{C}^{2.5.a}$     & $>0$     & $=0$     & $>0$     & $>0$     & $>0$     & $=0$     & $=0$     & $=0$     & $=0$ \\
$\mathcal{C}^{2.4}$       & $>0$     & $>0$     & $>0$     & $>0$     & $=0$     & $=0$     & $=0$     & $=0$     & $=0$ \\
$\mathcal{C}^{2.3.a}$     & $>0$     & $>0$     & $>0$     & $>0$     & $>0$     & $=0$     & $=0$     & $=0$     & $=0$ \\
$\mathcal{C}^{2.2.a}$     & $>0$     & $>0$     & $>0$     & $>0$     & $=0$     & $>0$     & $>0$     & $=0$     & $=0$ \\
$\mathcal{C}^{2.1}$       & $>0$     & $>0$     & $>0$     & $>0$     & $>0$     & $>0$     & $>0$     & $=0$     & $=0$ \\
\hline
$\mathcal{C}^\text{W}$    & $>0$     & $>0$     & $>0$     & $>0$     & $>0$     & $>0$     & $>0$     & $>0$     & $=0$ \\
$\mathcal{C}^\text{GHZ}$  & $>0$     & $>0$     & $>0$     & $>0$     & $>0$     & $>0$     & $>0$     & $>0$     & $>0$    
\end{tabular}
\caption{PSS classes of mixed three-qubit states given in table \ref{tab:mixClasses}
identified by the vanishing of the mixed indicator functions
(convex roof extension of the indicator functions (\ref{eq:newPureLUinvs})).}
\label{tab:mixIdent}
\end{table*}
%%%%%%%%%%%%%%%%%%%%%%%%%%%%%%%%%%%%%%%%
Because of their vanishing properties, we call the convex roof extension of pure indicator functions
\emph{mixed indicator functions}.

Note that the convex roof extension is a nonlinear operation:
$\cnvroof{(f_1+f_2)}\neq \cnvroof{f_1}+\cnvroof{f_2}$.
However, an inequality holds,
for example, $s_a=g_b+g_c$ and 
$\cnvroof{s}_a=\cnvroof{(g_b+g_c)}\geq\cnvroof{g}_b+\cnvroof{g}_c$,
so $\cnvroof{s}_a$ can be nonzero even if both $\cnvroof{g}_b$ and $\cnvroof{g}_c$ are zero.
This is why we could identify $21$ classes of mixed states
by the use of the convex roof extension of functions
which identify only $6$ classes of state vectors.
%On the other hand,
%the observations (\ref{eq:vanishing})
%give conditions for the vanishing of the mixed indicator functions
%through the inclusion scheme in figure \ref{fig:incl}.
%In general, it is hard to give conditions for the vanishing of convex roofs,
%but in our case, these follows easily from the construction.
%%(See in Appendix \ref{}.)
%
On the other hand, if a classification does not involve all the PS(S) subsets,
then, through (\ref{eq:vanishing}), 
we have to use only some of the indicator functions,
for example, $y$, $s_a$ and $t$ for the classification obtained by Seevinck and Uffink \cite{SeevinckUffinkMixSep},
$y$ and $s_a$ for the classification obtained by D\"ur, Cirac and Tarrach \cite{DurCiracTarrachBMixSep},
and $y$, $t$ and $\tau^2$ for the classification obtained by Ac\'in, Bru\ss{}, Lewenstein and Sanpera \cite{Acinetal3QBMixClass}.

%*******************************************************************************
%*******************************************************************************
\section{Examples}
\label{sec:Xmpl}

%\subsection{General considerations}
%\label{subsec:Xmpl:General}
At this point, the most important question is whether all of the PSS classes in (\ref{eq:Classes}) are nonempty.
Of course, this can be checked by the use of (\ref{eq:vanishing}),
but calculating convex roof extensions symbolically is a hard problem.
Here we give considerations apart from convex roofs.

The classes given by Seevinck and Uffink in \cite{SeevinckUffinkMixSep} are nonempty, 
which are $\mathcal{C}^{3}$,
$\mathcal{C}^{2.8}$,
$\mathcal{C}^{2.7.a}$,
the unions $\mathcal{C}^{2.6.a}\cup\mathcal{C}^{2.5.a}$,
the union $\mathcal{C}^{2.4}
\cup\mathcal{C}^{2.3.1}\cup\mathcal{C}^{2.3.2}\cup\mathcal{C}^{2.3.3}
\cup\mathcal{C}^{2.2.1}\cup\mathcal{C}^{2.2.2}\cup\mathcal{C}^{2.2.3}
\cup\mathcal{C}^{2.1}$,
and $\mathcal{C}^\text{W}\cup\mathcal{C}^\text{GHZ}$.
Both of the classes $\mathcal{C}^\text{W}$ and $\mathcal{C}^\text{GHZ}$ are nonempty \cite{Acinetal3QBMixClass}.
On the other hand, the pure sets (\ref{eq:Psets}) are contained by the following classes:
$\mathcal{P}^{1|2|3}   \subset\mathcal{C}^3$,
$\mathcal{P}^{a|bc}    \subset\mathcal{C}^{2.5.a}$,
$\mathcal{P}^\text{W}  \subset\mathcal{C}^\text{W}$,
$\mathcal{P}^\text{GHZ}\subset\mathcal{C}^\text{GHZ}$,
so we have additionally that $\mathcal{C}^{2.5.a}$ is nonempty.
%In the next subsection,
In the next paragraphs,
we construct states contained in classes $\mathcal{C}^{2.2.a}$ and $\mathcal{C}^{2.1}$.
This justifies the use of $b|ac$-$c|ab$-separable sets in the classification
(since we can distinguish between $\mathcal{C}^{2.2.a}$ and $\mathcal{C}^{2.1}$ by the use of these),
although the nonemptiness of
$\mathcal{C}^{2.6.a}$,
$\mathcal{C}^{2.4}$, and 
$\mathcal{C}^{2.3.a}$
has not been shown yet.

%******************************************************************************
%\subsection{Explicit examples}
%\label{subsec:Xmpl:Explicit}

From the point of view of ``mixtures of extremal points,'' 
it is easy to check that
the bipartite subsystems are separable for states in some PS subsets as follows:
\begin{align*}
\varrho&\in\mathcal{D}^{1|2|3}
&\Longrightarrow&
&\text{$\varrho_{23}$ sep. and $\varrho_{13}$ sep. and $\varrho_{12}$ sep.}\\
\varrho&\in\mathcal{D}^{a|bc}
&\Longrightarrow&
&\text{\phantom{$\varrho_{23}$ sep. and } $\varrho_{ac}$ sep. and $\varrho_{ab}$ sep.}\\
\varrho&\in\mathcal{D}^{\twoprt{b|ac}{c|ab}}
&\Longrightarrow& 
&\text{$\varrho_{bc}$ sep. \phantom{and $\varrho_{13}$ sep. and $\varrho_{12}$ sep.}}
\end{align*}
Unfortunately, the reverse implications are not true. 
For example, for the standard GHZ state (\ref{eq:GHZ}),
all bipartite subsystems are separable,
although $\cket{\text{GHZ}}\bra{\text{GHZ}}\notin\mathcal{D}^{1|2|3}$.
However, the negation of the implications above will turn out to be useful:
\begin{align*}
\varrho&\notin\mathcal{D}^{1|2|3}
&\Longleftarrow&
&\text{$\varrho_{23}$ ent. or $\varrho_{13}$ ent. or $\varrho_{12}$ ent.}\\
\varrho&\notin\mathcal{D}^{a|bc}
&\Longleftarrow&
&\text{\phantom{$\varrho_{23}$ ent. or } $\varrho_{ac}$ ent. or $\varrho_{ab}$ ent.}\\
\varrho&\notin\mathcal{D}^{\twoprt{b|ac}{c|ab}}
&\Longleftarrow& 
&\text{$\varrho_{bc}$ ent. \phantom{or $\varrho_{13}$ ent. or $\varrho_{12}$ ent.}}
\end{align*}
The entanglement of two-qubit subsystems can be easily checked,
for example, by the Peres-Horodecki criterion \cite{PeresCrit,HorodeckiPosMapWitness}:
\begin{equation}
\label{eq:PPT}
\text{$\omega$ separable}\quad\Longleftrightarrow\quad
\ptransp{\omega}{1}\geq 0.
\end{equation}
Here $\omega\in\mathcal{D}(\mathcal{H}^b\otimes\mathcal{H}^c)$, 
and $\ptransp{\;}{1}$ means transposition on the first subsystem,
which, although it is basis-dependent, the positivity of $\ptransp{\omega}{1}$ is not.
The $\Leftarrow$ implication in (\ref{eq:PPT}) holds only for qubit-qubit or qubit-qutrit systems.

Now, take a $\varrho\in\mathcal{D}^{\twoprt{2|13}{3|12}}$. 
Then $\varrho_{23}$ is always separable,
but if both  $\varrho_{12}$, and $\varrho_{13}$ are entangled, then 
by the above observations we have
$\varrho\notin\mathcal{D}^{1|23}$,
$\varrho\notin\mathcal{D}^{2|13}$, $\varrho\notin\mathcal{D}^{3|12}$,
moreover,
$\varrho\notin\mathcal{D}^{\twoprt{1|23}{2|13}}$, and
$\varrho\notin\mathcal{D}^{\twoprt{1|23}{3|12}}$.
This singles out exactly one class from Table \ref{tab:mixClasses}, namely $\mathcal{C}^{2.2.1}$.
So if we can mix a state $\varrho$ from $\mathcal{P}^{1|2|3}$, $\mathcal{P}^{b|ac}$, and $\mathcal{P}^{c|ab}$,
whose $\varrho_{ab}$ and $\varrho_{ac}$ subsystems are entangled,
then $\varrho\in\mathcal{C}^{2.2.a}$.
For example, such a state is the uniform mixture of projectors to the 
$\cket{0}_b\otimes\cket{\text{B}}_{ac}$ and
$\cket{0}_c\otimes\cket{\text{B}}_{ab}$
vectors:
\begin{equation*}
\frac12 \cket{0}\bra{0}_b\otimes\cket{\text{B}}\bra{\text{B}}_{ac}+ 
\frac12 \cket{0}\bra{0}_c\otimes\cket{\text{B}}\bra{\text{B}}_{ab}
\in\mathcal{C}^{2.2.a}, 
\end{equation*}
where $\cket{\text{B}}$ is the usual Bell state (\ref{eq:B}).

Now, take a $\varrho\in\mathcal{D}^\text{$2$-sep}$. 
Then if the states of all the two-qubit subsystems are entangled,
by the above observations we have
$\varrho\notin\mathcal{D}^{1|23}$,
$\varrho\notin\mathcal{D}^{2|13}$, $\varrho\notin\mathcal{D}^{3|12}$,
moreover,
$\varrho\notin\mathcal{D}^{\twoprt{2|13}{3|12}}$, 
$\varrho\notin\mathcal{D}^{\twoprt{1|23}{3|12}}$, and
$\varrho\notin\mathcal{D}^{\twoprt{1|23}{2|13}}$.
This singles out exactly one class from Table \ref{tab:mixClasses}, namely $\mathcal{C}^{2.1}$.
So if we can mix a state $\varrho$ from $\mathcal{P}^{1|2|3}$, $\mathcal{P}^{1|23}$, $\mathcal{P}^{2|13}$, and $\mathcal{P}^{3|12}$,
whose all two-qubit subsystems are entangled,
then $\varrho\in\mathcal{C}^{2.1}$.
For example, such a state is the mixture of projectors to the
previous two vectors together with
$\cket{1}_a\otimes\cket{\text{B}}_{bc}$:
\begin{equation*}
\begin{split}
 \frac14 \cket{0}\bra{0}_b\otimes\cket{\text{B}}\bra{\text{B}}_{ac}+ 
 \frac14 \cket{0}\bra{0}_c\otimes\cket{\text{B}}\bra{\text{B}}_{ab}\\
+\frac12 \cket{1}\bra{1}_a\otimes\cket{\text{B}}\bra{\text{B}}_{bc}
\in\mathcal{C}^{2.1}. 
\end{split}
\end{equation*}

%******************************************************************************
%\subsection{Numerical examples ??? ZOLI ???}
%\label{subsec:Xmpl:Numerical}

%******************************************************************************
%******************************************************************************
\section{Generalizations I. -- Three subsystems}
\label{sec:GenThreePart}

The considerations written out in detail in Secs.~\ref{sec:Pure} and \ref{sec:Mixed}
contain the main ideas which will be generalized in this and in the next sections.
In this section we break up with qubits, and consider tripartite systems 
composed from \emph{subsystems of arbitrary dimensions.}
Obviously, this has no influence on the PS sets and PS classes,
given in Secs.~\ref{subsec:Mixed:Subsets} and \ref{subsec:Mixed:Classes},
the only question is about the construction of mixed-state indicator functions of Sec.~\ref{subsec:Mixed:CRoof}.
The generalization to \emph{arbitrary number of subsystems} is left to the next section.

%******************************************************************************
\subsection{Pure state indicator functions 
for tripartite systems from the FTS approach}
\label{subsec:GenThreePart:FTS}

To get the necessary and sufficient conditions for the PS classes in the tripartite case,
we need the generalizations of the pure-state indicator functions in (\ref{eq:newPureLUinvs:y})--(\ref{eq:newPureLUinvs:t}).
Apart from continuity,
the main---and only---requirement for these is
to satisfy the vanishing requirements for pure states given in Table \ref{tab:pureSLOCC3}
(apart from the column $\tau^2$, 
and for row $\mathcal{V}^{123}$ instead of rows $\mathcal{V}^\text{W}$ and $\mathcal{V}^\text{GHZ}$).
Then
their convex roof extensions satisfy the vanishing requirements for mixed states given in Table \ref{tab:mixIdent}
(apart from the column $\cnvroof{{\tau^2}}$, 
and for row $\mathcal{C}^{1}$ instead of rows $\mathcal{C}^\text{W}$ and $\mathcal{C}^\text{GHZ}$),
since in (\ref{eq:vanishing:y})--(\ref{eq:vanishing:t})
we have used only the vanishing-requirements for pure states.

The pure-state indicator functions of (\ref{eq:newPureLUinvs}) have been obtained in the FTS approach,
which works only for the qubit case.
However, some parts of the definitions can be generalized.
To do this, our basic quantities will be the local entropies 
$s_a(\psi)=S_q(\pi_a)$ instead of the functions $g_a(\psi)$ given in (\ref{eq:newPureLUinvs:ga}),
since the former ones are defined for all dimensions.
[Here we use the notation for the projector $\pi=\cket{\psi}\bra{\psi}$,
and $\pi_a=\tr_{bc}(\pi)$.]
The most basic quantum entropy is the \emph{von Neumann entropy},
\begin{equation}
\label{eq:Neumann}
S(\varrho)=-\tr\bigl[\varrho\ln(\varrho)\bigr],
\end{equation}
having the strongest properties among all entropies.
The \emph{Tsallis entropy,} sometimes called $q$-entropy, in the quantum case is defined as
\begin{subequations}
\begin{equation}
\label{eq:Tsallis}
S_q(\varrho)=\frac{1}{1-q}\bigl[\tr(\varrho^q)-1\bigr],\qquad q>0,
\end{equation}
which is a nonadditive generalization of the von Neumann entropy:
$\lim_{q\to1}S_q(\varrho)=S(\varrho)$.
Again, as in (\ref{eq:pureLUinvs:sa}),
we can use the concurrence-squared, which is the normalized Tsallis entropy of parameter $2$:
\begin{equation}
\label{eq:conc2}
C^2(\varrho)=\frac{d}{d-1}S_2(\varrho)=\frac{d}{d-1}\bigl[1-\tr(\varrho^2)\bigr],
\end{equation}
\end{subequations}
if we prefer to deal with \emph{polynomials} in the $\psi^{ijk}$ and $\cc{(\psi^{ijk})}$ coefficients.
This is the nontrivial polynomial of the lowest degree which is also an entropy, that is, Schur-concave, 
so tells us something about mixedness.
[In (\ref{eq:conc2}), $d$ is the dimension of the Hilbert space on which $\varrho$ acts,
so the prefactor $\frac{d}{d-1}$ ensures that $0\leq C^2(\varrho)\leq1$.]

Obviously, for all Tsallis entropies of the subsystems,
$s_a(\psi)=S_q(\pi_a)$ fulfils the corresponding column of Table \ref{tab:pureSLOCC3},
since it vanishes if and only if the subsystem is pure,
which means the separability of that subsystem from the rest of the system
if the whole system is in pure state.
From (\ref{eq:newPureLUinvs:sa}) and (\ref{eq:newPureLUinvs:ga}),
it turns out that $y$, given in (\ref{eq:newPureLUinvs:y}),
is just the average of the local entropies $y=\frac13(s_1+s_2+s_3)$,
vanishing if and only if no entanglement is present.
This works well not only for qubits, so we can keep this definition of $y$. 

The functions $g_a$ in (\ref{eq:newPureLUinvs:ga}) can also be expressed by the local entropies (\ref{eq:newPureLUinvs:sa})
for qubits as $g_a=\frac12(s_b+s_c-s_a)$.
Can this definition be kept for subsystems of arbitrary dimensions?
For $\mathcal{V}^{1|2|3}$, obviously $g_a=0$.
For $\mathcal{V}^{a|bc}$, the subsystem $a$ can be separated from the others
so the subsystems $a$ and $bc$ are in pure states, $s_a=0$ and $s_b=s_c\neq0$,
from which $g_a\neq0$ and $g_b=g_c=0$.
So the first five rows of the $g_a$ columns of Table \ref{tab:pureSLOCC3} is fulfilled.
For the last row, we need that $g_a>0$
when genuine tripartite entanglement is present.
This is the problematic point.
%First note that, for tripartite pure states, 
%the functions $g_a$ are related to the failure of additivity of the entropies of the bipartite subsystems.
This question can be traced back to the subadditivity of the Tsallis entropies.
Raggio's conjecture \cite{RaggioTsallis} 
about that is twofold: For $q>1$,
\begin{subequations}
\label{eq:Raggio}
\begin{align}
\label{eq:Raggio:subadd}
S_q(\varrho) &\leq S_q(\varrho_1) + S_q(\varrho_2),\\
\label{eq:Raggio:add}
S_q(\varrho) &= S_q(\varrho_1) + S_q(\varrho_2) 
\;\Longleftrightarrow\;
\left\{\begin{aligned}
&\varrho = \varrho_1\otimes \varrho_2,\\
&\text{$\varrho_1$ or $\varrho_2$ pure}.
\end{aligned}\right.
\end{align}
\end{subequations}
[Note that for $0<q<1$, there is no definite relation between $S_q(\varrho)$ and $S_q(\varrho_1) + S_q(\varrho_2)$.]
Both statements hold for the classical scenario \cite{RaggioTsallis},
which can be modeled in the quantum scenario 
by density matrices being LU equivalent to diagonal ones.
The first part (\ref{eq:Raggio:subadd}) of the conjecture
has been proven by Audenaert \cite{AudenaertTsallisSubadd}.
This guarantees the non-negativity of the functions $g_a$:
For pure states,
$S_q(\pi_a)=S_q(\pi_{bc})\leq S_q(\pi_{b}) + S_q(\pi_{c})$,
so $0\leq \frac12(s_b+s_c-s_a) = g_a$.
On the other hand, (\ref{eq:Raggio:add}) is exactly what we need:
$\cket{\psi}\in\mathcal{V}^{123}$ if and only if
neither of its subsystems are pure, which means that 
there is subadditivity in a strict sense,
so $0 < \frac12(s_b+s_c-s_a) = g_a$.
The $\Leftarrow$ implication in (\ref{eq:Raggio:add}) holds obviously,
but the whole second part (\ref{eq:Raggio:add}) of the conjecture,
to our knowledge, has not been proven yet.
A very little side result of our work is that 
Raggio's conjecture holds
for the very restricted case of two-qubit mixed states which are, at the most, of rank 2.

We note that the von Neumann entropies ($q\to1$) of the subsystems \emph{are not suitable} for the role of the functions $s_a$,
if we want to write the functions $g_a$ by that as $\frac12(s_b+s_c-s_a)$,
since the von Neumann entropy is additive for product states 
without any reference to the purity of the subsystems: % in the equality condition of subadditivity:
\begin{subequations}
\label{eq:NeumannProp}
\begin{align}
\label{eq:NeumannProp:subadd}
S(\varrho) &\leq S(\varrho_1) + S(\varrho_2),\\
\label{eq:NeumannProp:add}
S(\varrho) &= S(\varrho_1) + S(\varrho_2) 
\;\Longleftrightarrow\;
\varrho = \varrho_1\otimes \varrho_2.
\end{align}
\end{subequations}
Indeed, it is easy to construct a tripartite state,
which is not separable under any partition, but has vanishing $g_a$ (defined by the von Neumann entropy).
For example, let $\dim\mathcal{H}^a=4$,
then for the state
\begin{equation*}
\cket{\psi}=\frac12\bigl(\cket{000}+\cket{101}+\cket{210}+\cket{311}\bigr)
\end{equation*}
$\pi_{23}=\pi_2\otimes\pi_3$, so $g_1(\psi)=\frac12\bigl(S(\pi_2)+S(\pi_3)-S(\pi_1)\bigr)=0$,
while $S(\pi_1)=\ln4$, and $S(\pi_2)=S(\pi_3)=\ln2$, so neither of the subsystems are pure,
the state is genuinely tripartite entangled.

The \emph{R\'enyi entropy} is defined as
\begin{equation}
\label{eq:Renyi}
S^\text{R}_q(\varrho)=\frac{1}{1-q}\ln\bigl[\tr(\varrho^q)\bigr],\qquad q>0,
\end{equation}
which is another generalization of the von Neumann entropy:
$\lim_{q\to1}S^R_q(\varrho)=S(\varrho)$,
having the advantage of additivity:
\begin{equation}
S^\text{R}_q(\varrho)=S^\text{R}_q(\varrho_1)+S^\text{R}_q(\varrho_2) 
\quad\Longleftarrow\quad \varrho = \varrho_1\otimes \varrho_2.
\end{equation}
This is an advantage when entanglement is studied in the asymptotic regime, 
when the state is present in multiple copies
and properties are investigated against the number of copies. 
Again, this advantage is a disadvantage from our point of view,
the R\'enyi entropies of the subsystems \emph{are not suitable} for the role of the functions $s_a$,
if we want to write the functions $g_a$ by that as $\frac12(s_b+s_c-s_a)$.
Moreover, subadditivity does not hold for R\'enyi entropy,
so the non-negativity of the functions $g_a$ defined by R\'enyi entropies
is not even guaranteed.
(For further properties and references on the quantum entropies,
see, for example, \cite{BengtssonZyczkowski,OhyaPetzQEntr,Petzfdivergence,FuruichiTsallis}.)

%******************************************************************************
\subsection{Pure state indicator functions 
for tripartite systems outside the FTS approach}
\label{subsec:GenThreePart:nFTS}

Fortunately, it is easy to define the pure-state indicator functions
of three subsystems of arbitrary dimensions
without the issues of equality in the subadditivity of $q$-entropies.
Again, the basic quantities are the local entropies,
and we use a ``multiplicative'' definition for the functions $g_a$
instead of the ``additive'' one, which came from the FTS approach,
\begin{subequations}
\label{eq:genNewPureLUinvs}
\begin{align}
y(\psi)     &= s_1(\psi) + s_2(\psi) + s_3(\psi),\\
s_a(\psi)   &= S_q(\pi_a),\\
g_a(\psi)   &= s_b(\psi)s_c(\psi),\\
t(\psi)     &= s_1(\psi)s_2(\psi)s_3(\psi).
\end{align}
\end{subequations}
These functions obviously reproduce the relevant part of Table \ref{tab:pureSLOCC3},
so, by (\ref{eq:vanishing:y})--(\ref{eq:vanishing:t}), 
their convex roof extensions reproduce Table \ref{tab:mixIdent}
for the identification of the PS classes of the tripartite case given in Table \ref{tab:mixClasses}.
The structure of the formulas above give us a hint 
for the generalization for arbitrary number of subsystems of arbitrary dimensions: 
Te just have to play a game with 
statements like ``being zero,'' with the logical connectives ``and'' and ``or,''
parallel to the addition and multiplication,
and also parallel to the set-theoretical inclusion, union, and intersection.

%******************************************************************************
%******************************************************************************
\section{Generalizations II. -- Partial separability of multipartite systems}
\label{sec:Gen}

In the previous sections,
we have followed a didactic treatment
in order to illustrate the main concept;
now it is high time to turn to abstract definitions
to handle the PS classification and criteria
for arbitrary number of subsystems of arbitrary dimensions.

For $n$ subsystems, the set of the labels of the subsystems is $L=\{1,2,\dots,n\}$.
Let $\alpha=L_1|L_2|\dots|L_k$ denote a $k$-partite split,
that is, a partition of the labels
into $k$ disjoint non-empty sets $L_r$,
where $L_1\cup L_2\cup\dots\cup L_k=L$.
For two partitions, $\beta$ and $\alpha$,
$\beta$ is contained in $\alpha$,
denoted as $\beta\preceq\alpha$,
if $\alpha$ can be obtained from $\beta$ by joining some---maybe neither---of the parts of $\beta$.
This defines a partial order on the partitions.
[It is easy to see from the definition that 
$\alpha\preceq\alpha$ (reflexivity);
if $\gamma\preceq\beta$ and $\beta\preceq\alpha$ then $\gamma\preceq\alpha$ (transitivity);
if $\beta\preceq\alpha$ and $\alpha\preceq\beta$ then $\alpha=\beta$ (antisymmetry).]
For example, for the tripartite case $1|2|3\preceq a|bc\preceq 123$.
Since 
there is a greatest and a smallest element
(the full $n$-partite split and the trivial partition without split, respectively,
$1|2|\dots|n\preceq\alpha\preceq12\dots n$,)
the set of partitions of $L$ for $\preceq$ forms a bounded lattice.

%******************************************************************************
\subsection{PS subsets in general}
\label{subsec:Gen:PSsubsets}

The first point is the generalization of the PS subsets $\mathcal{D}^\text{\dots}$.
Let $\mathcal{P}^\alpha$ be the set of pure states
which are separable under the partition $\alpha=L_1|L_2|\dots|L_k$,
but not separable under any $\beta\prec\alpha$.
Then the PS subset of \emph{$\alpha$-separable states} is
\begin{subequations}
\label{eq:genDsets}
\begin{equation}
\label{eq:genDsets:alpha}
\mathcal{D}^\alpha = \Conv \bigcup_{\beta\preceq\alpha}\mathcal{P}^\beta,
\end{equation}
which is a special case %($l=1$) 
of the PS subsets of \emph{$\vs{\alpha}$-separable states}
\begin{equation}
\label{eq:genDsets:alphal}
\mathcal{D}^{\vs{\alpha}} = \Conv \bigcup_{\alpha\in\vs{\alpha}} \bigcup_{\beta\preceq\alpha}\mathcal{P}^\beta
\equiv \Conv \bigcup_{\alpha\in\vs{\alpha}} \mathcal{D}^{\alpha},
\end{equation}
\end{subequations}
with the \emph{label} $\vs{\alpha}$ being an arbitrary \emph{set} of partitions.
[In the writing we omit the $\{\dots\}$ set brackets, as was seen in, e.g.,(\ref{eq:Dsets:bcacab}).]
The set of $k$-separable states $\mathcal{D}^\text{$k$-sep}$ arises as a special case
where the $\alpha$ elements of $\vs{\alpha}$ are all the possible $k$-partite splits.
Note that in general, the $\alpha$ partitions are not required to be $k$-partite splits for the same $k$.
This freedom can not be seen in the case of three subsystems.

The $\mathcal{P}^\alpha$ sets are not closed if and only if $\alpha$ is not the full $n$-partite split $1|2|\dots|n$,
but $\cup_{\beta\preceq\alpha}\mathcal{P}^\beta$ is closed,
so the sets $\mathcal{D}^{\vs{\alpha}}$ are closed, and convex by construction.
Note that different $\vs{\alpha}$ labels can give rise to the same $\mathcal{D}^{\vs{\alpha}}$ sets;
in other words,
the $\vs{\alpha}\mapsto \mathcal{D}^{\vs{\alpha}}$ ``labeling map'' defined by (\ref{eq:genDsets:alphal})
is surjective but not injective.
For the full PS classification we need all the possible \emph{different} $\mathcal{D}^{\vs{\alpha}}$ sets.
%So first we work out a general framework for the labellin 
Because of the nontrivial structure of the lattice of partitions,
obtaining all the different PS sets is also a nontrivial task.
We cannot provide a closed formula for that, but only an algorithm.
Before we do this, we need some constructions.

First, observe that
if $\beta\preceq\alpha$ then $\mathcal{D}^\beta\subseteq\mathcal{D}^\alpha$
[from definition (\ref{eq:genDsets:alpha}), and the transitivity of $\preceq$],
from which it follows that 
for the labels $\vs{\beta}$ and $\vs{\alpha}$,
if for every $\beta\in\vs{\beta}$ there is an $\alpha\in\vs{\alpha}$ for which $\beta\preceq\alpha$ 
then $\mathcal{D}^{\vs{\beta}}\subseteq\mathcal{D}^{\vs{\alpha}}$.
[From definition (\ref{eq:genDsets:alphal}). We will prove the reverse too.]
These observations motivate the extension of %the partial order 
$\preceq$ from the partitions to the labels:
\begin{equation}
\label{eq:labelreldef}
\vs{\beta}\preceq\vs{\alpha}
\qquad\overset{\text{def.}}{\Longleftrightarrow}\qquad
\forall \beta\in\vs{\beta}, \exists \alpha\in\vs{\alpha}:\; \beta\preceq\alpha.
\end{equation}
Note that, at this point,
the relation $\preceq$ on the labels is not a partial order,
only the reflexivity and the transitivity properties hold for that.
The antisymmetry property fails, which is the consequence of
that the definition (\ref{eq:labelreldef}) was motivated by the inclusion of the PS sets,
and different $\vs{\alpha}$s can lead to the same PS set.
%rather than....
Independently of this problem, which will be handled later, the following is true:
\begin{equation}
\label{eq:relationEquiv}
\vs{\beta}\preceq\vs{\alpha} \qquad\Longleftrightarrow\qquad
\mathcal{D}^{\vs{\beta}}\subseteq\mathcal{D}^{\vs{\alpha}}.
\end{equation}
For the proof, see Appendix \ref{appsubsec:Gen:Incl}.
Again, note that the relations $\preceq$ and $\subseteq$
are defined on nonisomorphic sets,
so (\ref{eq:relationEquiv}) does not contradict the fact
that the latter is a partial order while the former is not.

The next step is to define 
those labels for which $\preceq$ will be a partial order.
A label $\vs{\alpha}$ is called a \emph{proper label} if
\begin{equation}
\label{eq:properlabel}
%\forall \alpha,\alpha'\in\vs{\alpha}:\qquad
%\text{if $\alpha\neq\alpha'$ then $\alpha\npreceq\alpha'$}.
\forall \alpha,\alpha'\in\vs{\alpha},\; \alpha\neq\alpha'\quad\Longrightarrow\qquad \alpha\npreceq\alpha'.
\end{equation}
On the set of proper labels, the relation $\preceq$ defined in (\ref{eq:labelreldef})
is a partial order.
For the proof, see Appendix \ref{appsubsec:Gen:pOrder}.
A corollary is that
the set of proper labels for $\preceq$ forms a bounded lattice,
its greatest and smallest elements are the one-element labels 
of full $n$-partite split and the trivial partition without split, respectively:
$1|2|\dots|n\preceq\vs{\alpha}\preceq12\dots n$.

Is it true that every PS subset can be labeled by proper label?
Do different proper labels lead to different PS subsets?
In other words,
is the $\vs{\alpha}\mapsto \mathcal{D}^{\vs{\alpha}}$ ``labelling map''
from the \emph{set of proper labels} to the set of PS subsets
an isomorphism?
The injectivity is the $\Leftarrow$ implication from the observation, that
for $\vs{\alpha}$, $\vs{\beta}$, proper labels
\begin{equation}
\label{eq:propInj}
\vs{\beta}=\vs{\alpha}
\qquad\Longleftrightarrow\qquad
\mathcal{D}^{\vs{\beta}}=\mathcal{D}^{\vs{\alpha}}.
\end{equation}
For the proof, see Appendix \ref{appsubsec:Gen:propInj}.
%To see the injectivity,
If $\vs{\beta}$ is a label, 
then we can obtain a unique proper label from that,
if we drop every $\beta\in\vs{\beta}$ for which there is a $\beta'\in\vs{\beta}$ for which $\beta\preceq\beta'$.
The remaining partitions form a proper label which we denote $\vs{\alpha}$,
and the partitions which have been dropped out form a label which we denote $\vs{\gamma}$.
Then $\vs{\beta}=\vs{\alpha}\vs{\gamma}$, which means the union of labels $\vs{\alpha}$ and $\vs{\gamma}$.
(We omit the union sign too.)
Our next observation is useful for this case.
For the general labels $\vs{\alpha}$ and $\vs{\gamma}$,
\begin{equation}
\label{eq:propPart}
\vs{\gamma}\preceq\vs{\alpha}
\qquad\Longleftrightarrow\qquad
\mathcal{D}^{\vs{\alpha}\vs{\gamma}}=\mathcal{D}^{\vs{\alpha}},
\end{equation}
which means that when we obtain a proper label $\vs{\alpha}$
from a general label $\vs{\beta}$,
as was done above,
both of these lead to the same PS subset.
For the proof, see Appendix \ref{appsubsec:Gen:propPart}.
Since all PS subsets arise from general labels,
the above shows that they arise also from proper labels,
which is the surjectivity of the labeling by proper labels.

Now we have that the set of proper labels is isomorphic to the set of PS subsets.
The former one is much easier to handle.
Moreover, (\ref{eq:relationEquiv}) states now that
the lattice of $\vs{\alpha}$ proper labels with respect to the partial order $\preceq$
is isomorphic to
the lattice of $\mathcal{D}^{\vs{\alpha}}$ PS subsets with respect to the partial order $\subseteq$.
(This lattice is the generalization of the ``inclusion hierarchy'' in Fig.~\ref{fig:incl}.)
To get all the PS subsets, we have to obtain all the proper labels.
A brute-force method for this
is to form all the $\vs{\beta}$ labels (all the subsets of the set of all partitions),
then obtain the proper labels $\vs{\alpha}$ as before 
($\vs{\beta}=\vs{\alpha}\vs{\gamma}$)
and keep the different proper labels obtained in this way.
A much more sophisticated algorithm is given in Appendix \ref{appsubsec:Gen:Alg}.

%******************************************************************************
\subsection{PS classes in general}
\label{subsec:Gen:PSclasses}

The second point is the generalization of the PS classes $\mathcal{C}^\text{\dots}$,
which are the possible non-trivial intersections of the PS subsets $\mathcal{D}^\text{\dots}$.
Constructing these requires direct calculations for a given $n$, as was done in Sec.~\ref{subsec:Mixed:Classes}.

Let us divide the set of proper labels into two disjoint subsets, $\vvs{\alpha}$ and $\vvs{\beta}$;
then all the possible intersections of PS subsets can be labeled by such a pair,
which is called \emph{class label}, as
% lehetne altalanosabban: 
% tetszoleges setek, aztan kulonbozok adnak azonos classokat, aztan mi lesz
\begin{equation}
\mathcal{C}^{\vvs{\alpha},\vvs{\beta}}=
\bigcap_{\vs{\alpha}\in\vvs{\alpha}} \cmpl{\mathcal{D}^{\vs{\alpha}}}\cap \bigcap_{\vs{\beta}\in\vvs{\beta}}\mathcal{D}^{\vs{\beta}}.
\end{equation}
It can happen that $\mathcal{C}^{\vvs{\alpha},\vvs{\beta}}=\emptyset$ \emph{by construction,}
under which we mean that its emptiness follows from the inclusion hierarchy of PS subsets.
For example, if $\mathcal{D}^{\vs{\beta}}\subseteq\mathcal{D}^{\vs{\alpha}}$ 
for some $\vs{\alpha}\in\vvs{\alpha}$ and $\vs{\beta}\in\vvs{\beta}$, then the intersection above is identically empty.
The PS(S) classes for three subsystems in Sec.~\ref{subsec:Mixed:Classes} 
were obtained by the use of this observation.
In this general framework, this observation is formulated as follows:
\begin{equation*}
\begin{split}
\mathcal{C}^{\vvs{\alpha},\vvs{\beta}}=\emptyset
\;&\overset{\text{(i)}}{\Longleftrightarrow}\;
\cmpl{\bigcup_{\vs{\alpha}\in\vvs{\alpha}} \mathcal{D}^{\vs{\alpha}} } \cap \bigcap_{\vs{\beta}\in\vvs{\beta}}\mathcal{D}^{\vs{\beta}}=\emptyset\\
\;&\overset{\text{(ii)}}{\Longleftrightarrow}\;
\bigcap_{\vs{\beta}\in\vvs{\beta}}\mathcal{D}^{\vs{\beta}} \subseteq \bigcup_{\vs{\alpha}\in\vvs{\alpha}} \mathcal{D}^{\vs{\alpha}}\\
\;&\overset{\text{(iii)}}{\Longleftarrow}\;
\exists \vs{\alpha}\in\vvs{\alpha}, \exists\vs{\beta}\in\vvs{\beta}\;:\;\mathcal{D}^{\vs{\beta}}\subseteq\mathcal{D}^{\vs{\alpha}}\\
\;&\overset{\text{(iv)}}{\Longleftrightarrow}\;
\exists \vs{\alpha}\in\vvs{\alpha}, \exists\vs{\beta}\in\vvs{\beta}\;:\; \vs{\beta}\preceq \vs{\alpha}.
\end{split}
\end{equation*}
[\textit{Equivalence (i)} comes from De Morgan's law $\cmpl{A}\cap\cmpl{B}=\cmpl{A\cup B}$.
\textit{Equivalence (ii)} comes from the identity $B\subseteq A\;\Leftrightarrow\;B\cap\cmpl{A}\equiv B\setminus A=\emptyset$.
\textit{Implication (iii)} comes from $B\subseteq A\;\Rightarrow\; B\cap B'\subseteq A\cup A'$.
%$\bigcap_{\vs{\beta}\in\vvs{\beta}}\mathcal{D}^{\vs{\beta}} 
%\subseteq \mathcal{D}^{\vs{\beta}}
%\subseteq \mathcal{D}^{\vs{\alpha}}
%\subseteq \bigcup_{\vs{\alpha}\in\vvs{\alpha}} \mathcal{D}^{\vs{\alpha}}$.
\textit{Equivalence (iv)} is (\ref{eq:relationEquiv}).]

Implication (iii) is the point which makes it possible 
to formulate the emptiness of PS classes by the use of \emph{labels only}. 
That is still a question
whether implication (iii) can be replaced with a stronger one,
which leads to a condition involving only labels again.
(The problem is that we have no interpretations of $\cap$ and $\cup$ in the language of labels.)
\emph{Our first conjecture is that 
implication (iii) above is the strongest one which leads to a condition involving only labels.}

Summarizing, we have
\begin{subequations}
\begin{equation}
\label{eq:emptyByConstruction}
\mathcal{C}^{\vvs{\alpha},\vvs{\beta}}=\emptyset 
\quad\Longleftarrow\quad
\exists \vs{\alpha}\in\vvs{\alpha}, \exists\vs{\beta}\in\vvs{\beta}\;:\; \vs{\beta}\preceq \vs{\alpha}.
\end{equation}
If the right-hand side holds, then we say, according to the conjecture above,
that $\mathcal{C}^{\vvs{\alpha},\vvs{\beta}}$ is \emph{empty by construction}.
Since this implication is only one way,
it could happen that $\mathcal{C}^{\vvs{\alpha},\vvs{\beta}}=\emptyset$
for such class label $\vvs{\alpha},\vvs{\beta}$ for which the right-hand side does not hold.
However, we think that this cannot happen:
\emph{Our second conjecture is that there is an equivalence in (\ref{eq:emptyByConstruction});}
that is, all the PS classes which are not empty by construction are nonempty.
[This implies the first conjecture above,
but it can still happen that implication (iii) can be replaced by a stronger condition,
so the first conjecture is false.
Then the (\ref{eq:emptyByConstruction}) definition of the emptiness-by-construction changes,
and the second conjecture concerns this new definition.]
The motivation of this is the same as
in the tripartite case, (see at the end of Sec.~\ref{subsec:Mixed:Classes}),
where the PS classes conjectured to be non-empty
were obtained under the same assumptions.

An advantage of the formulation by the labeling constructions
is---roughly speaking---that by the use of that
``we have separated the \emph{algebraic} and the \emph{geometric part}'' of the problem 
of nonemptiness of the classes.
At this point, it seems that 
we have tackled all the \emph{algebraic} issues of the question,
and these conjectures cannot be proven without the investigation of the \emph{geometry} of $\mathcal{D}$,
more precisely, the geometry of the different kinds of $\mathcal{P}^{\alpha}$ extremal points.

The negation of (\ref{eq:emptyByConstruction}) leads to
\begin{equation}
\label{eq:nonEmptyByConstruction}
\mathcal{C}^{\vvs{\alpha},\vvs{\beta}}\neq\emptyset 
\quad\Longrightarrow\quad
\forall \vs{\alpha}\in\vvs{\alpha},
\forall \vs{\beta}\in\vvs{\beta}\;:\; \vs{\beta}\npreceq \vs{\alpha},
\end{equation}
\end{subequations}
so if we obtain all $\vvs{\alpha},\vvs{\beta}$ class-labels
for which the right-hand side of this holds (``\emph{non-emptiness-by-construction}'')
then we will have all the nonempty classes,
together with some empty classes if the second conjecture does not hold.
Because of the nontrivial structure of the lattice of proper labels,
obtaining all the classlabels leading to
nonempty-by-construction classes
is also a nontrivial task.
%A final note for this subsection is that
The number of all the partitions of $n$ grows rapidly \cite{oeisA000110,oeisA000041},
which is only the number of the PS subsets of $\alpha$-separability $\mathcal{D}^{\alpha}$.
So the number of all the PS subsets $\mathcal{D}^{\vs{\alpha}}$ grows more rapidly,
and the number of all the PS classes $\mathcal{C}^{\vvs{\alpha},\vvs{\beta}}$ grows even more rapidly.
However, at least, it is finite.

%******************************************************************************
\subsection{Indicator functions in general}
\label{subsec:Gen:Indicators}

The third point is the generalization of the indicator functions.
Let $F:\mathcal{D}(\mathcal{H}^K)\to\field{R}$ be a continuous function
for all $K\subset L$, that is, for all---also composite---subsystems.
The only condition on $F$ is 
\begin{equation}
\label{eq:Fprop}
F(\varrho)\geq0, \quad\text{with equality if and only if $\varrho$ is pure},
\end{equation}
for example, the von Neumann entropy or any Tsallis or R\'enyi entropies are suitable.
(Note that the additional requirements of the features of
LU invariance, convexity, Schur-concavity, %entanglement monotonity, 
additivity, being homogeneous polynomial, etc.,
are only optional; 
they will not have been needed for the construction.)
For all $K\subset L$ subsystems,
let the following functions on pure states be defined:
\begin{equation}
\label{eq:fK}
\begin{split}
f_K&:\mathcal{P}\longrightarrow \field{R},\\
f_K&(\pi) = F(\pi_K),
\end{split}
\end{equation}
where again, $\pi=\cket{\psi}\bra{\psi}$, and $\pi_K=\tr_{\pcmpl{K}}(\pi)$, with $\pcmpl{K}=L\setminus K$.
Then, for the $k$-partite split $\alpha=L_1|L_2|\dots|L_k$,
$f_{L_r}$ identifies the bipartite split $L_r|\pcmpl{L_r}$, (where $\pcmpl{L_r}=L\setminus L_r$,) as
\begin{equation}
f_{L_r}(\pi)=0 \quad\Longleftrightarrow\quad 
\pi\in \bigcup_{\beta\preceq L_r|\pcmpl{L_r}}\mathcal{P}^\beta,
\end{equation}
which is the consequence of (\ref{eq:Fprop}).
Note that $\alpha$ is the greatest element which is smaller than $L_r|\pcmpl{L_r}$ for all $r$.
Then the function
\begin{equation}
\label{eq:genIndicators}
f_\alpha(\pi)=
\sum_{r=1}^k f_{L_r}(\pi),
\end{equation}
has the ability to identify the $k$-partite split $\alpha$ as
\begin{equation}
\label{eq:genIndicatorsdef}
f_\alpha(\pi)=0 \quad\Longleftrightarrow\quad 
\pi\in \bigcup_{\beta\preceq \alpha}\mathcal{P}^\beta.
\end{equation}
All non-negative $f_\alpha$ functions satisfying (\ref{eq:genIndicatorsdef})
are called \emph{$\alpha$-indicator functions},
not only the ones defined in (\ref{eq:genIndicators}).
The generalization of (\ref{eq:genIndicators}) for more-than-one partitions,
that is, for all labels, is defined as
\begin{equation}
\label{eq:genIndicatorsl}
f_{\vs{\alpha}}(\pi)= \prod_{\alpha\in\vs{\alpha}} f_{\alpha}(\pi),
\end{equation}
being the generalization of (\ref{eq:genNewPureLUinvs}).
It vanishes exactly for the convenient $\mathcal{P}^{\alpha}$s
\begin{equation}
\label{eq:genIndicatorsldef}
f_{\vs{\alpha}}(\pi)=0
\quad\Longleftrightarrow\quad 
\pi\in\bigcup_{\alpha\in\vs{\alpha}} \bigcup_{\beta\preceq\alpha}\mathcal{P}^\beta.
\end{equation}
All non-negative $f_{\vs{\alpha}}$ functions satisfying (\ref{eq:genIndicatorsldef})
are called \emph{$\vs{\alpha}$-indicator functions},
not only the ones defined in (\ref{eq:genIndicatorsl}).
For example, the functions in (\ref{eq:newPureLUinvs})
were not constructed by (\ref{eq:genIndicatorsl}),
but still satisfy (\ref{eq:genIndicatorsldef}),
which is equivalent to the relavant part of Table \ref{tab:pureSLOCC3} for the three-qubit case. 
Now, the vanishing of their convex roof extension 
\begin{equation*}
\cnvroof{f}_{\vs{\alpha}}(\varrho)=
\min\sum_i p_i f_{\vs{\alpha}}(\pi_i)
\end{equation*}
%(for all pure state decompositions $\sum_i p_i \pi_i=\varrho$)
identifies the PS sets:
\begin{equation}
\label{eq:genvanishing}
\cnvroof{f}_{\vs{\alpha}}(\varrho)=0
\quad\Longleftrightarrow\quad 
\varrho\in \mathcal{D}^{\vs{\alpha}},
\end{equation}
being the generalization of (\ref{eq:vanishing}).
Indeed,
$\cnvroof{f}_{\vs{\alpha}}(\varrho)=0$
if and only if there exists a decomposition $\varrho=\sum_ip_i\pi_i$
such that $f_{\vs{\alpha}}(\pi_i)=0$ for all $i$
($f_{\vs{\alpha}}$ is non-negative),
which means that $\pi_i\in\bigcup_{\alpha\in\vs{\alpha}} \bigcup_{\beta\preceq\alpha}\mathcal{P}^\beta$
(\ref{eq:genIndicatorsldef}),
which means that $\varrho\in\mathcal{D}^{\vs{\alpha}}$.

%******************************************************************************
\subsection{Entanglement-monotone indicator functions in general}
\label{subsec:Gen:monIndicators}

There is a possibility to choose indicator functions so that 
they obey some axioms of \emph{entanglement measures} \cite{HorodeckiEntMeas}.
The most fundamental one of them is the monotonicity under LOCC \cite{HorodeckiEntMeas,VidalEntMon}.
A $\mu:\mathcal{D}\to\field{R}$ is \emph{(nonincreasing) monotone under LOCC} 
if
\begin{subequations}
\begin{equation}
\label{eq:meas:mon}
\mu\bigl(\Lambda(\varrho)\bigr) \leq \mu(\varrho)
\end{equation}
for any LOCC transformation $\Lambda$,
which expresses that entanglement can not increase by the use of local operations
and classical communication.
A $\mu:\mathcal{D}\to\field{R}$ is \emph{nonincreasing on average under LOCC} 
if
\begin{equation}
\label{eq:meas:average}
\sum_j p_j \mu(\varrho_j) \leq \mu(\varrho),
\end{equation}
where the LOCC is constituted as $\Lambda=\sum_j \Lambda_j$,
where the $\Lambda_j$s are the parts of the LOCC realizing the outcomes of selective measurements,
and $\varrho_j=\frac1{p_j}\Lambda_j(\varrho)$
with $p_j=\tr[\Lambda_j(\varrho)]$.
This latter condition is stronger than the former one
if the function is \emph{convex}:
\begin{equation}
\label{eq:meas:conv}
\mu\Bigl(\sum_j p_j \varrho_j\Bigr) \leq \sum_j p_j \mu(\varrho_j)
\end{equation}
\end{subequations}
for all ensemble $\{(p_j,\varrho_j)\}$,
which expresses that entanglement can not increase by mixing.
A $\mu:\mathcal{D}\to\field{R}$ is \emph{entanglement-monotone}
if (\ref{eq:meas:average}) and (\ref{eq:meas:conv}) hold for that \cite{VidalEntMon}.
There is common agreement \cite{Horodecki4}
that LOCC-monotonity (\ref{eq:meas:mon}) is the only necessary postulate
for a function to be an \emph{entanglement measure.}
However, the stronger condition (\ref{eq:meas:average}) 
is often satisfied too,
and it is often easier to prove.
This holds also for our case.

If $\mu$ is defined only for pure states
$\mu:\mathcal{P}\to\field{R}$,
then only (\ref{eq:meas:average}) makes sense, whose restriction is
\begin{equation}
\label{eq:averagePure}
\sum_i p_i \mu(\pi_i) \leq \mu(\pi).
\end{equation}
Here $\{(p_i,\pi_i)\}$ is the pure ensemble
generated by all the Kraus operators of all $\Lambda_j$s
from the input state $\pi$.
(Not all $\pi_i$ members of the ensemble are accessible physically,
only the outcomes of the LOCC, which are formed by partial mixtures of this ensemble \cite{HorodeckiEntMeas}.
Mathematically, however, we can use the pure ensemble, which make the construction much more simple.)
If we have such a function $\mu:\mathcal{P}\to\field{R}$,
%as has been shown by Vidal \cite{VidalEntMon},
(\ref{eq:meas:average}) holds for its convex roof extension \cite{VidalEntMon,HorodeckiEntMeas}:
\begin{equation}
\label{eq:averageConvRoof}
\sum_i p_i \mu(\pi_i) \leq \mu(\pi)
\quad\Longrightarrow\quad
\sum_i p_i \cnvroof{\mu}(\varrho_i) \leq \cnvroof{\mu}(\varrho).
\end{equation}
Since the convex roof extension of a function is convex [Eq.~\ref{eq:meas:conv}]
(moreover, it is the largest convex function taking the same values for pure states as the original function does, \cite{UhlmannOptimalDecomp}),
$\cnvroof{\mu}(\varrho)$ is also entanglement-monotone.

\setcounter{txtitem}{0}

Now, we construct indicator functions which are entanglement-monotone.
(These are denoted $m$ in contrast with the general $f$s.)
This is carried out in four steps.

\txtitem{} It has also been shown in \cite{VidalEntMon,HorodeckiEntMeas} that 
if $F:\mathcal{D}(\mathcal{H}^K)\to\field{R}$ is unitary invariant and concave,
then the $f_K$ functions defined in (\ref{eq:fK}) are non increasing on average for pure states,
that is, obey (\ref{eq:averagePure}).
So let
\begin{equation}
\label{eq:mK}
m_K(\pi) = M(\pi_K)
\end{equation}
with $M:\mathcal{D}(\mathcal{H}^K)\to\field{R}$ \emph{vanishing if and only if the state is pure,} as before, 
but now we demand also \emph{unitary invariance} and \emph{concavity.}
The von Neumann entropy (\ref{eq:Neumann}), 
the Tsallis entropies (\ref{eq:Tsallis}) for all $q>0$,
and the R\'enyi entropies (\ref{eq:Renyi}) for all $0<q<1$
are known to be concave \cite{BengtssonZyczkowski}, and all of them are unitary invariant.

\txtitem{} Clearly, the functions obeying (\ref{eq:averagePure}) form a cone;
that is, their sums and multiples by non-negative real numbers also obey (\ref{eq:averagePure}),
so we can conclude that the sums of the functions $m_K$ also obey (\ref{eq:averagePure}).
Here, instead of the original sums in (\ref{eq:genIndicators}),
we introduce the \emph{arithmetic mean} of the $m_{L_r}$ functions,
\begin{equation}
\label{eq:monIndicators}
m_\alpha(\pi)=\frac1k \sum_{r=1}^k m_{L_r}(\pi),
\end{equation}
which are also indicator functions, since they obey (\ref{eq:genIndicatorsdef}).
[The factor $1/k$ is not really important,
but the next step, and 
in the three-qubit case $y=1/3(s_1+s_2+s_3)$ from
(\ref{eq:newPureLUinvs:y})--(\ref{eq:newPureLUinvs:sa}) 
motivate the use of mean values.]

\txtitem{} The only problem we face here is that
the set of functions obeying (\ref{eq:averagePure}) is not closed under multiplication, 
which is the case of the $f_{\vs{\alpha}}$ functions of (\ref{eq:genIndicatorsl}).
This is related to the fact that the product of two concave functions is not concave in general.
Moreover, a recent result of Eltschka \textit{et.~al.}~suggests
that functions obeying (\ref{eq:averagePure}) 
cannot be of arbitrary high degree 
(see Theorem I in \cite{EltschkaetalEntMon},
concerning a special class of functions),
so we make a trial of such a combination which does not change the degree,
but still fulfils the conditions (\ref{eq:genIndicatorsldef}).
The \emph{geometric mean} will be proven to be suitable, 
which is just a root of the product given in (\ref{eq:genIndicatorsl})
\begin{equation}
\label{eq:monIndicatorsl}
m_{\vs{\alpha}}(\pi)= \Bigl[\prod_{\alpha\in\vs{\alpha}} m_{\alpha}(\pi)\Bigr]^{1/l},
\end{equation}
where $l=\abs{\vs{\alpha}}$, the number of $m_{\alpha}$s in the product.
These functions obviously obey (\ref{eq:genIndicatorsldef})
and also (\ref{eq:averagePure}), which latter is proven in Appendix \ref{appsubsec:Gen:geomMean}.

\txtitem{} Now, the function $m_{\vs{\alpha}}(\pi)$ of (\ref{eq:monIndicatorsl}) 
is nonincreasing on average for pure states (\ref{eq:averagePure})
so 
\begin{equation*}
\cnvroof{m}_{\vs{\alpha}}(\varrho)=
\min\sum_i p_i m_{\vs{\alpha}}(\pi_i)
\end{equation*}
is also non increasing on average (\ref{eq:meas:average})
[because of (\ref{eq:averageConvRoof})], 
so it is entanglement-monotone
and also identifies the PS subsets
\begin{equation}
\cnvroof{m}_{\vs{\alpha}}(\varrho)=0
\quad\Longleftrightarrow\quad 
\varrho\in \mathcal{D}^{\vs{\alpha}},
\end{equation}
as in (\ref{eq:genvanishing}).

%******************************************************************************
%******************************************************************************
\section{Summary and remarks}
\label{sec:Sum}

In this paper we have constructed the complete PS classification of multipartite quantum states
by the PS classes arising from the PS subsets (\ref{eq:genDsets:alphal}),
together with necessary and sufficient conditions for the identification of the PS classes
through the necessary and sufficient conditions for the identification of the PS subsets (\ref{eq:genvanishing}) 
by indicator functions arising as convex roof extensions of the pure-state indicator functions (\ref{eq:genIndicatorsl}).
The indicator functions can be constructed so as to be entanglement-monotone (Sec.~\ref{subsec:Gen:monIndicators}).
A side track is the PSS classification of three-qubit states,
(Sec.~\ref{sec:Mixed}),
where a different set of indicator functions 
has been obtained (\ref{eq:newPureLUinvs}), (\ref{eq:vanishing}) 
by the use of the FTS approach of three-qubit pure-state entanglement.

\setcounter{txtitem}{0}

Now, we list some remarks and open questions, first about the general case.
\txtitem{} As was mentioned before,
this PS classification scheme is an extension of the classification based on $k$-separability and $\alpha_k$-separability
given by Seevinck and Uffink \cite{SeevinckUffinkMixSep},
which is the extension of the classification dealing only with $\alpha_k$-separability
given by D\"ur and Cirac \cite{DurCiracTarrach3QBMixSep,DurCiracTarrachBMixSep}.
\txtitem{} The nonemptiness of the new classes was only conjectured.
More fully, we could not give necessary and sufficient condition for the nonemptiness of the PS classes
in the purely algebraic language of labels.
Probably, methods from geometry or calculus would be needed to solve this puzzle
(Sec.~\ref{subsec:Gen:PSclasses}).
\txtitem{} In close connection with this,
a further geometry-related conjecture could be drafted about the nonempty classes:
They are of nonzero measure.
It is known in the bipartite case that
the set of separable states is of nonzero measure \cite{Acinetal3QBMixClass,BengtssonZyczkowski}, %ChenDjokovicSemialg},
which can motivate this conjecture.
\txtitem{} We have given also the necessary and sufficient criteria of the classes.
This was done by convex roof extension,
which is a general method for the identification of convex subsets,
having advantages and disadvantages.
\txtitem{} First of all, convex roof extensions are hard to calculate.
However, necessary and sufficient criteria for the detection of convex subsets 
seem always to be hard to calculate,
since they always contain an optimization problem,
such as finding a suitable witness,
or positive map \cite{HorodeckiPosMapWitness}, 
or symmetric extension \cite{Dohertycrit1,Dohertycrit2,Dohertycrit3},
or local spin measurements \cite{SeevinckUffinkMixSep},
or detection vector \cite{HuberkCrit,HuberCrit2}, 
or local bases \cite{GuhneSevinckCrit}, etc.
(The latter three are for only necessary but not sufficient criteria.
For further references, see the reference lists of \cite{SzalaySepCrit,GuhneTothEntDet}.)
These optimization problems have no solutions in a closed form in general cases.
%(The convex roof extension can be carried out explicitly in the two-qubit case \cite{HillWoottersConc,WoottersConc},
%but not for subsystems of arbitrary dimensions.)
\txtitem{} Another disadvantage of convex roof extensions
is that this is a ``clearly theoretical'' method,
under which we mean that
the full tomography of the state is needed, then the criteria are applied by computer.
The majority of the other known criteria share this disadvantage.
Exceptions are the criteria by witnesses \cite{HorodeckiPosMapWitness}
and by local spin measurements \cite{SeevinckUffinkMixSep} 
(only necessary but not sufficient),
where the criteria can be used in the laboratory, by the tuning of measurement settings.
However, the optimization still has to be done by the measuring apparatus.
\txtitem{} An advantage of the convex roof extension
is that it works independently of the dimensions of the subsystems,
so the criteria by that work for arbitrary dimensions.
However, the numerical optimization depends strongly on the rank of the state,
which can be high if the dimension is high,
resulting slow convergence.
\txtitem{} The greatest advantage of the criteria given by convex roof constructions---at least for us---is that
they have a very transparent structure;
they reflect clearly the structure of the PS classes by construction [see (\ref{eq:genvanishing})].

Now, we turn to some remarks about the three-qubit case,
which is although particular but very important.
\txtitem{} First, note that the FTS approach of three-qubit entanglement 
\cite{BorstenetalFreudenthal3QBEnt}
is coming from the famous Black Hole/Qubit Correspondence \cite{BorstenDuffLevayBHQB}.
The FTS approach has turned out to be fruitful also in some other 
fields of quantum entanglement \cite{PeterPetiFTS,PetiPeterFTS}.
There are some advantages of the FTS approach in the three-qubit case,
although, as we have seen, criteria for the PS subsets can be obtained without the use of that.
\txtitem{} Since the convex roof extensions of polynomials can be known to be semialgebraic functions \cite{PetiPriv,ChenDjokovicSemialg},
it can be useful to use LU-invariant homogeneous polynomials for the identification of the classes.
Then we have polynomials of this kind 
from (\ref{eq:newPureLUinvs}) coming from the FTS approach,
and from (\ref{eq:genNewPureLUinvs}) with the Tsallis entropy for $q=2$ coming from the general constructions.
The former ones are of lower degree, which may lead to more simple convex roof extensions.
\txtitem{} Moreover, 
this holds also for the functions $g_a$ in the general tripartite case
if Raggio's conjecture holds (Sec.~\ref{subsec:GenThreePart:FTS}).
\txtitem{} A little side result of our work is that
Raggio's conjecture holds
for two-qubit mixed states which are, at the most, of rank $2$.
\txtitem{} An interesting question is 
as to whether all pure-state indicator functions can be obtained
without products of local entropies,
but using only linear combinations of them.
\txtitem{} We note that there are also recent attempts to study the general structure
of LU invariant homogeneous polynomials \cite{HWLUA,HWWLUA,PetiLUA1,PetiLUA23,SzDeg6}.
Looking for convex roof extensions in the language of LU-invariant polynomials
would be an interesting direction of research.
\txtitem{} As a disadvantage of the FTS approach,
we have to mention that
some of the indicator functions coming from the FTS approach 
are not nonincreasing on average (\ref{eq:averagePure}), 
namely $g_a$ and $t$ given in (\ref{eq:newPureLUinvs:ga}) and (\ref{eq:newPureLUinvs:t}).
[Counterexamples for (\ref{eq:averagePure}) can be constructed for these functions by direct calculation.]

Finally, we would like to 
summarize some
arguments for the relevance of the extension of the
Seevinck-Uffink classification.
\txtitem{} We can get back the classification given by Seevinck and Uffink
if we simply forget about the sets $\varrho\in\mathcal{D}^{\twoprt{b|ac}{c|ab}}$,
and the functions $\cnvroof{g}_a(\varrho)$.
However, the appearance of the $g_a(\psi)$ polynomials is natural
in the light of the formulas (\ref{eq:newPureLUinvs:y}), (\ref{eq:newPureLUinvs:sa}), and (\ref{eq:newPureLUinvs:ga}).
This motivates the introduction of the sets $\varrho\in\mathcal{D}^{\twoprt{b|ac}{c|ab}}$
to the classification.
\txtitem{} The $g_a$ functions are interesting in themselves (see Appendix \ref{appsubsec:explicit:WoottersConc}).
For all non-GHZ vectors, they coincide with
the Wootters concurrence-squared of two-qubit subsystems (\ref{eq:Conc3QB}).
However, note that the Wootters concurrences of two-qubit subsystems
are not suitable for being indicator functions,
since they can be zero also for GHZ-type vectors,
so they do not fulfill the last row of $g_a$ columns of Table \ref{tab:pureSLOCC3}.
For example, for the usual GHZ state (\ref{eq:GHZ}),
the Wootters concurrences of two-qubit subsystems are zero.
\txtitem{} In Sec.~\ref{sec:Xmpl} 
%In subsection \ref{subsec:Xmpl:Explicit} 
we have shown states
which are definitely in classes that are different in the extended classification.
This is another reason for using also the sets $\varrho\in\mathcal{D}^{\twoprt{b|ac}{c|ab}}$
in the classification.

In closing, there is an important question, which can be of research interest as well.
\txtitem{} The PS classification is about the following issue:
``From which kinds of pure entangled states can a given state be \emph{mixed}?''
Another issue, 
which is equivalently important from the point of view of quantum computation
but which we have not dealt with, is 
``Which kinds of pure entangled states can be \emph{distilled out} from a given state?''
What can be said about the latter?

%******************************************************************************

\begin{acknowledgments}
We thank P\'eter L\'evay and P\'eter Vrana for helpful discussions
and D\'enes Petz for the reference of some papers.
This work was  supported by the New Hungary Development Plan
(Project ID: T\'AMOP-4.2.1/B-09/1/KMR-2010-0002).
\end{acknowledgments}

\appendix
%******************************************************************************
\section{For the three-qubit invariants}
\label{appsec:explicit}

In this appendix, we list some features of the 
LU-invariant homogeneous polynomials given in (\ref{eq:newPureLUinvs}) for three-qubit vectors.

\subsection{In the standard basis}
\label{appsubsec:explicit:stdLU}

In \cite{Sudbery3qb},
the following set of algebraically independent LU-invariant homogeneous polynomials is given
for three-qubit state vectors:
\begin{subequations}
\label{eq:canonPureLUinvs}
\begin{align}
I_0(\psi) &= \tr(\pi)\equiv \norm{\psi}^2,\\
I_a(\psi) &= \tr(\pi_a^2),\\
I_4(\psi) &=3\tr\bigl[(\pi_b\otimes\pi_c)\pi_{bc}\bigr] - \tr(\pi_b^3) - \tr(\pi_c^3),\\
I_5(\psi) &= \abs{\Det(\psi)}^2,
\end{align}
\end{subequations}
where $\pi=\cket{\psi}\bra{\psi}$.
Here $I_4$ is the Kempe invariant \cite{Kempe3qb} 
(the same for all different $b,c\in\{1,2,3\}$ labels),
arising in connection with hidden nonlocality.
We can alternatively form the set of LU-invariant polynomials of (\ref{eq:newPureLUinvs}) as
\begin{subequations}
\begin{align}
n   &= I_0,\\
y   &= 2I_0^2-\frac23\bigl(I_1+I_2+I_3\bigr),\\
s_a &= 2\bigl(I_0^2-I_a\bigr),\\
g_a &= I_0^2+I_a-I_b-I_c,\\
t   &= \frac83 I_4 + \frac{10}{3}I_0^3 - 2I_0\bigl(I_1+I_2+I_3\bigr),\\
\tau^2&= 4I_5.
\end{align}
\end{subequations}
Obviously, these are not independent.

\subsection{For the LU canonical form}
\label{appsubsec:explicit:LUcanon}
In \cite{AcinetalGenSchmidt3QB,Acinetal3QBPureCanon},
the following \emph{LU-canonical form} (``generalized Schmidt decomposition'') is obtained
for normalized pure three-qubit state vectors,
\begin{subequations}
\label{eq:LUcanonPure}
\begin{equation}
\label{eq:LUcanonPure:state}
\begin{split}
\cket{\psi_\text{Sch}}=
 \sqrt{\eta_0}\cket{000}
+\ee^{i\alpha}\sqrt{\eta_1}\cket{100}
+\sqrt{\eta_2}&\cket{101}\\
+\sqrt{\eta_3}\cket{110}
+\sqrt{\eta_4}&\cket{111}
\end{split}
\end{equation}
with the phase $0\leq\alpha\leq\pi$
and the amplitudes $\eta_i\geq0$, 
the latter being normalized:
\begin{equation}
\label{eq:LUcanonPure:norm}
\sum_{i=0}^4\eta_i=1.
\end{equation}
\end{subequations}
The LU-invariant homogeneous polynomials of (\ref{eq:newPureLUinvs})
for this canonical form
can be written in a more convenient way
using another set of independent LU-invariant homogeneous polynomials  
\cite{AcinetalGenSchmidt3QB,Acinetal3QBPureCanon},
which is equivalent to that of (\ref{eq:canonPureLUinvs}).
This set calculated for the canonical form (\ref{eq:LUcanonPure}) is
\begin{subequations}
%\label{eq:puremeasurescanon}
\begin{align}
J_1 &= \Delta,\\
J_2 &= \eta_0\eta_2,\\
J_3 &= \eta_0\eta_3,\\
J_4 &= \eta_0\eta_4,\\
J_5 &= \eta_0\bigl(\Delta+\eta_2\eta_3-\eta_1\eta_4\bigr),
\end{align}
\end{subequations}
where $\Delta=\abs{\sqrt{\eta_1\eta_4}\ee^{i\alpha} - \sqrt{\eta_2\eta_3}}^2$.
Calculating the polynomials of (\ref{eq:newPureLUinvs})
for the canonical form (\ref{eq:LUcanonPure}),
we can identify these invariants as
\begin{subequations}
\label{eq:puremeasurescanon}
\begin{align}
y       &= 4J_4 + \frac83\bigl(J_1+J_2+J_3\bigr),\\
s_a     &= 4\bigl(J_4+J_1+J_2+J_3-J_a\bigr),\\
g_a     &= 2J_4 + 4J_a,\\
\label{eq:puremeasurescanon:t}
t       &= 4J_4+8J_5,\\
\label{eq:puremeasurescanon:tau}
\tau^2  &= (4J_4)^2.
\end{align}
\end{subequations}

\subsection{Connections with the Wootters concurrence and fidelity of two-qubit subsystems}
\label{appsubsec:explicit:WoottersConc}

The concurrence $c(\chi)$ for a two-qubit pure state $\cket{\chi}\in\mathcal{H}^1\otimes\mathcal{H}^2$ is
given by the concurrence (\ref{eq:conc2}) of the one-qubit subsystems as
\begin{equation}
\label{eq:conc22}
c(\chi)=C\bigl[\tr_b\bigl(\cket{\chi}\bra{\chi}\bigr)\bigr]  =\vert\bracket{\tilde{\chi}\vert\chi}\vert,
\end{equation}
where $\bra{\tilde{\chi}}%=\sigma_2\otimes\sigma_2\cket{\chi}
=-\varepsilon\otimes\varepsilon\cket{\chi}$.
%where 
%$\cc{\chi}$ is the complex conjugation of the components calculated in the computation basis $\{\cket{ij}\}$,
%and 
%$\sigma_2=-i\varepsilon$.
[Note that 
$\varepsilon \in \Lin(\mathcal{H}^a\to\mathcal{H}^{a*})$,
so $\cket{\tilde{\chi}}=\bra{\tilde{\chi}}^\dagger=-\varepsilon_{ii'}\varepsilon_{jj'}\cc{\chi^{i'j'}}\cket{ij}$,
leading to the usual expression of the spin flip \cite{WoottersConc} with $\sigma_2=-i\varepsilon$.
On the other hand, note the familiar expression: 
$\bracket{\tilde{\chi}\vert\chi}=-\varepsilon_{ii'}\varepsilon_{jj'}\chi^{ij}\chi^{i'j'}$.]
Its square is the local entropy (\ref{eq:pureLUinvs:sa}):
$c^2(\chi)=s_a(\chi)=4\det(\tr_b\cket{\chi}\bra{\chi})$,
this time for the two-qubit pure state,
$a,b\in\{1,2\}$, $a\neq b$.

In the convex roof extension of (\ref{eq:conc22}), called Wootters concurrence,
the minimization can be carried out explicitly
\cite{HillWoottersConc,WoottersConc},
and for a two-qubit mixed state $\omega$
it is given by the famous formula
\begin{equation}
\label{eq:WConc}
\cnvroof{c}(\omega)=
\max\{0,\lambda^\downarrow_1-\lambda^\downarrow_2-\lambda^\downarrow_3-\lambda^\downarrow_4\},
\end{equation}
where $\lambda^\downarrow_i$s
are the decreasingly ordered eigenvalues
of the positive matrix $\sqrt{\sqrt{\omega}\tilde{\omega}\sqrt{\omega}}$
(being the same as the square root of the eigenvalues of the non-Hermitian matrix $\omega\tilde{\omega}$),
and the spin-flipped state is
$\tilde{\omega}=\varepsilon\otimes\varepsilon\cc{\omega}\varepsilon\otimes\varepsilon$.

If the two-qubit mixed state
for which the Wootters concurrence-squared is calculated 
is reduced from a pure three-qubit state $\pi=\cket{\psi}\bra{\psi}$ as
$\omega=\pi_{bc}=\tr_a\bigl(\cket{\psi}\bra{\psi}\bigr)$,
as was investigated in \cite{CKWThreetangle},
then $\pi_{bc}$ is, at the most, of rank $2$, 
and $\bigl[\cnvroof{c}(\pi_{bc})\bigr]^2=(\lambda_1-\lambda_2)^2
=\tr\bigl(\pi_{bc}\tilde{\pi}_{bc}\bigr)-2\lambda_1\lambda_2$.
One can check that
\begin{equation}
\label{eq:Conc3QB1}
\tr\bigl(\pi_{bc}\tilde{\pi}_{bc}\bigr)=\tr\bigl[\gamma_a(\psi)^\dagger\gamma_a(\psi)\bigr],
\end{equation}
which is just $g_a(\psi)$ [see in (\ref{eq:newPureLUinvs:ga})],
and 
\begin{equation}
\label{eq:Conc3QB2}
\lambda_1\lambda_2= \abs{\det[\gamma_a(\psi)]}=\abs{\Det(\psi)},
\end{equation}
(see \cite{CKWThreetangle}).
The concurrence is then given by
\begin{equation}
\label{eq:Conc3QB}
\bigl[\cnvroof{c}(\pi_{bc})\bigr]^2= g_a(\psi)-\frac12 \tau(\psi).
\end{equation}
The CKW (Coffmann-Kundu-Wootters) equality \cite{CKWThreetangle}
about entanglement monogamy,
\begin{equation}
\label{eq:CKW}
s_a(\psi)=\bigl[ \cnvroof{c}(\pi_{ab}) \bigr]^2+ \bigl[\cnvroof{c}(\pi_{ac})\bigr]^2+\tau(\psi),
\end{equation}
is then equivalent to (\ref{eq:newPureLUinvs:sa}).

The roof extension relates the concurrence
with another important quantity, the fidelity \cite{UhlmannFidelityConcurrence}.
The \emph{fidelity between two density matrices} $\omega$ and $\sigma$
is $F(\omega,\sigma)=\tr\sqrt{\sqrt{\omega}\sigma\sqrt{\omega}}$,
which is the square root of the transition probability,
and it is in connection with distances and distinguishability measures on the space of density matrices \cite{BengtssonZyczkowski}.
The \emph{fidelity of a state} with respect the spin flip is $F(\omega,\tilde{\omega})$, 
which is just the \emph{concave roof extension} of $c(\chi)$:
\begin{equation}
%\mathcal{F}(\omega)=F(\omega,\tilde{\omega})=\lambda_1+\lambda_2+\lambda_3+\lambda_4.
\cncroof{c}(\omega)=F(\omega,\tilde{\omega})=\lambda_1+\lambda_2+\lambda_3+\lambda_4
\end{equation}
in the two-qubit case.
[The concave roof extension is the maximization of the weighted average over the decompositions
instead of the minimization (\ref{eq:cnvroofext}).]
Again, for the mixed states of two-qubit subsystems 
arising from a three-qubit system being in a pure state,
the fidelity is 
%$\mathcal{F}^2(\pi_{bc})
$\bigl[\cncroof{c}(\pi_{bc})\bigr]^2
=(\lambda_1+\lambda_2)^2=\tr\bigl(\pi_{bc}\tilde{\pi}_{bc}\bigr)+2\lambda_1\lambda_2$.
Using (\ref{eq:Conc3QB1}) and (\ref{eq:Conc3QB2}),
it is of the form similar to the concurrence (\ref{eq:Conc3QB})
\begin{equation}
\label{eq:Fid3QB}
%\mathcal{F}^2(\pi_{bc})
\bigl[\cncroof{c}(\pi_{bc})\bigr]^2= g_a(\psi)+\frac12 \tau(\psi).
\end{equation}
Then, using (\ref{eq:newPureLUinvs:sa}),
we get a CKW-like equality for the fidelities,
\begin{equation}
\label{eq:CKWFid}
%s_a(\psi)= \mathcal{F}^2(\pi_{ab})+\mathcal{F}^2(\pi_{ac})-\tau(\psi).
s_a(\psi)= \bigl[ \cncroof{c}(\pi_{ab}) \bigr]^2+ \bigl[\cncroof{c}(\pi_{ac})\bigr]^2-\tau(\psi).
\end{equation}
On the other hand, from (\ref{eq:Conc3QB}) and (\ref{eq:Fid3QB}), 
$g_a(\psi)$ is just the average of the concave and convex roofs
and $\tau(\psi)$ is their difference
\begin{align}
g_a(\psi)&=\frac12\bigl( \bigl[\cncroof{c}(\pi_{bc})\bigr]^2+\bigl[\cnvroof{c}(\pi_{bc})\bigr]^2 \bigr),\\
\tau(\psi)&= \bigl[\cncroof{c}(\pi_{bc})\bigr]^2- \bigl[\cnvroof{c}(\pi_{bc})\bigr]^2.
\end{align}
Hence, on the zero-measured set of non-GHZ pure states,
the convex and concave roof extensions of $c(\chi)$ are equal,
and both of them are equal to $g_a(\psi)$.

%******************************************************************************
\subsection{The ranges of the invariants}
\label{appsubsec:explicit:ranges}
In this appendix, we show that the LU-invariant polynomials
of Eq.~(\ref{eq:newPureLUinvs})
range from $0$ to $1$ if the state vector is normalized.
[Otherwise, the maximum of the functions
goes by the corresponding power of the norm:
$0\leq y(\psi),s_a(\psi),g_a(\psi) \leq n^2(\psi)$,
$0\leq t(\psi) \leq n^3(\psi)$,
$0\leq \tau^2(\psi) \leq n^4(\psi)$.]

It is well known that $s_a(\psi)\leq1$, and this upper bound can be achieved.
Indeed, using the inequality of arithmetic and geometric means (AM-GM inequality),
$s_a(\psi)=4\det(\pi_a)=4\lambda_1\lambda_2\leq4\bigl[\frac12(\lambda_1+\lambda_2)\bigr]^2=1$ 
(for the eigenvalues of $\pi_a$),
which is saturated if and only if $\lambda_1=\lambda_2=\frac12$,
for example, for $\cket{\text{B}}_{ab}\otimes\cket{0}_c$,
where $\cket{\text{B}}$ is a Bell state of (\ref{eq:B}).

From (\ref{eq:newPureLUinvs:sa})
it is clear that $g_a(\psi)\leq s_b(\psi)\leq1$.
To show that the $g_a(\psi)\leq1$ upper bound can be achieved,
we note that, for example, $g_a(\cket{0}_a\otimes\cket{\text{B}}_{bc})=1$.

From (\ref{eq:newPureLUinvs:y}) and (\ref{eq:newPureLUinvs:sa}),
one can find that $y(\psi)=1/3\bigl(s_1(\psi)+s_3(\psi)+s_3(\psi)\bigr)\leq1$.
To show that the $y(\psi)\leq1$ upper bound can be achieved,
we note that, for example, 
for the usual normalized GHZ state (\ref{eq:GHZ})
$s_a(\text{GHZ})=1$.

It is known \cite{CKWThreetangle} that $\tau^2(\psi)\leq1$,
and this upper bound is achieved; for example, $\tau^2(\text{GHZ})=1$.

The proof of the $t(\psi)\leq1$ inequality is lengthy but straightforward.
$t$ is LU invariant,
so it is enough to calculate (\ref{eq:puremeasurescanon:t})
given for the LU-canonical form (\ref{eq:LUcanonPure}):
\begin{equation*}
t(\psi_\text{Sch})=
4\eta_0\bigl[\eta_4+4\sqrt{\eta_2\eta_3}\bigl(\sqrt{\eta_2\eta_3}-\sqrt{\eta_1\eta_4}\cos\alpha\bigr)\bigr].
\end{equation*}
First we note that for state vectors of genuine tripartite entanglement
$t(\psi)\neq0$ (see in Table \ref{tab:pureSLOCC3}),
so $\eta_0\neq0$.
On the other hand, $\tau(\psi)=4\eta_0\eta_4$ [see in (\ref{eq:puremeasurescanon:tau})],
so $\eta_4$ decides to which class the tripartite-entangled state belongs:
If $\cket{\psi_\text{Sch}}\in\mathcal{V}^\text{W}$, then $\eta_4=0$, and
if $\cket{\psi_\text{Sch}}\in\mathcal{V}^\text{GHZ}$, then $\eta_4\neq0$.

For $\cket{\psi_\text{Sch}}\in\mathcal{V}^\text{W}$, ($\eta_0\neq0$, $\eta_4=0$) we have
\begin{equation*}
\begin{split}
t(\psi_\text{Sch})&=16\eta_0\eta_2\eta_3
\leq 16\frac{1}{3^3}(\eta_0+\eta_2+\eta_3)^3\\
&=\frac{16}{27}(1-\eta_1)^3
\leq \frac{16}{27}=t(\text{W})<1,
\end{split}
\end{equation*}
where we have used the AM-GM inequality, and the (\ref{eq:LUcanonPure:norm}) normalization of the state.
The first two inequalities allow equalities for $\eta_0=\eta_2=\eta_3$,
and $\lambda_1=0$, respectively,
which results in a state $\frac1{\sqrt{3}}\bigl(\cket{000}+\cket{101}+\cket{110}\bigr)$
being LU equivalent to the usual normalized W state (\ref{eq:W}).

For $\cket{\psi_\text{Sch}}\in\mathcal{V}^\text{GHZ}$, ($\eta_0\neq0$, $\eta_4\neq0$) we have to make further distinctions.

If either $\eta_2=0$ or $\eta_3=0$, then
\begin{equation*}
\begin{split}
t(\psi_\text{Sch})&=4\eta_0\eta_4
\leq 4\frac{1}{2^2}(\eta_0+\eta_4)^2\\
&=(1-\eta_1-\eta_2-\eta_3)^2
\leq 1 =t(\text{GHZ})
\end{split}
\end{equation*}
by AM-GM inequality and normalization (\ref{eq:LUcanonPure:norm}).
The inequalities can be saturated for $\eta_1=\eta_2=\eta_3$,
in which case $\cket{\psi_\text{Sch}}=\cket{GHZ}$.

If $\eta_1=0$, then
\begin{equation*}
\begin{split}
t(\psi_\text{Sch})&=4\eta_0\bigl[\eta_4+4\eta_2\eta_3\bigr]\\
&\leq 4\eta_0\bigl[\eta_4+(\eta_2+\eta_3)^2\bigr]
\leq 4\eta_0\bigl[\eta_4+\eta_2+\eta_3\bigr],
\end{split}
\end{equation*}
where we used again the AM-GM inequality,
and that $x^2\leq x$ for all $0\leq x\leq1$ with the possibility of equality,
if and only if $x=0$ or $1$.
$\eta_2+\eta_3\neq1$ since the state is normalized and $\eta_0\neq0$.
So we have that the inequalities are equalities if and only if $\eta_2=\eta_3=0$,
which results that $t(\psi_\text{Sch})\leq 4\eta_0\eta_4$, which is maximized by the GHZ state, as in the previous case.

If $\eta_1\neq0$, $\eta_2\neq0$, and $\eta_3\neq0$, then
\begin{equation*}
\begin{split}
t(\psi_\text{Sch})&=4\eta_0\bigl[\eta_4+4\sqrt{\eta_2\eta_3}\bigl(\sqrt{\eta_2\eta_3}-\sqrt{\eta_1\eta_4}\cos\alpha\bigr)\bigr]\\
&\leq 4\eta_0\bigl[\eta_4+ 4\eta_2\eta_3 + 4\sqrt{\eta_2\eta_3\eta_1\eta_4})\bigr]\\
&\leq 4\eta_0\bigl[\eta_4+ (\eta_2+\eta_3)^2 + \frac14(\eta_2+\eta_3+\eta_1+\eta_4)^2\bigr]
\end{split}
\end{equation*}
where the inequalities are equalities for $\alpha=\pi$,
and $\eta_2=\eta_3=\eta_1=\eta_4$.
This results in $t(\psi_\text{Sch})\leq4\eta_0\eta_4[1+8\eta_4]$.
Using the normalization (\ref{eq:LUcanonPure:norm}) in this case: $\eta_0+4\eta_4=1$, 
we can find the maximum at $\eta_0=\frac1{6}\bigl(5-\sqrt7\bigr)$.
This is achieved for the state
\begin{equation*}
\begin{split}
\cket{\psi_\text{m}}=
&\sqrt{ \frac1{6}(5-\sqrt7) }\cket{000} +\\
&\sqrt{ \frac1{6}(1+\sqrt7) }\frac12\bigl(-\cket{100}+\cket{101}+\cket{110}+\cket{111} \bigr)
\end{split}
\end{equation*}
For this state $t(\psi_\text{m})=\frac{1}{54}(10+7\sqrt7)=0.52815\dots <1$
meaning that this is only a local maximum.

%******************************************************************************
\section{For the PS classification in general}
\label{appsec:Gen}

In this appendix, we prove some statements
used in the general PS classification.

\subsection{Inclusion}
\label{appsubsec:Gen:Incl}
Here we prove (\ref{eq:relationEquiv}) in the following steps:
\begin{equation*}
\begin{split}
\mathcal{D}^{\vs{\beta}}\subseteq\mathcal{D}^{\vs{\alpha}}
\quad&\overset{\text{(i)}}{\Longleftrightarrow}\quad
          \Conv \bigcup_{\beta\in\vs{\beta}}   \bigcup_{\delta\preceq\beta }\mathcal{P}^\delta
\subseteq \Conv \bigcup_{\alpha\in\vs{\alpha}} \bigcup_{\gamma\preceq\alpha}\mathcal{P}^\gamma\\
&\overset{\text{(ii)}}{\Longleftrightarrow}\quad
          \bigcup_{\beta\in\vs{\beta}}   \bigcup_{\delta\preceq\beta }\mathcal{P}^\delta
\subseteq \bigcup_{\alpha\in\vs{\alpha}} \bigcup_{\gamma\preceq\alpha}\mathcal{P}^\gamma\\
&\overset{\text{(iii)}}{\Longleftrightarrow}\quad
\forall \beta\in\vs{\beta}, \forall \delta\preceq\beta, \exists \alpha\in\vs{\alpha}: \delta\preceq\alpha\\
&\overset{\text{(iv)}}{\Longleftrightarrow}\quad
\forall \beta\in\vs{\beta}, \exists \alpha\in\vs{\alpha}:\; \beta\preceq\alpha\\
&\overset{\text{(v)}}{\Longleftrightarrow}\quad
\vs{\beta}\preceq\vs{\alpha}.
\end{split}
\end{equation*}

The \textit{equivalences (i) and (v)} are by definition (\ref{eq:genDsets:alphal}) and (\ref{eq:labelreldef}), respectively.

\textit{Equivalence (ii)} is the only one where 
it comes into the picture that the story is about quantum states.
The $\overset{\text{(ii)}}{\Longleftarrow}$ implication holds, since it is true, in general, that
$\Conv B \subseteq \Conv A \Leftarrow B \subseteq A$.
However, to the $\overset{\text{(ii)}}{\Longrightarrow}$ implication 
we have to use some special properties coming from geometry.
Obviously, for the extremal points,
\begin{equation*}
\begin{split}
            \Extr\Conv \bigcup_{\beta\in\vs{\beta}}   \bigcup_{\delta\preceq\beta}\mathcal{P}^\delta
      &\subseteq \Conv \bigcup_{\beta\in\vs{\beta}}   \bigcup_{\delta\preceq\beta}\mathcal{P}^\delta\\
&\qquad\subseteq \Conv \bigcup_{\alpha\in\vs{\alpha}} \bigcup_{\gamma\preceq\alpha }\mathcal{P}^\gamma,
\end{split}
\end{equation*}
so $\pi\in\Extr\Conv\bigcup_{\beta\in\vs{\beta}} \bigcup_{\delta\preceq\beta}\mathcal{P}^\delta$
is also the element of $\Conv \bigcup_{\alpha\in\vs{\alpha}} \bigcup_{\gamma\preceq\alpha  }\mathcal{P}^\gamma$.
However, $\pi$ is a projector of rank $1$,
so it is extremal also in $\Conv \bigcup_{\alpha\in\vs{\alpha}} \bigcup_{\gamma\preceq\alpha  }\mathcal{P}^\gamma$.
This holds for all $\pi$,
so we have that
\begin{equation*}
          \Extr\Conv\bigcup_{\beta\in\vs{\beta}}   \bigcup_{\delta\preceq\beta}\mathcal{P}^\delta
\subseteq \Extr\Conv\bigcup_{\alpha\in\vs{\alpha}} \bigcup_{\gamma\preceq\alpha }\mathcal{P}^\gamma.
\end{equation*}
Any $A$ sets of projectors of rank $1$ 
have the property that $A=\Extr\Conv A$;
that is, they are all extreme points of their convex hulls,
which leads to the $\overset{\text{(ii)}}{\Longrightarrow}$ implication.
%so $\Extr\Conv\bigcup_{j'=1}^{l'} \bigcup_{\delta\preceq\beta_{j'}}\mathcal{P}^\delta = \bigcup_{j'=1}^{l'} \bigcup_{\delta\preceq\beta_{j'}}\mathcal{P}^\delta$
%and $\Extr\Conv\bigcup_{j=1}^l     \bigcup_{\gamma\preceq\alpha_j  }\mathcal{P}^\gamma = \bigcup_{\gamma\preceq\alpha_j  }\mathcal{P}^\gamma$.

\textit{Equivalence (iii)} comes from set algebra.
To see the $\overset{\text{(iii)}}{\Longrightarrow}$ implication,
we note that the $\mathcal{P}^{\dots}$ sets are disjoint,
so every $\mathcal{P}^\delta$ on the left-hand side of the inclusion
appears on the right-hand side as a $\mathcal{P}^\gamma$,
which means that $\forall \beta\in\vs{\beta}$, $\forall \delta\preceq\beta$, $\exists \alpha\in\vs{\alpha}$ so that $\delta\preceq\alpha$.
To see the $\overset{\text{(iii)}}{\Longleftarrow}$ implication,
from the condition $\exists\alpha$ so that $\mathcal{P}^\delta\subseteq \bigcup_{\gamma\preceq\alpha  }\mathcal{P}^\gamma$,
but for different $\delta\preceq\beta$s there may exist different $\alpha$s,
so $\bigcup_{\delta\preceq\beta}\mathcal{P}^\delta\subseteq \bigcup_{\alpha\in\vs{\alpha}}\bigcup_{\gamma\preceq\alpha  }\mathcal{P}^\gamma$,
which holds for all $\beta\in\vs{\beta}$.

\textit{Equivalence (iv)} comes from the properties of partial ordering:
$\preceq$ is reflexive on partitions, that is, $\beta\preceq\beta$,
so the $\overset{\text{(iv)}}{\Longrightarrow}$ implication follows from the $\delta=\beta$ choice.
On the other hand,
$\preceq$ is transitive on partitions, which is just the $\overset{\text{(iv)}}{\Longleftarrow}$ implication:
For all $\delta$,
if $\delta\preceq\beta$ and $\beta\preceq\alpha$ then $\delta\preceq\alpha$.

%******************************************************************************
\subsection{Partial order on proper labels}
\label{appsubsec:Gen:pOrder}
We show that the relation (\ref{eq:labelreldef})
is a partial order on the set of proper labels (\ref{eq:properlabel}).

\textit{Reflexivity on labels:}
We need that $\vs{\alpha}\preceq\vs{\alpha}$, 
which means by definition (\ref{eq:labelreldef})
$\forall\alpha\in\vs{\alpha}$ $\exists \alpha'\in\vs{\alpha}$
for which $\alpha\preceq\alpha'$.
This holds with the $\alpha'=\alpha$ choice,
since $\preceq$ is reflexive on partitions, that is, $\alpha\preceq\alpha$.

\textit{Transitivity on labels:}
Suppose that $\vs{\beta}\preceq\vs{\alpha}$ and $\vs{\gamma}\preceq\vs{\beta}$,
so by definition (\ref{eq:labelreldef})
$\forall\gamma\in\vs{\gamma}$ $\exists \beta\in\vs{\beta}$,
for which $\gamma\preceq\beta$,
and
for this $\beta$ $\exists \alpha\in\vs{\alpha}$,
for which $\beta\preceq\alpha$.
Since $\preceq$ is transitive on partitions, we have that
$\forall\gamma\in\vs{\gamma}$ $\exists \alpha\in\vs{\alpha}$
for which $\gamma\preceq\alpha$,
which is $\vs{\gamma}\preceq\vs{\alpha}$  by definition (\ref{eq:labelreldef})

\textit{Antisymmetry on proper labels:}
Let $\vs{\alpha}$ and $\vs{\beta}$ be proper labels.
Suppose that $\vs{\beta}\preceq\vs{\alpha}$ and $\vs{\alpha}\preceq\vs{\beta}$,
so by definition (\ref{eq:labelreldef})
$\forall\beta\in\vs{\beta}$ $\exists \alpha\in\vs{\alpha}$
for which $\beta\preceq\alpha$,
and
for this $\alpha$ $\exists \beta'\in\vs{\beta}$
for which $\alpha\preceq\beta'$.
Since $\preceq$ is transitive on partitions, we have that
$\beta\preceq\beta'$.
This can be true only if $\beta=\beta'$,
since $\vs{\beta}$ is a proper label,
so we have that $\beta\preceq\alpha$ and $\alpha\preceq\beta$.
Since $\preceq$ is antisymmetric on partitions, we have that
$\forall\beta\in\vs{\beta}$, $\exists \alpha\in\vs{\alpha}$
for which $\alpha=\beta$, which means that $\vs{\beta}\subseteq\vs{\alpha}$.
Interchanging the roles of $\vs{\alpha}$ and $\vs{\beta}$,
we have that $\vs{\alpha}\subseteq\vs{\beta}$.
Since $\subseteq$ is antisymmetric on sets, we have that
$\vs{\beta}=\vs{\alpha}$.

%******************************************************************************
\subsection{For the injectivity of labeling with proper labels}
\label{appsubsec:Gen:propInj}

Let $\vs{\alpha}$ and $\vs{\beta}$ be proper labels.
Here we prove (\ref{eq:propInj}) in the following steps:
\begin{equation*}
\begin{split}
\mathcal{D}^{\vs{\beta}}=\mathcal{D}^{\vs{\alpha}}
%\;&\overset{\text{\mtitem{impl:Ddef}}}{\Longleftrightarrow}\;
%\;&\overset{\text{\txtitem{}\label{impl:Ddef}}}{\Longleftrightarrow}\;
\quad&\overset{\text{(i)}}{\Longleftrightarrow}\quad
\mathcal{D}^{\vs{\beta}}\subseteq\mathcal{D}^{\vs{\alpha}}
\;\;\text{and}\;\;
\mathcal{D}^{\vs{\alpha}}\subseteq\mathcal{D}^{\vs{\beta}}\\
&\overset{\text{(ii)}}{\Longleftrightarrow}\quad
\vs{\beta}\preceq\vs{\alpha}
\;\;\text{and}\;\;
\vs{\alpha}\preceq\vs{\beta}\\
&\overset{\text{(iii)}}{\Longleftrightarrow}\quad
\vs{\beta}=\vs{\alpha}.
\end{split}
\end{equation*}
\textit{Equivalence (i)} is the antisymmetry of $\subseteq$ on sets,
\textit{equivalence (ii)} is (\ref{eq:relationEquiv}) on labels,
\textit{equivalence (iii)} is the antisymmetry of $\preceq$ on proper labels.

%******************************************************************************
\subsection{For the surjectivity of labeling with proper labels}
\label{appsubsec:Gen:propPart}

Here we prove (\ref{eq:propPart}) in the following steps:
\begin{equation*}
\begin{split}
\mathcal{D}^{\vs{\alpha}\vs{\gamma}}=\mathcal{D}^{\vs{\alpha}}
%\;&\overset{\text{\mtitem{impl:Ddef}}}{\Longleftrightarrow}\;
%\;&\overset{\text{\txtitem{}\label{impl:Ddef}}}{\Longleftrightarrow}\;
\quad&\overset{\text{(i)}}{\Longleftrightarrow}\quad
\mathcal{D}^{\vs{\alpha}\vs{\gamma}}\subseteq\mathcal{D}^{\vs{\alpha}}
\;\;\text{and}\;\;
\mathcal{D}^{\vs{\alpha}}\subseteq\mathcal{D}^{\vs{\alpha}\vs{\gamma}}\\
&\overset{\text{(ii)}}{\Longleftrightarrow}\quad
\vs{\alpha}\vs{\gamma}\preceq\vs{\alpha}
\;\;\text{and}\;\;
\vs{\alpha}\preceq\vs{\alpha}\vs{\gamma}\\
&\overset{\text{(iii)}}{\Longleftrightarrow}\quad
\vs{\gamma}\preceq\vs{\alpha}.
\end{split}
\end{equation*}
\textit{Equivalence (i)} is the antisymmetry of $\subseteq$ on sets,
\textit{equivalence (ii)} is (\ref{eq:relationEquiv}) on labels,
($\vs{\alpha}\preceq\vs{\alpha}\vs{\gamma}$ holds always)
\textit{equivalence (iii)} is from the observation that
$\vs{\beta}\preceq\vs{\alpha}$ and $\vs{\beta}'\preceq\vs{\alpha}$
if and only if $\vs{\beta}\vs{\beta}'\preceq\vs{\alpha}$, 
which can be easily seen from the definition (\ref{eq:labelreldef}).

%******************************************************************************
\subsection{Algorithm generating PS sets}
\label{appsubsec:Gen:Alg}

To obtain an efficient algorithm generating the proper labels of
all PS subsets,
it is necessary to consider the labels as \emph{$l$-tuples of partitions} instead of \emph{sets of partitions.}
In this case, $\vs{\alpha}=\alpha_1,\alpha_2,\dots,\alpha_l$,
so the order of the elements is considered to be fixed when an $l$-tuple is given,
and the $\alpha_j$s are different for different $j$s.
[The $(\dots)$ $l$-tuple brackets are omitted.
Note that, contrary to \cite{SeevinckUffinkMixSep}, 
the lower index of the partitions $\alpha_j$ here does not refer to the $k$ number of $L_r$ sets in $\alpha_j$.]
Using ordered structure has further advantages beyond the obvious one that
a computer stores data sequentially, so implementing sets would mean additional difficulty.
Now the algorithm is the following.

\begin{enumerate}
\item{} [\textit{Initialization}] 
Fix an order of the partitions,
this will define a lexicographical ordering for $l$-tuples of partitions.
This is denoted by $<$.
(This is to avoid obtaining an $l$-tuple more than once
and to make the algorithm more optimized.)
\item{} [\textit{Level $1$}] Using this ordering,
we have all the $1$-tuples of partitions ordered lexicographically.
\item{} [\textit{Induction step:} obtaining the $l+1$-tuples of partitions (level $l+1$)
from the $l$-tuples of partitions (level $l$)]
To every $\vs{\alpha}=\alpha_1,\alpha_2,\dots,\alpha_l$ $l$-tuples 
(coming in lexicographically ordered sequence)
we have to append 
any such partition $\alpha_{l+1}$
(coming in lexicographically ordered sequence) that\\
(i) $\alpha_{l+1}\npreceq \alpha_j$ and $\alpha_j\npreceq\alpha_{l+1}$ for all $j=1,2,\dots,l$, and\\
(ii) $\alpha_{l+1}>\alpha_l$. 
[Because of the lexicographical order $<$, it is enough to consider only the last ($l$th) partition.] \\
Then the resulting $\vs{\alpha}=\alpha_1,\alpha_2,\dots,\alpha_l,\alpha_{l+1}$ $l+1$-tuples, 
and also the partitions in every such $l+1$-tuple are ordered lexicographically.
\end{enumerate}
The algorithm stops when no new partition can be appended to any of the $l$-tuples,
which comes in finite steps, since the number of all the partitions is finite.

%******************************************************************************
\subsection{Geometric means}
\label{appsubsec:Gen:geomMean}
Here we prove that if the functions 
$\mu_j:\mathcal{P}\to\field{R}$ ($j=1,2,\dots,q$)
are non-negative and nonincreasing on average,
\begin{subequations}
\begin{align}
\mu_j(\pi)&\geq0,\\
\label{eq:condAverage}
\sum_{i=1}^m p_i \mu_j(\pi_i) &\leq \mu_j(\pi),
\end{align}
\end{subequations}
then their geometric mean 
\begin{equation*}
\mu=(\mu_1\mu_2\dots\mu_q)^{1/q} 
\end{equation*}
is also non-negative (trivially) and nonincreasing on average
\begin{subequations}
\begin{align}
\mu(\pi)&\geq0,\\
\sum_{i=1}^m p_i \mu(\pi_i) &\leq \mu(\pi).
\end{align}
\end{subequations}
(We use this for functions defined on pure states,
although the following proof does not use that,
so the statement holds also for functions defined on all states.)

To obtain this, we will need a Cauchy-Bunyakowski-Schwarz-like inequality,
for non-negative vectors $\ve{x}^{(j)}\in\field{R}^m$, $x_i^{(j)}\geq0$:
\begin{equation}
\label{eq:CBSlike}
\sum_{i=1}^m  x_i^{(1)} x_i^{(2)}\dots x_i^{(q)} \leq \norm{\ve{x}^{(1)}}_q \norm{\ve{x}^{(2)}}_q\dots \norm{\ve{x}^{(q)}}_q,
\end{equation}
where the usual $q$-norm is 
\begin{equation}
\norm{\ve{x}}_q=\Bigl[\sum_{i=1}^m x_i^q\Bigr]^{1/q}.
\end{equation}
Indeed,
if $\ve{x}^{(j)}=\vs{0}$ for some $j$, then the inequality holds trivially,
or else
\begin{equation*}
\begin{split}
&\sum_{i=1}^m \frac{x_i^{(1)}}{\norm{\ve{x}^{(1)}}_q} \frac {x_i^{(2)}}{\norm{\ve{x}^{(2)}}_q} \dots\frac{x_i^{(q)}}{\norm{\ve{x}^{(q)}}_q}\\
\equiv&\sum_{i=1}^m \left[\frac{(x_i^{(1)})^q}{\norm{\ve{x}^{(1)}}_q^q} \frac{(x_i^{(2)})^q}{\norm{\ve{x}^{(2)}}_q^q}\dots
\frac{(x_i^{(q)})^q}{\norm{\ve{x}^{(q)}}_q^q}  \right]^{1/q}\\
\leq&\sum_{i=1}^m \frac1q\left[ \frac{(x_i^{(1)})^q}{\norm{\ve{x}^{(1)}}_q^q}+\frac{(x_i^{(2)})^q}{\norm{\ve{x}^{(2)}}_q^q}+\dots+
\frac{(x_i^{(q)})^q}{\norm{\ve{x}^{(q)}}_q^q} \right] = 1,
\end{split}
\end{equation*}
where the inequality follows
from the inequality of the arithmetic and geometric means, applied to all terms in the sum.

Using this,
\begin{equation*}
\begin{split}
 &\sum_{i=1}^m p_i\mu(\pi_i)
=\sum_{i=1}^m p_i \bigl[\mu_1(\pi_i)\mu_2(\pi_i)\dots \mu_q(\pi_i) \bigr]^{1/q}\\
=&\sum_{i=1}^m \bigl[p_i\mu_1(\pi_i)\bigr]^{1/q} \bigl[p_i\mu_2(\pi_i)\bigr]^{1/q}\dots \bigl[p_i\mu_q(\pi_i)\bigr]^{1/q}\\
\leq &\Bigl[\sum_{i=1}^mp_i\mu_1(\pi_i)\Bigr]^{1/q} \Bigl[\sum_{i=1}^mp_i\mu_2(\pi_i)\Bigr]^{1/q} \dots \Bigl[\sum_{i=1}^mp_i\mu_q(\pi_i)\Bigr]^{1/q}\\
\leq &\bigl[\mu_1(\pi)\big]^{1/q} \bigl[\mu_2(\pi)\bigr]^{1/q} \dots \bigl[\mu_q(\pi)\bigr]^{1/q}
= \mu(\pi),
\end{split}
\end{equation*}
where the first inequality is (\ref{eq:CBSlike}) for $x_i^{(j)}=\bigl[p_i\mu_j(\pi_i)\bigr]^{1/q}$
and the second inequality is the condition (\ref{eq:condAverage}).

\bibliography{mix3qb}

%Merlin.mbs v4.21 2009-07-09.
\begin{thebibliography}{10}%
\makeatletter
\providecommand \@ifxundefined [1]{%
 \ifx #1\undefined \expandafter \@firstoftwo
 \else \expandafter \@secondoftwo
\fi
}%
\providecommand \@ifnum [1]{%
 \ifnum #1\expandafter \@firstoftwo
 \else \expandafter \@secondoftwo
\fi
}%
\providecommand \enquote [1]{``#1''}%
\providecommand \bibnamefont  [1]{#1}%
\providecommand \bibfnamefont [1]{#1}%
\providecommand \citenamefont [1]{#1}%
\providecommand\href[0]{\@sanitize\@href}%
\providecommand\@href[1]{\endgroup\@@startlink{#1}\endgroup\@@href}%
\providecommand\@@href[1]{#1\@@endlink}%
\providecommand \@sanitize [0]{\begingroup\catcode`\&12\catcode`\#12\relax}%
\@ifxundefined \pdfoutput {\@firstoftwo}{%
 \@ifnum{\z@=\pdfoutput}{\@firstoftwo}{\@secondoftwo}%
}{%
 \providecommand\@@startlink[1]{\leavevmode\special{html:<a href="#1">}}%
 \providecommand\@@endlink[0]{\special{html:</a>}}%
}{%
 \providecommand\@@startlink[1]{%
  \leavevmode
  \pdfstartlink
   attr{/Border[0 0 1 ]/H/I/C[0 1 1]}%
   user{/Subtype/Link/A<</Type/Action/S/URI/URI(#1)>>}%
  \relax
 }%
 \providecommand\@@endlink[0]{\pdfendlink}%
}%
\providecommand \url  [0]{\begingroup\@sanitize \@url }%
\providecommand \@url [1]{\endgroup\@href {#1}{\urlprefix}}%
\providecommand \urlprefix [0]{URL }%
\providecommand \Eprint[0]{\href }%
\@ifxundefined \urlstyle {%
  \providecommand \doi [1]{doi:\discretionary{}{}{}#1}%
}{%
  \providecommand \doi [0]{doi:\discretionary{}{}{}\begingroup
  \urlstyle{rm}\Url }%
}%
\providecommand \doibase [0]{http://dx.doi.org/}%
\providecommand \Doi[1]{\href{\doibase#1}}%
\providecommand \bibAnnote [3]{%
  \BibitemShut{#1}%
  \begin{quotation}\noindent
    \textsc{Key:}\ #2\\\textsc{Annotation:}\ #3%
  \end{quotation}%
}%
\providecommand \bibAnnoteFile [2]{%
  \IfFileExists{#2}{\bibAnnote {#1} {#2} {\input{#2}}}{}%
}%
\providecommand \typeout [0]{\immediate \write \m@ne }%
\providecommand \selectlanguage [0]{\@gobble}%
\providecommand \bibinfo [0]{\@secondoftwo}%
\providecommand \bibfield [0]{\@secondoftwo}%
\providecommand \translation [1]{[#1]}%
\providecommand \BibitemOpen[0]{}%
\providecommand \bibitemStop [0]{}%
\providecommand \bibitemNoStop [0]{.\EOS\space}%
\providecommand \EOS [0]{\spacefactor3000\relax}%
\providecommand \BibitemShut [1]{\csname bibitem#1\endcsname}%
%</preamble>
\bibitem{BengtssonZyczkowski}%
  \BibitemOpen
  \bibfield{author}{%
  \bibinfo {author} {\bibfnamefont{I.}~\bibnamefont{Bengtsson}}\ and\ \bibinfo
  {author} {\bibfnamefont{K.}~\bibnamefont{\r{Z}yczkowski}},\ }%
  \emph{\bibinfo {title} {Geometry of Quantum States: An Introduction to
  Quantum Entanglement}}\ (\bibinfo {publisher} {Cambridge University Press},\
  \bibinfo {address} {New York, NY, USA},\ \bibinfo {year} {2006})%
  \bibAnnoteFile{NoStop}{BengtssonZyczkowski}%
\bibitem{Horodecki4}%
  \BibitemOpen
  \bibfield{author}{%
  \bibinfo {author} {\bibfnamefont{R.}~\bibnamefont{Horodecki}}, \bibinfo
  {author} {\bibfnamefont{P.}~\bibnamefont{Horodecki}}, \bibinfo {author}
  {\bibfnamefont{M.}~\bibnamefont{Horodecki}},\ and\ \bibinfo {author}
  {\bibfnamefont{K.}~\bibnamefont{Horodecki}},\ }%
  \bibfield{journal}{%
  \Doi{10.1103/RevModPhys.81.865}{\bibinfo {journal} {Rev. Mod. Phys.}}\ }%
  \textbf{\bibinfo {volume} {81}},\ \bibinfo {pages} {865} (\bibinfo {month}
  {Jun}\ \bibinfo {year} {2009})%
  \bibAnnoteFile{NoStop}{Horodecki4}%
\bibitem{Schrodinger}%
  \BibitemOpen
  \bibfield{author}{%
  \bibinfo {author} {\bibfnamefont{E.}~\bibnamefont{Schr\"odinger}},\ }%
  \bibfield{journal}{%
  \bibinfo {journal} {Naturwissenschaften}\ }%
  \textbf{\bibinfo {volume} {23}},\ \bibinfo {pages} {807} (\bibinfo {year}
  {1935})%
  \bibAnnoteFile{NoStop}{Schrodinger}%
\bibitem{Schrodinger2}%
  \BibitemOpen
  \bibfield{author}{%
  \bibinfo {author} {\bibfnamefont{E.}~\bibnamefont{Schr\"odinger}},\ }%
  \bibfield{journal}{%
  \Doi{10.1017/S0305004100013554}{\bibinfo {journal} {Math. Proc. Camb. Phil.
  Soc.}}\ }%
  \textbf{\bibinfo {volume} {31}},\ \bibinfo {pages} {555} (\bibinfo {year}
  {1935})%
  \bibAnnoteFile{NoStop}{Schrodinger2}%
\bibitem{NielsenChuang}%
  \BibitemOpen
  \bibfield{author}{%
  \bibinfo {author} {\bibfnamefont{M.~A.}\ \bibnamefont{Nielsen}}\ and\
  \bibinfo {author} {\bibfnamefont{I.~L.}\ \bibnamefont{Chuang}},\ }%
  \emph{\bibinfo {title} {Quantum Computation and Quantum Information}},\
  \bibinfo {edition} {1st}\ ed.\ (\bibinfo {publisher} {Cambridge University
  Press},\ \bibinfo {year} {2000})%
  \bibAnnoteFile{NoStop}{NielsenChuang}%
\bibitem{SeevinckUffinkMixSep}%
  \BibitemOpen
  \bibfield{author}{%
  \bibinfo {author} {\bibfnamefont{M.}~\bibnamefont{Seevinck}}\ and\ \bibinfo
  {author} {\bibfnamefont{J.}~\bibnamefont{Uffink}},\ }%
  \bibfield{journal}{%
  \Doi{10.1103/PhysRevA.78.032101}{\bibinfo {journal} {Phys. Rev. A}}\ }%
  \textbf{\bibinfo {volume} {78}},\ \bibinfo {pages} {032101} (\bibinfo {month}
  {Sep}\ \bibinfo {year} {2008})%
  \bibAnnoteFile{NoStop}{SeevinckUffinkMixSep}%
\bibitem{DurCiracTarrach3QBMixSep}%
  \BibitemOpen
  \bibfield{author}{%
  \bibinfo {author} {\bibfnamefont{W.}~\bibnamefont{D\"ur}}, \bibinfo {author}
  {\bibfnamefont{J.~I.}\ \bibnamefont{Cirac}},\ and\ \bibinfo {author}
  {\bibfnamefont{R.}~\bibnamefont{Tarrach}},\ }%
  \bibfield{journal}{%
  \Doi{10.1103/PhysRevLett.83.3562}{\bibinfo {journal} {Phys. Rev. Lett.}}\ }%
  \textbf{\bibinfo {volume} {83}},\ \bibinfo {pages} {3562} (\bibinfo {month}
  {Oct}\ \bibinfo {year} {1999})%
  \bibAnnoteFile{NoStop}{DurCiracTarrach3QBMixSep}%
\bibitem{DurCiracTarrachBMixSep}%
  \BibitemOpen
  \bibfield{author}{%
  \bibinfo {author} {\bibfnamefont{W.}~\bibnamefont{D\"ur}}\ and\ \bibinfo
  {author} {\bibfnamefont{J.~I.}\ \bibnamefont{Cirac}},\ }%
  \bibfield{journal}{%
  \Doi{10.1103/PhysRevA.61.042314}{\bibinfo {journal} {Phys. Rev. A}}\ }%
  \textbf{\bibinfo {volume} {61}},\ \bibinfo {pages} {042314} (\bibinfo {month}
  {Mar}\ \bibinfo {year} {2000})%
  \bibAnnoteFile{NoStop}{DurCiracTarrachBMixSep}%
\bibitem{DurVidalCiracSLOCC3QB}%
  \BibitemOpen
  \bibfield{author}{%
  \bibinfo {author} {\bibfnamefont{W.}~\bibnamefont{D\"ur}}, \bibinfo {author}
  {\bibfnamefont{G.}~\bibnamefont{Vidal}},\ and\ \bibinfo {author}
  {\bibfnamefont{J.~I.}\ \bibnamefont{Cirac}},\ }%
  \bibfield{journal}{%
  \Doi{10.1103/PhysRevA.62.062314}{\bibinfo {journal} {Phys. Rev. A}}\ }%
  \textbf{\bibinfo {volume} {62}},\ \bibinfo {pages} {062314} (\bibinfo {month}
  {Nov}\ \bibinfo {year} {2000})%
  \bibAnnoteFile{NoStop}{DurVidalCiracSLOCC3QB}%
\bibitem{Acinetal3QBMixClass}%
  \BibitemOpen
  \bibfield{author}{%
  \bibinfo {author} {\bibfnamefont{A.}~\bibnamefont{Ac\'in}}, \bibinfo {author}
  {\bibfnamefont{D.}~\bibnamefont{Bru\ss{}}}, \bibinfo {author}
  {\bibfnamefont{M.}~\bibnamefont{Lewenstein}},\ and\ \bibinfo {author}
  {\bibfnamefont{A.}~\bibnamefont{Sanpera}},\ }%
  \bibfield{journal}{%
  \Doi{10.1103/PhysRevLett.87.040401}{\bibinfo {journal} {Phys. Rev. Lett.}}\
  }%
  \textbf{\bibinfo {volume} {87}},\ \bibinfo {pages} {040401} (\bibinfo {month}
  {Jul}\ \bibinfo {year} {2001})%
  \bibAnnoteFile{NoStop}{Acinetal3QBMixClass}%
\bibitem{BorstenetalFreudenthal3QBEnt}%
  \BibitemOpen
  \bibfield{author}{%
  \bibinfo {author} {\bibfnamefont{L.}~\bibnamefont{Borsten}}, \bibinfo
  {author} {\bibfnamefont{D.}~\bibnamefont{Dahanayake}}, \bibinfo {author}
  {\bibfnamefont{M.~J.}\ \bibnamefont{Duff}}, \bibinfo {author}
  {\bibfnamefont{W.}~\bibnamefont{Rubens}},\ and\ \bibinfo {author}
  {\bibfnamefont{H.}~\bibnamefont{Ebrahim}},\ }%
  \bibfield{journal}{%
  \Doi{10.1103/PhysRevA.80.032326}{\bibinfo {journal} {Phys. Rev. A}}\ }%
  \textbf{\bibinfo {volume} {80}},\ \bibinfo {pages} {032326} (\bibinfo {month}
  {Sep}\ \bibinfo {year} {2009})%
  \bibAnnoteFile{NoStop}{BorstenetalFreudenthal3QBEnt}%
\bibitem{CKWThreetangle}%
  \BibitemOpen
  \bibfield{author}{%
  \bibinfo {author} {\bibfnamefont{V.}~\bibnamefont{Coffman}}, \bibinfo
  {author} {\bibfnamefont{J.}~\bibnamefont{Kundu}},\ and\ \bibinfo {author}
  {\bibfnamefont{W.~K.}\ \bibnamefont{Wootters}},\ }%
  \bibfield{journal}{%
  \Doi{10.1103/PhysRevA.61.052306}{\bibinfo {journal} {Phys. Rev. A}}\ }%
  \textbf{\bibinfo {volume} {61}},\ \bibinfo {pages} {052306} (\bibinfo {month}
  {Apr}\ \bibinfo {year} {2000})%
  \bibAnnoteFile{NoStop}{CKWThreetangle}%
\bibitem{OsborneVerstraeteMonogamy}%
  \BibitemOpen
  \bibfield{author}{%
  \bibinfo {author} {\bibfnamefont{T.~J.}\ \bibnamefont{Osborne}}\ and\
  \bibinfo {author} {\bibfnamefont{F.}~\bibnamefont{Verstraete}},\ }%
  \bibfield{journal}{%
  \Doi{10.1103/PhysRevLett.96.220503}{\bibinfo {journal} {Phys. Rev. Lett.}}\
  }%
  \textbf{\bibinfo {volume} {96}},\ \bibinfo {pages} {220503} (\bibinfo {month}
  {Jun}\ \bibinfo {year} {2006})%
  \bibAnnoteFile{NoStop}{OsborneVerstraeteMonogamy}%
\bibitem{PeterGeom3QBEnt}%
  \BibitemOpen
  \bibfield{author}{%
  \bibinfo {author} {\bibfnamefont{P.}~\bibnamefont{L\'evay}},\ }%
  \bibfield{journal}{%
  \Doi{10.1103/PhysRevA.71.012334}{\bibinfo {journal} {Phys. Rev. A}}\ }%
  \textbf{\bibinfo {volume} {71}},\ \bibinfo {pages} {012334} (\bibinfo {month}
  {Jan}\ \bibinfo {year} {2005})%
  \bibAnnoteFile{NoStop}{PeterGeom3QBEnt}%
\bibitem{BennettetalMixedStates}%
  \BibitemOpen
  \bibfield{author}{%
  \bibinfo {author} {\bibfnamefont{C.~H.}\ \bibnamefont{Bennett}}, \bibinfo
  {author} {\bibfnamefont{D.~P.}\ \bibnamefont{DiVincenzo}}, \bibinfo {author}
  {\bibfnamefont{J.~A.}\ \bibnamefont{Smolin}},\ and\ \bibinfo {author}
  {\bibfnamefont{W.~K.}\ \bibnamefont{Wootters}},\ }%
  \bibfield{journal}{%
  \Doi{10.1103/PhysRevA.54.3824}{\bibinfo {journal} {Phys. Rev. A}}\ }%
  \textbf{\bibinfo {volume} {54}},\ \bibinfo {pages} {3824} (\bibinfo {month}
  {Nov}\ \bibinfo {year} {1996})%
  \bibAnnoteFile{NoStop}{BennettetalMixedStates}%
\bibitem{UhlmannFidelityConcurrence}%
  \BibitemOpen
  \bibfield{author}{%
  \bibinfo {author} {\bibfnamefont{A.}~\bibnamefont{Uhlmann}},\ }%
  \bibfield{journal}{%
  \Doi{10.1103/PhysRevA.62.032307}{\bibinfo {journal} {Phys. Rev. A}}\ }%
  \textbf{\bibinfo {volume} {62}},\ \bibinfo {pages} {032307} (\bibinfo {month}
  {Aug}\ \bibinfo {year} {2000})%
  \bibAnnoteFile{NoStop}{UhlmannFidelityConcurrence}%
\bibitem{UhlmannConvRoofs}%
  \BibitemOpen
  \bibfield{author}{%
  \bibinfo {author} {\bibfnamefont{A.}~\bibnamefont{Uhlmann}},\ }%
  \bibfield{journal}{%
  \Doi{10.3390/e12071799}{\bibinfo {journal} {Entropy}}\ }%
  \textbf{\bibinfo {volume} {12}},\ \bibinfo {pages} {1799} (\bibinfo {month}
  {July}\ \bibinfo {year} {2010})%
  \bibAnnoteFile{NoStop}{UhlmannConvRoofs}%
\bibitem{WernerSep}%
  \BibitemOpen
  \bibfield{author}{%
  \bibinfo {author} {\bibfnamefont{R.~F.}\ \bibnamefont{Werner}},\ }%
  \bibfield{journal}{%
  \Doi{10.1103/PhysRevA.40.4277}{\bibinfo {journal} {Phys. Rev. A}}\ }%
  \textbf{\bibinfo {volume} {40}},\ \bibinfo {pages} {4277} (\bibinfo {month}
  {Oct}\ \bibinfo {year} {1989})%
  \bibAnnoteFile{NoStop}{WernerSep}%
\bibitem{SzalaySepCrit}%
  \BibitemOpen
  \bibfield{author}{%
  \bibinfo {author} {\bibfnamefont{{\relax Sz}.}~\bibnamefont{{Sz}alay}},\ }%
  \bibfield{journal}{%
  \Doi{10.1103/PhysRevA.83.062337}{\bibinfo {journal} {Phys. Rev. A}}\ }%
  \textbf{\bibinfo {volume} {83}},\ \bibinfo {pages} {062337} (\bibinfo {month}
  {Jun}\ \bibinfo {year} {2011})%
  \bibAnnoteFile{NoStop}{SzalaySepCrit}%
\bibitem{GuhneTothEntDet}%
  \BibitemOpen
  \bibfield{author}{%
  \bibinfo {author} {\bibfnamefont{O.}~\bibnamefont{G\"uhne}}\ and\ \bibinfo
  {author} {\bibfnamefont{G.}~\bibnamefont{T\'oth}},\ }%
  \bibfield{journal}{%
  \Doi{DOI: 10.1016/j.physrep.2009.02.004}{\bibinfo {journal} {Physics
  Reports}}\ }%
  \textbf{\bibinfo {volume} {474}},\ \bibinfo {pages} {1 } (\bibinfo {year}
  {2009})%
  \bibAnnoteFile{NoStop}{GuhneTothEntDet}%
\bibitem{BennettetalEquivalences}%
  \BibitemOpen
  \bibfield{author}{%
  \bibinfo {author} {\bibfnamefont{C.~H.}\ \bibnamefont{Bennett}}, \bibinfo
  {author} {\bibfnamefont{S.}~\bibnamefont{Popescu}}, \bibinfo {author}
  {\bibfnamefont{D.}~\bibnamefont{Rohrlich}}, \bibinfo {author}
  {\bibfnamefont{J.~A.}\ \bibnamefont{Smolin}},\ and\ \bibinfo {author}
  {\bibfnamefont{A.~V.}\ \bibnamefont{Thapliyal}},\ }%
  \bibfield{journal}{%
  \Doi{10.1103/PhysRevA.63.012307}{\bibinfo {journal} {Phys. Rev. A}}\ }%
  \textbf{\bibinfo {volume} {63}},\ \bibinfo {pages} {012307} (\bibinfo {month}
  {Dec}\ \bibinfo {year} {2000})%
  \bibAnnoteFile{NoStop}{BennettetalEquivalences}%
\bibitem{LindenPopescuOnMultipartEnt}%
  \BibitemOpen
  \bibfield{author}{%
  \bibinfo {author} {\bibfnamefont{N.}~\bibnamefont{Linden}}\ and\ \bibinfo
  {author} {\bibfnamefont{S.}~\bibnamefont{Popescu}},\ }%
  \bibfield{journal}{%
  \bibinfo {journal} {Fortschr. Phys.}\ }%
  \textbf{\bibinfo {volume} {46}},\ \bibinfo {pages} {567} (\bibinfo {year}
  {1998})%
  \bibAnnoteFile{NoStop}{LindenPopescuOnMultipartEnt}%
\bibitem{AcinetalGenSchmidt3QB}%
  \BibitemOpen
  \bibfield{author}{%
  \bibinfo {author} {\bibfnamefont{A.}~\bibnamefont{Ac\'in}}, \bibinfo {author}
  {\bibfnamefont{A.}~\bibnamefont{Andrianov}}, \bibinfo {author}
  {\bibfnamefont{L.}~\bibnamefont{Costa}}, \bibinfo {author}
  {\bibfnamefont{E.}~\bibnamefont{Jan\'e}}, \bibinfo {author}
  {\bibfnamefont{J.~I.}\ \bibnamefont{Latorre}},\ and\ \bibinfo {author}
  {\bibfnamefont{R.}~\bibnamefont{Tarrach}},\ }%
  \bibfield{journal}{%
  \Doi{10.1103/PhysRevLett.85.1560}{\bibinfo {journal} {Phys. Rev. Lett.}}\ }%
  \textbf{\bibinfo {volume} {85}},\ \bibinfo {pages} {1560} (\bibinfo {month}
  {Aug}\ \bibinfo {year} {2000})%
  \bibAnnoteFile{NoStop}{AcinetalGenSchmidt3QB}%
\bibitem{Acinetal3QBPureCanon}%
  \BibitemOpen
  \bibfield{author}{%
  \bibinfo {author} {\bibfnamefont{A.}~\bibnamefont{Ac\'in}}, \bibinfo {author}
  {\bibfnamefont{A.}~\bibnamefont{Andrianov}}, \bibinfo {author}
  {\bibfnamefont{E.}~\bibnamefont{Jan\'e}},\ and\ \bibinfo {author}
  {\bibfnamefont{R.}~\bibnamefont{Tarrach}},\ }%
  \bibfield{journal}{%
  \bibinfo {journal} {J. Phys. A}\ }%
  \textbf{\bibinfo {volume} {34}},\ \bibinfo {pages} {6725} (\bibinfo {year}
  {2001})%
  \bibAnnoteFile{NoStop}{Acinetal3QBPureCanon}%
\bibitem{Sudbery3qb}%
  \BibitemOpen
  \bibfield{author}{%
  \bibinfo {author} {\bibfnamefont{A.}~\bibnamefont{Sudbery}},\ }%
  \bibfield{journal}{%
  \bibinfo {journal} {J. Phys. A}\ }%
  \textbf{\bibinfo {volume} {34}},\ \bibinfo {pages} {643} (\bibinfo {year}
  {2001})%
  \bibAnnoteFile{NoStop}{Sudbery3qb}%
\bibitem{Kempe3qb}%
  \BibitemOpen
  \bibfield{author}{%
  \bibinfo {author} {\bibfnamefont{J.}~\bibnamefont{Kempe}},\ }%
  \bibfield{journal}{%
  \Doi{10.1103/PhysRevA.60.910}{\bibinfo {journal} {Phys. Rev. A}}\ }%
  \textbf{\bibinfo {volume} {60}},\ \bibinfo {pages} {910} (\bibinfo {year}
  {1999})%
  \bibAnnoteFile{NoStop}{Kempe3qb}%
\bibitem{VerstraeteetalSLOCC4QB}%
  \BibitemOpen
  \bibfield{author}{%
  \bibinfo {author} {\bibfnamefont{F.}~\bibnamefont{Verstraete}}, \bibinfo
  {author} {\bibfnamefont{J.}~\bibnamefont{Dehaene}}, \bibinfo {author}
  {\bibfnamefont{B.}~\bibnamefont{De~Moor}},\ and\ \bibinfo {author}
  {\bibfnamefont{H.}~\bibnamefont{Verschelde}},\ }%
  \bibfield{journal}{%
  \Doi{10.1103/PhysRevA.65.052112}{\bibinfo {journal} {Phys. Rev. A}}\ }%
  \textbf{\bibinfo {volume} {65}},\ \bibinfo {pages} {052112} (\bibinfo {month}
  {Apr}\ \bibinfo {year} {2002})%
  \bibAnnoteFile{NoStop}{VerstraeteetalSLOCC4QB}%
\bibitem{ChterentalDjokovicSLOCC4QB}%
  \BibitemOpen
  \bibfield{author}{%
  \bibinfo {author} {\bibfnamefont{O.}~\bibnamefont{Chterental}}\ and\ \bibinfo
  {author} {\bibfnamefont{D.~Z.}\ \bibnamefont{Djokovic}},\ }%
  in\ \emph{\bibinfo {booktitle} {Linear Algebra Research Advances}}\ (\bibinfo
  {publisher} {Nova Science Publishers},\ \bibinfo {year} {2007})\
  Chap.~\bibinfo {chapter} {5}, pp.\ \bibinfo {pages} {133--167},\
  \Eprint{http://arxiv.org/abs/arXiv:quant-ph/0612184}{arXiv:quant-ph/0612184}%
  \bibAnnoteFile{NoStop}{ChterentalDjokovicSLOCC4QB}%
\bibitem{CayleyHDet}%
  \BibitemOpen
  \bibfield{author}{%
  \bibinfo {author} {\bibfnamefont{A.}~\bibnamefont{Cayley}},\ }%
  \bibfield{journal}{%
  \bibinfo {journal} {Camb. Math. J.}\ }%
  \textbf{\bibinfo {volume} {4}},\ \bibinfo {pages} {193} (\bibinfo {year}
  {1845})%
  \bibAnnoteFile{NoStop}{CayleyHDet}%
\bibitem{GelfandetalDiscriminants}%
  \BibitemOpen
  \bibfield{author}{%
  \bibinfo {author} {\bibfnamefont{I.~M.}\ \bibnamefont{Gel'fand}}, \bibinfo
  {author} {\bibfnamefont{M.~M.}\ \bibnamefont{Kapranov}},\ and\ \bibinfo
  {author} {\bibfnamefont{A.~V.}\ \bibnamefont{Zelevinsky}},\ }%
  \emph{\bibinfo {title} {Discriminants, resultants, and multidimensional
  determinants}}\ (\bibinfo {publisher} {Birkh{\"a}user, Boston},\ \bibinfo
  {year} {2008})%
  \bibAnnoteFile{NoStop}{GelfandetalDiscriminants}%
\bibitem{BennettetalUPB}%
  \BibitemOpen
  \bibfield{author}{%
  \bibinfo {author} {\bibfnamefont{C.~H.}\ \bibnamefont{Bennett}}, \bibinfo
  {author} {\bibfnamefont{D.~P.}\ \bibnamefont{DiVincenzo}}, \bibinfo {author}
  {\bibfnamefont{T.}~\bibnamefont{Mor}}, \bibinfo {author}
  {\bibfnamefont{P.~W.}\ \bibnamefont{Shor}}, \bibinfo {author}
  {\bibfnamefont{J.~A.}\ \bibnamefont{Smolin}},\ and\ \bibinfo {author}
  {\bibfnamefont{B.~M.}\ \bibnamefont{Terhal}},\ }%
  \bibfield{journal}{%
  \Doi{10.1103/PhysRevLett.82.5385}{\bibinfo {journal} {Phys. Rev. Lett.}}\ }%
  \textbf{\bibinfo {volume} {82}},\ \bibinfo {pages} {5385} (\bibinfo {month}
  {Jun}\ \bibinfo {year} {1999})%
  \bibAnnoteFile{NoStop}{BennettetalUPB}%
\bibitem{SchrodingerMixtureThm}%
  \BibitemOpen
  \bibfield{author}{%
  \bibinfo {author} {\bibfnamefont{E.}~\bibnamefont{Schr\"odinger}},\ }%
  \bibfield{journal}{%
  \Doi{10.1017/S0305004100019137}{\bibinfo {journal} {Math. Proc. Camb. Phil.
  Soc.}}\ }%
  \textbf{\bibinfo {volume} {32}},\ \bibinfo {pages} {446} (\bibinfo {year}
  {1936})%
  \bibAnnoteFile{NoStop}{SchrodingerMixtureThm}%
\bibitem{GisinMixtureThm}%
  \BibitemOpen
  \bibfield{author}{%
  \bibinfo {author} {\bibfnamefont{N.}~\bibnamefont{Gisin}},\ }%
  \bibfield{journal}{%
  \bibinfo {journal} {Helvetica Physica Acta}\ }%
  \textbf{\bibinfo {volume} {62}},\ \bibinfo {pages} {363} (\bibinfo {year}
  {1989})%
  \bibAnnoteFile{NoStop}{GisinMixtureThm}%
\bibitem{HughstonJozsaWoottersMixtureThm}%
  \BibitemOpen
  \bibfield{author}{%
  \bibinfo {author} {\bibfnamefont{L.~P.}\ \bibnamefont{Hughston}}, \bibinfo
  {author} {\bibfnamefont{R.}~\bibnamefont{Jozsa}},\ and\ \bibinfo {author}
  {\bibfnamefont{W.~K.}\ \bibnamefont{Wootters}},\ }%
  \bibfield{journal}{%
  \Doi{10.1016/0375-9601(93)90880-9}{\bibinfo {journal} {Phys. Lett. A}}\ }%
  \textbf{\bibinfo {volume} {183}},\ \bibinfo {pages} {14 } (\bibinfo {year}
  {1993})%
  \bibAnnoteFile{NoStop}{HughstonJozsaWoottersMixtureThm}%
\bibitem{UhlmannOptimalDecomp}%
  \BibitemOpen
  \bibfield{author}{%
  \bibinfo {author} {\bibfnamefont{A.}~\bibnamefont{Uhlmann}},\ }%
  \bibfield{journal}{%
  \bibinfo {journal} {Open Sys. Inf. Dyn.}\ }%
  \textbf{\bibinfo {volume} {5}},\ \bibinfo {pages} {209} (\bibinfo {year}
  {1998}),\ ISSN \bibinfo {issn} {1230-1612}%
  \bibAnnoteFile{NoStop}{UhlmannOptimalDecomp}%
\bibitem{PeresCrit}%
  \BibitemOpen
  \bibfield{author}{%
  \bibinfo {author} {\bibfnamefont{A.}~\bibnamefont{Peres}},\ }%
  \bibfield{journal}{%
  \Doi{10.1103/PhysRevLett.77.1413}{\bibinfo {journal} {Phys. Rev. Lett.}}\ }%
  \textbf{\bibinfo {volume} {77}},\ \bibinfo {pages} {1413} (\bibinfo {month}
  {Aug}\ \bibinfo {year} {1996})%
  \bibAnnoteFile{NoStop}{PeresCrit}%
\bibitem{HorodeckiPosMapWitness}%
  \BibitemOpen
  \bibfield{author}{%
  \bibinfo {author} {\bibfnamefont{M.}~\bibnamefont{Horodecki}}, \bibinfo
  {author} {\bibfnamefont{P.}~\bibnamefont{Horodecki}},\ and\ \bibinfo {author}
  {\bibfnamefont{R.}~\bibnamefont{Horodecki}},\ }%
  \bibfield{journal}{%
  \Doi{10.1016/S0375-9601(96)00706-2}{\bibinfo {journal} {Phys. Lett. A}}\ }%
  \textbf{\bibinfo {volume} {223}},\ \bibinfo {pages} {1 } (\bibinfo {year}
  {1996})%
  \bibAnnoteFile{NoStop}{HorodeckiPosMapWitness}%
\bibitem{RaggioTsallis}%
  \BibitemOpen
  \bibfield{author}{%
  \bibinfo {author} {\bibfnamefont{G.~A.}\ \bibnamefont{Raggio}},\ }%
  \bibfield{journal}{%
  \Doi{10.1063/1.530920}{\bibinfo {journal} {J. Math. Phys.}}\ }%
  \textbf{\bibinfo {volume} {36}},\ \bibinfo {pages} {4785} (\bibinfo {year}
  {1995})%
  \bibAnnoteFile{NoStop}{RaggioTsallis}%
\bibitem{AudenaertTsallisSubadd}%
  \BibitemOpen
  \bibfield{author}{%
  \bibinfo {author} {\bibfnamefont{K.~M.~R.}\ \bibnamefont{Audenaert}},\ }%
  \bibfield{journal}{%
  \Doi{10.1063/1.2771542}{\bibinfo {journal} {J. Math. Phys.}}\ }%
  \textbf{\bibinfo {volume} {48}},\ \bibinfo {eid} {083507} (\bibinfo {year}
  {2007})%
  \bibAnnoteFile{NoStop}{AudenaertTsallisSubadd}%
\bibitem{OhyaPetzQEntr}%
  \BibitemOpen
  \bibfield{author}{%
  \bibinfo {author} {\bibfnamefont{M.}~\bibnamefont{Ohya}}\ and\ \bibinfo
  {author} {\bibfnamefont{D.}~\bibnamefont{Petz}},\ }%
  \emph{\bibinfo {title} {Quantum Entropy and Its Use}},\ \bibinfo {edition}
  {1st}\ ed.\ (\bibinfo {publisher} {Springer Verlag},\ \bibinfo {year}
  {1993})%
  \bibAnnoteFile{NoStop}{OhyaPetzQEntr}%
\bibitem{Petzfdivergence}%
  \BibitemOpen
  \bibfield{author}{%
  \bibinfo {author} {\bibfnamefont{D.}~\bibnamefont{Petz}},\ }%
  \bibfield{journal}{%
  \Doi{10.3390/e12030304}{\bibinfo {journal} {Entropy}}\ }%
  \textbf{\bibinfo {volume} {12}},\ \bibinfo {pages} {304} (\bibinfo {year}
  {2010})%
  \bibAnnoteFile{NoStop}{Petzfdivergence}%
\bibitem{FuruichiTsallis}%
  \BibitemOpen
  \bibfield{author}{%
  \bibinfo {author} {\bibfnamefont{S.}~\bibnamefont{Furuichi}},\ }%
  in\ \emph{\bibinfo {booktitle} {Aspects of Optical Sciences and Quantum
  Information}}\ (\bibinfo {publisher} {Research Signpost},\ \bibinfo {year}
  {2007})\ pp.\ \bibinfo {pages} {1--86}%
  \bibAnnoteFile{NoStop}{FuruichiTsallis}%
\bibitem{oeisA000110}%
  \BibitemOpen
  \enquote{\bibinfo {title} {The {O}n-{L}ine {E}ncyclopedia of {I}nteger
  {S}equences, {A}000110},}\ \bibinfo {note} {Bell or exponential numbers: ways
  of placing n labeled balls into n indistinguishable boxes},\
  \url{http://oeis.org/A000110}%
  \bibAnnoteFile{NoStop}{oeisA000110}%
\bibitem{oeisA000041}%
  \BibitemOpen
  \enquote{\bibinfo {title} {The {O}n-{L}ine {E}ncyclopedia of {I}nteger
  {S}equences, {A}000041},}\ \bibinfo {note} {Number of partitions of n},\
  \url{http://oeis.org/A000041}%
  \bibAnnoteFile{NoStop}{oeisA000041}%
\bibitem{HorodeckiEntMeas}%
  \BibitemOpen
  \bibfield{author}{%
  \bibinfo {author} {\bibfnamefont{M.}~\bibnamefont{Horodecki}},\ }%
  \bibfield{journal}{%
  \bibinfo {journal} {Quant. Inf. Comp.}\ }%
  \textbf{\bibinfo {volume} {1}},\ \bibinfo {pages} {3} (\bibinfo {month}
  {May}\ \bibinfo {year} {2001})%
  \bibAnnoteFile{NoStop}{HorodeckiEntMeas}%
\bibitem{VidalEntMon}%
  \BibitemOpen
  \bibfield{author}{%
  \bibinfo {author} {\bibfnamefont{G.}~\bibnamefont{Vidal}},\ }%
  \bibfield{journal}{%
  \Doi{10.1080/09500340008244048}{\bibinfo {journal} {J. Mod. Opt.}}\ }%
  \textbf{\bibinfo {volume} {47}},\ \bibinfo {pages} {355} (\bibinfo {year}
  {2000})%
  \bibAnnoteFile{NoStop}{VidalEntMon}%
\bibitem{EltschkaetalEntMon}%
  \BibitemOpen
  \bibfield{author}{%
  \bibinfo {author} {\bibfnamefont{C.}~\bibnamefont{Eltschka}}, \bibinfo
  {author} {\bibfnamefont{T.}~\bibnamefont{Bastin}}, \bibinfo {author}
  {\bibfnamefont{A.}~\bibnamefont{Osterloh}},\ and\ \bibinfo {author}
  {\bibfnamefont{J.}~\bibnamefont{Siewert}},\ }%
  \bibfield{journal}{%
  \Doi{10.1103/PhysRevA.85.022301}{\bibinfo {journal} {Phys. Rev. A}}\ }%
  \textbf{\bibinfo {volume} {85}},\ \bibinfo {pages} {022301} (\bibinfo {month}
  {Feb}\ \bibinfo {year} {2012})%
  \bibAnnoteFile{NoStop}{EltschkaetalEntMon}%
\bibitem{Dohertycrit1}%
  \BibitemOpen
  \bibfield{author}{%
  \bibinfo {author} {\bibfnamefont{A.~C.}\ \bibnamefont{Doherty}}, \bibinfo
  {author} {\bibfnamefont{P.~A.}\ \bibnamefont{Parrilo}},\ and\ \bibinfo
  {author} {\bibfnamefont{F.~M.}\ \bibnamefont{Spedalieri}},\ }%
  \bibfield{journal}{%
  \Doi{10.1103/PhysRevLett.88.187904}{\bibinfo {journal} {Phys. Rev. Lett.}}\
  }%
  \textbf{\bibinfo {volume} {88}},\ \bibinfo {pages} {187904} (\bibinfo {month}
  {Apr}\ \bibinfo {year} {2002})%
  \bibAnnoteFile{NoStop}{Dohertycrit1}%
\bibitem{Dohertycrit2}%
  \BibitemOpen
  \bibfield{author}{%
  \bibinfo {author} {\bibfnamefont{A.~C.}\ \bibnamefont{Doherty}}, \bibinfo
  {author} {\bibfnamefont{P.~A.}\ \bibnamefont{Parrilo}},\ and\ \bibinfo
  {author} {\bibfnamefont{F.~M.}\ \bibnamefont{Spedalieri}},\ }%
  \bibfield{journal}{%
  \Doi{10.1103/PhysRevA.69.022308}{\bibinfo {journal} {Phys. Rev. A}}\ }%
  \textbf{\bibinfo {volume} {69}},\ \bibinfo {pages} {022308} (\bibinfo {month}
  {Feb}\ \bibinfo {year} {2004})%
  \bibAnnoteFile{NoStop}{Dohertycrit2}%
\bibitem{Dohertycrit3}%
  \BibitemOpen
  \bibfield{author}{%
  \bibinfo {author} {\bibfnamefont{A.~C.}\ \bibnamefont{Doherty}}, \bibinfo
  {author} {\bibfnamefont{P.~A.}\ \bibnamefont{Parrilo}},\ and\ \bibinfo
  {author} {\bibfnamefont{F.~M.}\ \bibnamefont{Spedalieri}},\ }%
  \bibfield{journal}{%
  \Doi{10.1103/PhysRevA.71.032333}{\bibinfo {journal} {Phys. Rev. A}}\ }%
  \textbf{\bibinfo {volume} {71}},\ \bibinfo {pages} {032333} (\bibinfo {month}
  {Mar}\ \bibinfo {year} {2005})%
  \bibAnnoteFile{NoStop}{Dohertycrit3}%
\bibitem{HuberkCrit}%
  \BibitemOpen
  \bibfield{author}{%
  \bibinfo {author} {\bibfnamefont{A.}~\bibnamefont{Gabriel}}, \bibinfo
  {author} {\bibfnamefont{B.~C.}\ \bibnamefont{Hiesmayr}},\ and\ \bibinfo
  {author} {\bibfnamefont{M.}~\bibnamefont{Huber}},\ }%
  \bibfield{journal}{%
  \bibinfo {journal} {Quant. Inf. Comp.}\ }%
  \textbf{\bibinfo {volume} {10}},\ \bibinfo {pages} {829} (\bibinfo {year}
  {2010})%
  \bibAnnoteFile{NoStop}{HuberkCrit}%
\bibitem{HuberCrit2}%
  \BibitemOpen
  \bibfield{author}{%
  \bibinfo {author} {\bibfnamefont{M.}~\bibnamefont{Huber}}, \bibinfo {author}
  {\bibfnamefont{F.}~\bibnamefont{Mintert}}, \bibinfo {author}
  {\bibfnamefont{A.}~\bibnamefont{Gabriel}},\ and\ \bibinfo {author}
  {\bibfnamefont{B.~C.}\ \bibnamefont{Hiesmayr}},\ }%
  \bibfield{journal}{%
  \Doi{10.1103/PhysRevLett.104.210501}{\bibinfo {journal} {Phys. Rev. Lett.}}\
  }%
  \textbf{\bibinfo {volume} {104}},\ \bibinfo {pages} {210501} (\bibinfo
  {month} {May}\ \bibinfo {year} {2010})%
  \bibAnnoteFile{NoStop}{HuberCrit2}%
\bibitem{GuhneSevinckCrit}%
  \BibitemOpen
  \bibfield{author}{%
  \bibinfo {author} {\bibfnamefont{O.}~\bibnamefont{G\"uhne}}\ and\ \bibinfo
  {author} {\bibfnamefont{M.}~\bibnamefont{Seevinck}},\ }%
  \bibfield{journal}{%
  \bibinfo {journal} {N. J. Phys.}\ }%
  \textbf{\bibinfo {volume} {12}},\ \bibinfo {pages} {053002} (\bibinfo {year}
  {2010})%
  \bibAnnoteFile{NoStop}{GuhneSevinckCrit}%
\bibitem{BorstenDuffLevayBHQB}%
  \BibitemOpen
  \bibfield{author}{%
  \bibinfo {author} {\bibfnamefont{L.}~\bibnamefont{Borsten}}, \bibinfo
  {author} {\bibfnamefont{M.~J.}\ \bibnamefont{Duff}},\ and\ \bibinfo {author}
  {\bibfnamefont{P.}~\bibnamefont{L\'evay}}}%
   (\bibinfo {year} {2012}),\
  \Eprint{http://arxiv.org/abs/1206.3166}{arXiv:1206.3166 [hep-th]}%
  \bibAnnoteFile{NoStop}{BorstenDuffLevayBHQB}%
\bibitem{PeterPetiFTS}%
  \BibitemOpen
  \bibfield{author}{%
  \bibinfo {author} {\bibfnamefont{P.}~\bibnamefont{L\'evay}}\ and\ \bibinfo
  {author} {\bibfnamefont{P.}~\bibnamefont{Vrana}},\ }%
  \bibfield{journal}{%
  \Doi{10.1103/PhysRevA.78.022329}{\bibinfo {journal} {Phys. Rev. A}}\ }%
  \textbf{\bibinfo {volume} {78}},\ \bibinfo {pages} {022329} (\bibinfo {month}
  {Aug}\ \bibinfo {year} {2008})%
  \bibAnnoteFile{NoStop}{PeterPetiFTS}%
\bibitem{PetiPeterFTS}%
  \BibitemOpen
  \bibfield{author}{%
  \bibinfo {author} {\bibfnamefont{P.}~\bibnamefont{Vrana}}\ and\ \bibinfo
  {author} {\bibfnamefont{P.}~\bibnamefont{L\'evay}},\ }%
  \bibfield{journal}{%
  \bibinfo {journal} {J. Phys. A}\ }%
  \textbf{\bibinfo {volume} {42}},\ \bibinfo {pages} {285303} (\bibinfo {year}
  {2009})%
  \bibAnnoteFile{NoStop}{PetiPeterFTS}%
\bibitem{PetiPriv}%
  \BibitemOpen
  \bibfield{author}{%
  \bibinfo {author} {\bibfnamefont{P.}~\bibnamefont{Vrana}},\ }%
  \bibinfo {note} {private communication}%
  \bibAnnoteFile{NoStop}{PetiPriv}%
\bibitem{ChenDjokovicSemialg}%
  \BibitemOpen
  \bibfield{author}{%
  \bibinfo {author} {\bibfnamefont{L.}~\bibnamefont{Chen}}\ and\ \bibinfo
  {author} {\bibfnamefont{D.~Z.}\ \bibnamefont{Djokovic}}}%
   (\bibinfo {year} {2012}),\
  \Eprint{http://arxiv.org/abs/1206.3775}{arXiv:1206.3775 [quant-ph]}%
  \bibAnnoteFile{NoStop}{ChenDjokovicSemialg}%
\bibitem{HWLUA}%
  \BibitemOpen
  \bibfield{author}{%
  \bibinfo {author} {\bibfnamefont{M.~W.}\ \bibnamefont{Hero}}\ and\ \bibinfo
  {author} {\bibfnamefont{J.~F.}\ \bibnamefont{Willenbring}},\ }%
  \bibfield{journal}{%
  \Doi{DOI: 10.1016/j.disc.2009.06.021}{\bibinfo {journal} {Discr. Math.}}\ }%
  \textbf{\bibinfo {volume} {309}},\ \bibinfo {pages} {6508 } (\bibinfo {year}
  {2009})%
  \bibAnnoteFile{NoStop}{HWLUA}%
\bibitem{HWWLUA}%
  \BibitemOpen
  \bibfield{author}{%
  \bibinfo {author} {\bibfnamefont{M.~W.}\ \bibnamefont{Hero}}, \bibinfo
  {author} {\bibfnamefont{J.~F.}\ \bibnamefont{Willenbring}},\ and\ \bibinfo
  {author} {\bibfnamefont{L.~K.}\ \bibnamefont{Williams}}}%
   (\bibinfo {year} {2009}),\
  \Eprint{http://arxiv.org/abs/0911.0222}{arXiv:0911.0222 [math.RT]}%
  \bibAnnoteFile{NoStop}{HWWLUA}%
\bibitem{PetiLUA1}%
  \BibitemOpen
  \bibfield{author}{%
  \bibinfo {author} {\bibfnamefont{P.}~\bibnamefont{Vrana}},\ }%
  \bibfield{journal}{%
  \bibinfo {journal} {J. Phys. A}\ }%
  \textbf{\bibinfo {volume} {44}},\ \bibinfo {pages} {115302} (\bibinfo {year}
  {2011})%
  \bibAnnoteFile{NoStop}{PetiLUA1}%
\bibitem{PetiLUA23}%
  \BibitemOpen
  \bibfield{author}{%
  \bibinfo {author} {\bibfnamefont{P.}~\bibnamefont{Vrana}},\ }%
  \bibfield{journal}{%
  \bibinfo {journal} {J. Phys. A}\ }%
  \textbf{\bibinfo {volume} {44}},\ \bibinfo {pages} {225304} (\bibinfo {year}
  {2011})%
  \bibAnnoteFile{NoStop}{PetiLUA23}%
\bibitem{SzDeg6}%
  \BibitemOpen
  \bibfield{author}{%
  \bibinfo {author} {\bibfnamefont{{\relax Sz}.}~\bibnamefont{{Sz}alay}},\ }%
  \bibfield{journal}{%
  \bibinfo {journal} {J. Phys. A}\ }%
  \textbf{\bibinfo {volume} {45}},\ \bibinfo {pages} {065302} (\bibinfo {year}
  {2012})%
  \bibAnnoteFile{NoStop}{SzDeg6}%
\bibitem{WoottersConc}%
  \BibitemOpen
  \bibfield{author}{%
  \bibinfo {author} {\bibfnamefont{W.~K.}\ \bibnamefont{Wootters}},\ }%
  \bibfield{journal}{%
  \Doi{10.1103/PhysRevLett.80.2245}{\bibinfo {journal} {Phys. Rev. Lett.}}\ }%
  \textbf{\bibinfo {volume} {80}},\ \bibinfo {pages} {2245} (\bibinfo {month}
  {Mar}\ \bibinfo {year} {1998})%
  \bibAnnoteFile{NoStop}{WoottersConc}%
\bibitem{HillWoottersConc}%
  \BibitemOpen
  \bibfield{author}{%
  \bibinfo {author} {\bibfnamefont{S.}~\bibnamefont{Hill}}\ and\ \bibinfo
  {author} {\bibfnamefont{W.~K.}\ \bibnamefont{Wootters}},\ }%
  \bibfield{journal}{%
  \Doi{10.1103/PhysRevLett.78.5022}{\bibinfo {journal} {Phys. Rev. Lett.}}\ }%
  \textbf{\bibinfo {volume} {78}},\ \bibinfo {pages} {5022} (\bibinfo {month}
  {Jun}\ \bibinfo {year} {1997})%
  \bibAnnoteFile{NoStop}{HillWoottersConc}%
\end{thebibliography}%

\end{document}